\documentclass[twocolumn, usenatbib]{mnras}
\usepackage{savesym}
\usepackage{graphicx}
\usepackage{longtable}
\usepackage{changepage}

\expandafter\let\csname equation*\endcsname\relax
  \expandafter\let\csname endequation*\endcsname\relax 
\usepackage{subfig}
\usepackage{amsmath}
\usepackage{amssymb}
\usepackage{verbatim}
\usepackage[yyyymmdd,hhmmss]{datetime}
\usepackage{array}
\usepackage{times}
\usepackage{xcolor}

\DeclareRobustCommand{\DE}[3]{#2}
\let\DEthebibliography\thebibliography
\def\thebibliography{\DeclareRobustCommand{\DE}[3]{##3}\DEthebibliography}

\newcommand{\chandra}{\textit{Chandra}}

\newcommand{\xmm}{{\it XMM-Newton}}

\title[Spin-plunge degeneracies in XRBs]{ A rapid black hole spin or emission from the plunging region?   }
\author [Andrew Mummery, et al.]{Andrew Mummery$^1$\thanks{E-mail:
andrew.mummery@physics.ox.ac.uk}, Jiachen Jiang$^2$, Adam Ingram$^3$, Andrew Fabian$^4$, Jake Rule$^5$ 
\\
$^1$Oxford Theoretical Physics, Beecroft Building,  Clarendon Laboratory, Parks Road, Oxford, OX1 3PU, United Kingdom \\
$^2${Department of Physics, University of Warwick, Gibbet Hill Road, Coventry CV4 7AL, United Kingdom}\\
$^3$School of Mathematics, Statistics and Physics, Newcastle University, Herschel Building, Newcastle upon Tyne, NE1 7RU, United Kingdom \\
$^4$Institute of Astronomy, Madingley Road, Cambridge CB3 0HA, United Kingdom \\
$^5$Oxford Astrophysics, Denys Wilkinson Building,  Keble Road, Oxford, OX1 3RH, United Kingdom
}

\date{}

\begin{document}

\pagerange{\pageref{firstpage}--\pageref{lastpage}} \pubyear{2024}

\maketitle

\label{firstpage}

\begin{abstract} 
Emission from within the plunging region of black hole accretion flows has recently been detected in two X-ray binary systems. There is, furthermore, a possible discrepancy between the inferred spins of gravitational wave and electromagnetically detected black holes. Motivated by these two results we demonstrate, using theoretical calculations, numerical simulations and observational data, that the inclusion of emission from within the innermost stable circular orbit (ISCO) results in a black hole with a low spin producing a thermal continuum X-ray spectrum that mimics that produced by a much more rapidly rotating black hole surrounded by a disk with no emission from within the ISCO.  We demonstrate this explicitly using the observed X-ray spectrum of a canonical soft-state high mass X-ray binary system M33 X-7. A vanishing ISCO temperature model requires a high spin $a_\bullet = 0.84\pm0.05$, as has been found previously in the literature. However, a disk around a Schwarzschild black hole can equally well (in fact slightly better) describe the data, provided that photons emitted from within the plunging region are included, and the ISCO stress is in line with that seen in numerical simulations of the accretion process. We then present an analysis of  two further soft-state X-ray binaries (MAXI J1820+070 and MAXI J0637$-$430)  which require the presence of intra-ISCO emission at high statistical significance. These two sources sit on the low-spin moderate-stress part of the degeneracy exhibited by M33 X-7, suggesting that when high quality data are available the high-spin low-stress region of parameter space is ruled out. We discuss how future advances in numerical simulations and data modeling will be essential to determining the spin of X-ray binary black holes which may well be systematically lower than current continuum fitting methods suggest. 
\end{abstract}

\begin{keywords}
accretion, accretion discs --- black hole physics --- X-rays: binaries
\end{keywords}
\noindent

\section{Introduction} 
Black holes are the  simplest macroscopic objects known to physics, completely described (in an astrophysical setting) by two numbers: their mass $M_\bullet$ and their angular momentum $J_\bullet$, or more typically their dimensionless spin parameter $a_\bullet \equiv c J_\bullet / GM_\bullet^2$, which satisfies $|a_\bullet|<1$ for a black hole. The mass parameter sets the large scale properties of the black hole's gravitational field, and so can be constrained in a number of ways not particularly sensitive to horizon-scale physics. The angular momentum of the black hole however only impacts the small radius properties of it's spacetime, with the leading order contributions to the metric scaling as $\sim a_\bullet M_\bullet^{1/2}/r^{3/2}$. This, naturally, makes spin inference more difficult than mass inference for astrophysical black hole systems, as a detailed understanding of the near-horizon, highly relativistic, behavior of the observed system is required. 

There are two well established electromagnetic approaches to constraining the spin of a black hole---the relativistic reflection \citep[e.g.,][or \citealt{Reynolds19} for a review]{Fabian1989} and continuum fitting \citep[][for a review]{McClintock14} methods. Both of these techniques are supposed to work in a similar manner: by measuring properties of the emission from the innermost regions of the accretion disk, one attempts to determine where ``the inner edge'' of the disk lies and, from that, infer the black hole's spin using a simple relationship between the innermost stable circular orbit (where it is {\it assumed} that the disk ``ends'') and the black hole spin \citep[e.g.][]{Bardeen72}. 

A possible problem with this framework is immediately apparent: what if the innermost stable circular orbit (ISCO) is not actually where the disk ends? Of course, in some sense it is obvious that the disk does not end at the ISCO. The accreting fluid inevitably travels from the ISCO onto the black hole itself, and it is undeniable that there is plasma between the ISCO and horizon. The relevant question, for both approaches, is whether this plasma plunging from the ISCO is detectable, and if so how important it is. 

The physics of relativistic reflection within this ``plunging region'' is complex, as it in effect probes a combination of the density and velocity of the disk,  and the external ionizing flux hitting the flow (assumed to originate from a hot corona). Pioneering calculations going back to \cite{Reynolds97} have shown that the plunging region may well be important for reflection studies.  On the other hand continuum fitting is a conceptually simpler approach, with the radius-dependent emitted luminosity of the flow the only property of the disk which contributes. For the remainder of this paper we shall focus exclusively on the continuum fitting approach. 

It is by now well known that the assumption that the continuum emission from a black hole accretion disk terminates at the ISCO is incorrect, and that conventional models which are fit to data are missing physics. The question remains what precise impact this extra emission has on spin inferences, as existing numerical simulations disagree (strongly) on the importance of this region for constraining black hole spins \citep{KrolikHawley02, Noble11, Zhu12, Lancova19}. 

Neglecting emission from within the plunging region is an approximation that goes back to the very first papers on relativistic accretion disk theory \citep{NovikovThorne73, PageThorne74, Thorne1974} where it was noted \citep[in a footnote in][]{Thorne1974} that neglecting this emission would be a bad approximation if magnetic fields played a role in the accretion process. We now know that magnetic fields are {\it essential} to the accretion process \citep{BalbusHawley91}, and numerous numerical simulations of the accretion process over the years have all found substantial emission from within the plunging region \citep[e.g.,][among many others]{HawleyKrolik01, Noble10, Penna10, Noble11, Schnittman16, Lancova19, Dhang25, Rule25}.

While the existence of bright emission from within the plunging region has been a known, numerical, fact for a long time, it is only recently that semi-analytical models (i.e., those models which can actually be feasibly fit to data) which reproduce the results of these simulations have become available \citep{MummeryBalbus2023, MummeryMori24, Mummery24Plunge, Mummery24PlungeB}. This has lead to the robust detection (at high statistical significance) of emission from within the plunging region in two (low mass) X-ray binary systems with sufficiently high quality data, namely MAXI J820+070 \citep[][where plunging region emission was first suspected in \citealt{Fabian20}]{Mummery24Plunge} and MAXI J0637$-$430 \citep{Mummery24PlungeB}. This semi-analytical model (named {\tt fullkerr} and publicly available) can be used to examine the robustness of black hole spin constraints from the continuum fitting approach. 

It is the purpose of this paper to look at the implications of these extended models in sources which do not necessarily have the data quality to unambiguously detect the plunging region, but may well still suffer from spectral contamination from this region. We shall demonstrate that in these systems there is a strong degeneracy between emission from the plunging region and the black hole spin, with disks around black holes with (much) lower spins and a bright plunging region able to reproduce X-ray data that requires -- in the no plunging region limit -- much more rapid rotation. While this argument has been made before \citep[e.g.][]{AgolKrolik00, KrolikHawley02, Noble11, Lancova19} in particular by \cite{Davis05, Davis06} who argued (correctly) that spin measurements were highly sensitive to the properties of the inner disk (but did not at that time have a model for the intra-ISCO emission), we believe that the combination of numerical, analytical and observational approaches used in this paper make the case for this effect convincingly. 

The layout of this paper is as follows. In Section \ref{physics} we discuss the physics of a black hole spin-plunging region degeneracy. We introduce the data (of the X-ray binaries M33 X-7, MAXI J1820+070 and MAXI J0637$-$430) and models used in this work in Section \ref{obs}, before presenting our results in Section \ref{sec:results}. We conclude with a discussion in Section \ref{conc}. 

Before moving on to our main results, we discuss the possible implications of this work for gravitational wave science. 

\subsection{Gravitational waves and X-rays: a black hole spin discrepancy?}
The study of astrophysical black holes is by now a field spanning multiple ``messengers'', with routine discoveries of new systems using both electromagnetic emission and gravitational waves. Some -- but by no means all  -- of the systems/parameters discovered by these different techniques should be directly comparable, certainly in terms of their population-level properties. The most promising comparison is between gravitational wave (GW) binary black hole systems, and those electromagnetically discovered systems in so-called high mass X-ray binary (HMXRB) systems, which contain a black hole surrounded by a large mass star. This large mass star could, in principle and after undergoing further stellar evolution, also become a black hole and the binary system could merge producing a gravitational wave source. Potential discrepancies between the observed properties of the black holes in HMXRBs and those in GW binaries may point to, among other things, a point of missing physics in our analysis of either HMXRB or GW systems. 

The main discrepancy between HMXRB populations and GW binaries lies in their inferred spin parameters\footnote{They also have systematically different black hole masses, but this can be explained by GW selection effects \citealt{Fishbach22} (which favor more massive and therefore ``brighter'' [in gravitational waves] systems) and the different metalicities of their host galaxies \citealt{Belczynski24}.}, with HMXRB systems generally inferred to be rapidly rotating $a_\bullet \gtrsim 0.8$ \citep[e.g.,][]{Liu08, Zhao21}, while the individual spins in GW sources is inferred to be low $a_\bullet \lesssim 0.2$ (with higher values in the tail of the distribution). The most easy to measure ``effective'' spin parameter of GW sources is 
\begin{equation}
    \vec \chi_{\rm eff} = {M_1 \vec a_1 \cos\theta_1 + M_2 \vec a_2 \cos\theta_2\over M_1 + M_2} ,
\end{equation}
where the subscripts $1/2$ denote the two black holes in the merging pair, and $\theta_i$ is the angle the angular momentum of the $i$th black hole makes with the total angular momentum of the binary system. The distribution of $|\vec \chi_{\rm eff}|$ is peaked at $\approx 0.05$ \citep{Abbott23, KAGRA:2021vkt}, with only a small fraction of sources satisfying $|\vec \chi_{\rm eff}| > 0.3$. The discrepancy between electromagnetic and gravitational wave spin inferences is claimed to be at the $>99.9\%$ level \citep{belczynski20, Fishbach22}. One natural explanation for this is that the individual black holes in GW binaries are slowly rotating, in contention with electromagnetic inferences.

A discrepancy like this is interesting, as it either teaches us something about the different stellar evolution pathways undertaken by those black holes which end up in HMXRB systems and those that end up as GW binaries, or it tells us something about how we are analysing the electromagnetic/gravitational radiation detected from these systems (or of course both). 

Those black holes in known HMXRBs are thought to be on different stellar evolution pathways than the majority of the discovered GW binary systems. As an example, \citet{liotine23} used population synthesis modelling alongside realistic detection thresholds for \chandra\ and Advanced LIGO to predict the probability of an (electromagnetically) detectable HMXRB becoming a (gravitationally) detectable LIGO source. They argued that due to observational selection effects, there is only a 3\% chance of detecting an HMXRB hosting a black hole with mass over $35 M_{\odot}$, and just a 0.6\% chance that such a system would evolve into a GW-detectable binary within a Hubble time. On the other hand, a recent re-analysis of the optical-UV emission from the Cyg X-1 system \citep{Ramachandran25} finds a new, lower, black hole mass and, importantly, they argue that Cyg X-1 \citep[a Galactic HMXRB with very high inferred spin, e.g.,][]{Zhao21} may well undergo a merger in the next $\sim 5$ Gyr \citep[although see][for discussions regarding the physical uncertainty in this statement]{Ramachandran25}.

Independent of any findings in this active area of research there is, likely, a limit to what can be explained (in terms of parameter discrepancies) by the different origins for these two populations. The physics of the supernova explosions themselves should be universal (between populations), and it is clear that both spin measurements (GW and X-ray) are probing natal spins. X-ray binary accretion cannot spin up their black holes, owing to the order one \citep[in reality $M_{\rm final}/M_{\rm initial} = \sqrt{6} -1\approx 1.45$][]{Bardeen1970} change in black hole mass required to spin up a Schwarzschild black hole to maximal rotation. Low mass XRBs do not have sufficient companion mass to change their spin by a meaningful fraction, while high mass companion stars do not live long enough to transfer enough material to their black hole companion \citep[the relevant timescale here is the][time $(\sim {\rm few} \times 10^7$ years) for Eddington limited accretion]{Salpeter64}. 

The results of this paper likely have important implications for how much the different spin parameters of X-ray bright HMXRBs and GW binaries (need be/are) explained by their different origins.

\section{The physical origin of a spin--plunging region degeneracy}\label{physics}
As we hope these results will be of interest to both the gravitational wave and electromagnetic communities we begin this section with an overview of the key points of physics involved, before presenting a detailed discussion of this physics in the context of observations. Those readers simply interested in the implications of the plunging region on observed X-ray spectra may wish to skip to section 2.2. 

\subsection{A physical overview}
It is potentially surprising, given their observational ubiquity, but the reason {\it why} accretion disks actually accrete was an unsolved problem in the community for decades after it was realized disk accretion was an important astrophysical process. 

First invoked by \cite{LyndenBell69},  formalized by \cite{SS73, LBP74} and extended into relativity by  \cite{NovikovThorne73,PageThorne74}, original disk models invoked a ``turbulently enhanced'' {\it viscosity} to explain the microphysics of the accretion process. It is simple to show that ordinary particulate viscosity is insufficient to drive the short observed timescales of the accretion process \citep{BalbusHawley98}, and some turbulent ``enhancement'' is clearly required. The problem was that no known linear instability existed (to initiate the turbulent state), as a fluid on circular orbits about a central gravitational potential trivially satisfies the classical Rayleigh criterion for stability. 

It turns out that the presence of magnetic fields, even in the extreme weak-field limit, linearly destabilize this equilibrium \citep{BalbusHawley91}. This magneto-rotational instability drives turbulent transfer of angular momentum, and dissipation in accretion flows (which ultimately sources the photon emission which we detect). It can be shown that on large scales and to leading order this magnetohydrodynamic turbulence acts much like a simple viscosity in how it redistributes angular momentum and dissipates energy \citep{BalbPap99}. This of course does not mean that MHD turbulence is a viscosity, and in different physical regimes a viscosity and MHD turbulence can have completely different physical properties, and completely different implications for the flow. 

Important points of difference include the conservation of global topological invariants in MHD (the magnetic field helicity), and the existence of flux freezing (Alfv\'ens theorem), which means that magnetic field lines are dragged with the flow. This leads to magnetic field amplification in places in which the flow is forced into small physical regions (because, loosely speaking, field strength $\times$ area is conserved). Magnetic fields can also act as an additional energy source if they are forced down to the resistive scale, where they can ``reconnect'' (or more generally dissipate) and release energy which directly heats the surrounding gas. 

The plunging region is precisely one of these physical regimes in which it is obvious, and extremely simple, to see that MHD turbulence and a viscosity (even an ``enhanced'' one) will have completely different properties \citep[this was first discussed in detail in][]{Krolik99, Gammie99}. Once circular orbits become unstable, the fluid loses rotational support and is rapidly accelerated towards the black hole. It is well known that a viscosity cannot transfer angular momentum across such a sharp transition (hence the original choice of neglecting this region), but magnetic fields redistribute angular momentum according to correlations in the product $B^rB^\phi$ \citep{BalbPap99}. There is absolutely no reason to suspect that a point of relativity (the destabilizing of circular orbits) will lead to a vanishing correlation in radial and azimuthal magnetic fields. 

In fact, the destabilizing of circular motion is likely to {\it enhance} the field strengths of both $B^r$ and $B^\phi$ \citep[][i.e., the exact opposite of what is often assumed]{Krolik99, Gammie99}. This is because the flow is channeled (by gravity) into an ever smaller region, dragging field lines with it (Alfv\'ens theorem). Forcing more magnetic field lines into a smaller region trivially enhances the magnetic field strength.  This channeling  is due both to the increasing radial acceleration as rotational support is removed, but also the growing vertical component of gravity close to the event horizon \citep{Abramowicz97} which crushes the flow towards the equatorial plane. Forcing the flow into a small region can also lead to magnetic reconnection \citep[as seen in numerical simulations][]{Penna10, Rule25} which further heats the gas. 

It is obvious therefore that classical viscous models will be, in effect, completely wrong in this region, as there is no way in which they can self consistently handle the growing field strength and large scale transport inherent to magnetic fields. The fact that numerical simulations confirm this repeatedly \citep{Noble10, Penna10,  Schnittman16, Lancova19, Dhang25, Rule25} is therefore not a surprise.

\begin{figure}
    \centering
    \includegraphics[width=\linewidth]{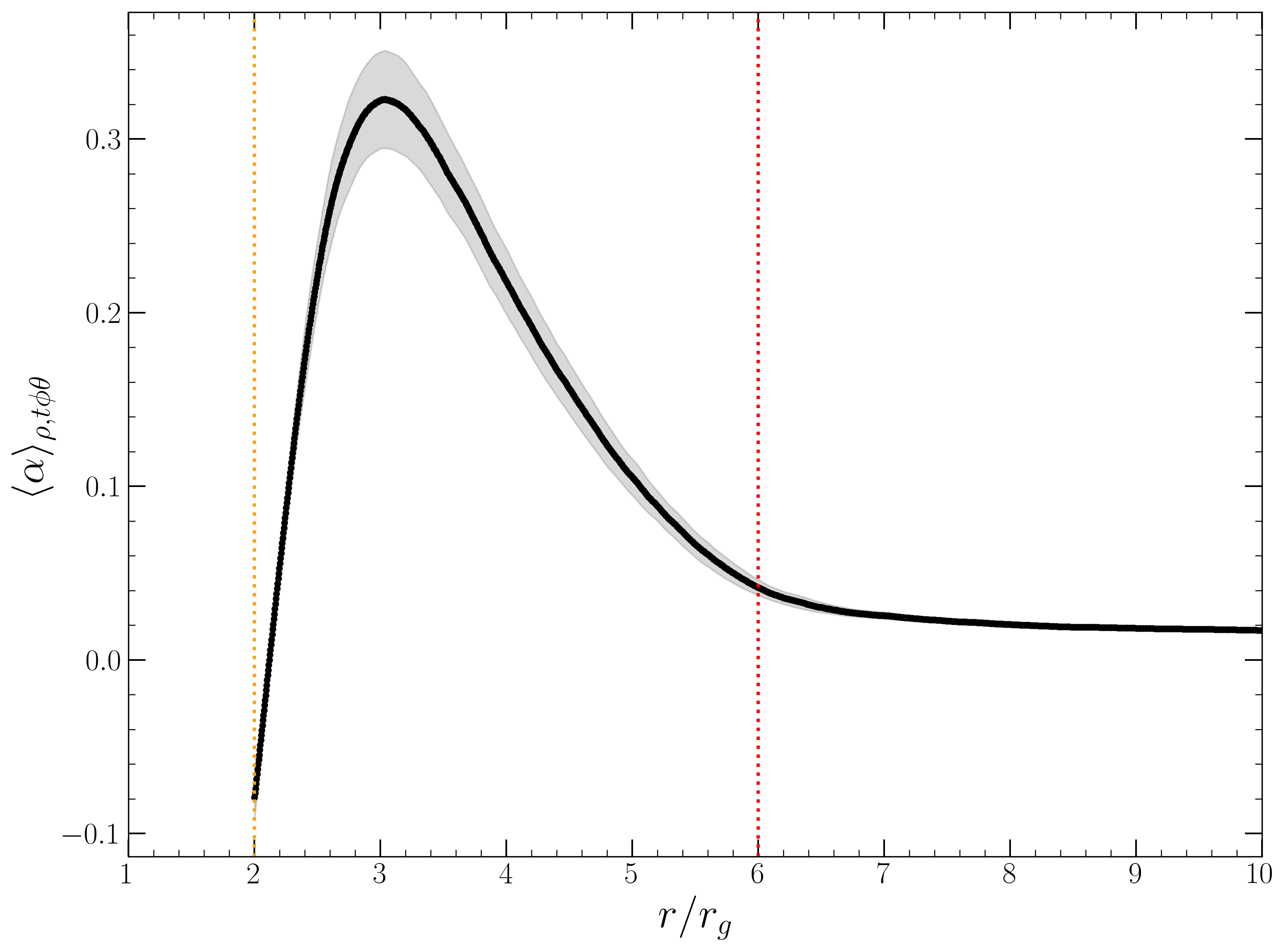}
    \includegraphics[width=\linewidth]{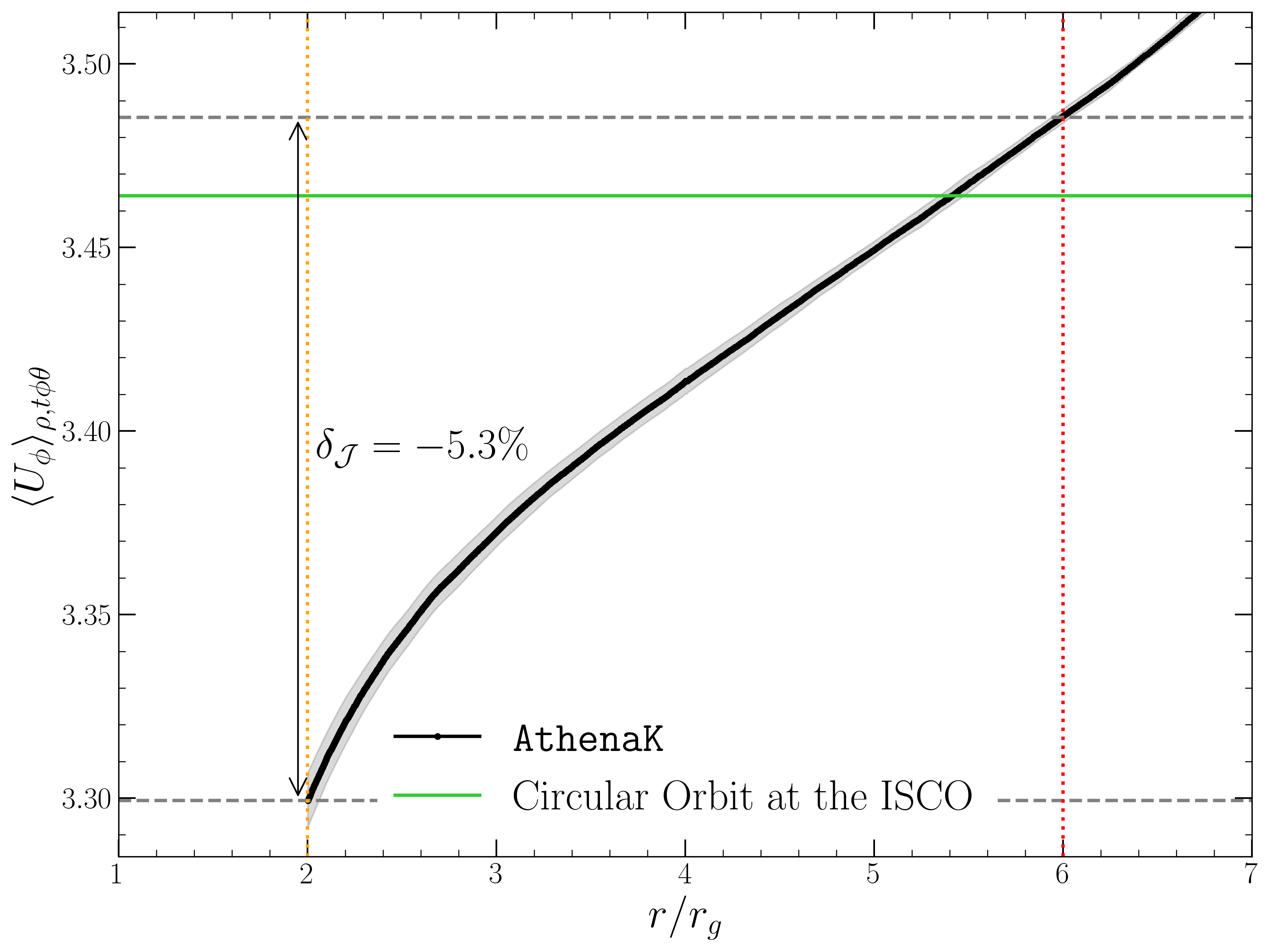}
    \caption{The physics of angular momentum transport within the ISCO, as seen in GRMHD simulations of thin disks. The plunging region is denoted by vertical dotted lines. In the lower panel we show the specific angular momentum of fluid elements, density weighted and then averaged over time and angles, as a function of radius. Turbulence transports $\sim 5\%$ of the ISCO angular momentum back to the main body of the disk. This relatively small angular momentum flux, when coupled to the very large shear of the disk in these relativistic regions, produces a large energy flux near the ISCO, leading to a significant amount of thermal emission being produced. In the upper panel we show the ``$\alpha$-parameter'' one would infer from the magnetic stresses acting on the disk, at each radius. Flux-freezing, a purely magnetohydrodynamic effect, leads to amplified magnetic stresses in this region as field lines are dragged into the plunging region by the rapidly accelerating flow, and an ``$\alpha$-parameter'' which grows by over an order of magnitude. The drop below zero of $\alpha$ at the smallest radii is a simple consequence of causality at an event horizon, as angular momentum can only be transported inwards. The behavior of accretion disk models within the plunging region which assume constant-$\alpha$ lack physical content. Figure reproduced from \citealt{Rule25}.   }
    \label{fig:physics}
\end{figure}

It has been recently argued \citep{Lasota24} that viscous so-called  ``slim $\alpha$-disk'' models should be used within the ISCO, despite their inability to reproduce the emission seen in any GRMHD simulations.  While, as we have discussed, viscous models have no real physical meaning in this regime, one must still verify why they they do not reproduce full General Relativistic Magnetohydrodynamic (GRMHD) simulations explicitly. The ``strength'' of the viscosity in classical models is characterised by a dimensionless number $\alpha$ \citep{SS73}, which is asserted to be a constant, and can be measured in simulations. We reproduce the results of the high-resolution GRMHD simulations of a thin disk about a Schwarzschild black hole presented in \cite{Rule25} in Figure \ref{fig:physics}. The $\alpha$ parameter is defined by the ratio of the $r\phi$ component of the magnetic stress tensor to the total pressure 
\begin{equation}
\alpha \equiv {\left\langle T^{r\phi}_{\rm mag} \right\rangle \over \left\langle P \right\rangle}  =  - {\left\langle B^rB^\phi \right\rangle\over 4\pi \left\langle P \right\rangle} ,
\end{equation}
where the angled brackets $\left\langle \cdot \right\rangle$, indicate an average over time and space, with all quantities evaluated in the rest frame of the fluid \citep[see][for the details of how this computation is performed]{Rule25}. In Figure \ref{fig:physics} we see the entirely predictable \citep{Krolik99} result that $\alpha$ grows by over an order of magnitude across the ISCO towards the horizon. Clearly then, the turbulent transfer of angular momentum within the ISCO is a highly non-trivial physical problem, one to which simple constant $\alpha$ models offer no insight. 

Fortunately, the amount of angular momentum transported within the ISCO is a computable quantity within GRMHD simulations, and therefore the energy flux back into the main body of the disk can be calibrated against these simulations, and used as physical input into spectral fitting models. In Figure \ref{fig:physics} we display the specific angular momentum of fluid elements, density weighted and then averaged over time and angles, as a function of radius. Turbulence transports $\sim 5\%$ of the ISCO angular momentum back to the main body of the disk. This is far higher than the amount which would be estimated in e.g., a slim disk model with the $\alpha$ parameter measured just outside of the ISCO. This relatively small angular momentum flux, when coupled to the very large shear of the disk in these relativistic regions, produces a large energy flux near the ISCO, leading to a significant amount of thermal emission being produced.

The results of Figure \ref{fig:physics} \citep[along with many other GRMHD studies][which find near-identical results]{Noble10, Noble11, Schnittman16, Lancova19, Dhang25} show that one cannot resort to $\alpha$-models, regardless of any added complexity, to compute the properties of accretion flows within the ISCO. The purpose of recently developed models \citep{MummeryBalbus2023,Mummery24Plunge} (which form the basis of the spectral fitting package {\tt fullkerr}) is to allow the angular momentum flux from within the ISCO to be a free, input, parameter of the theory. This model then computes the solutions of the disk thermodynamic equations, for a given angular momentum flux, which can then be compared to the data. It can be shown that these results reproduce the results of constant $\alpha$ models \citep[e.g.,][]{Potter21} when the angular momentum flux is taken to be very small (i.e., $\delta_{\cal J} \sim 10^{-4}$), these values are far lower than observed in any GRMHD simulation, however.

Ultimately, as there is angular momentum transport there will also be thermal emission emitted right down to the event horizon of black hole disks. If this plunging region emission makes up a reasonable fraction of the observed data in a given telescope band, and one tries to model this emission with a black hole disk which is artificially curtailed at the innermost stable circular orbit, then one will naturally infer a high black hole spin. This is simple to understand: the horizon for a Schwarzschild black hole ($r_{\rm HORIZON} = 2GM_\bullet/c^2$ for $a_\bullet = 0$) is located at the same location as the innermost stable circular orbit of a much more rapidly rotating black hole ($r_{\rm ISCO}=2GM_\bullet/c^2$ for $a_\bullet \approx 0.94$). Clearly a possible degeneracy may exist, and should be considered carefully now that analytical models for the plunging region thermodynamics exist which reproduce the properties of first principles GRMHD simulations.

\subsection{The emission from moderately stressed disks }
\begin{figure}
    \includegraphics[width=\linewidth]{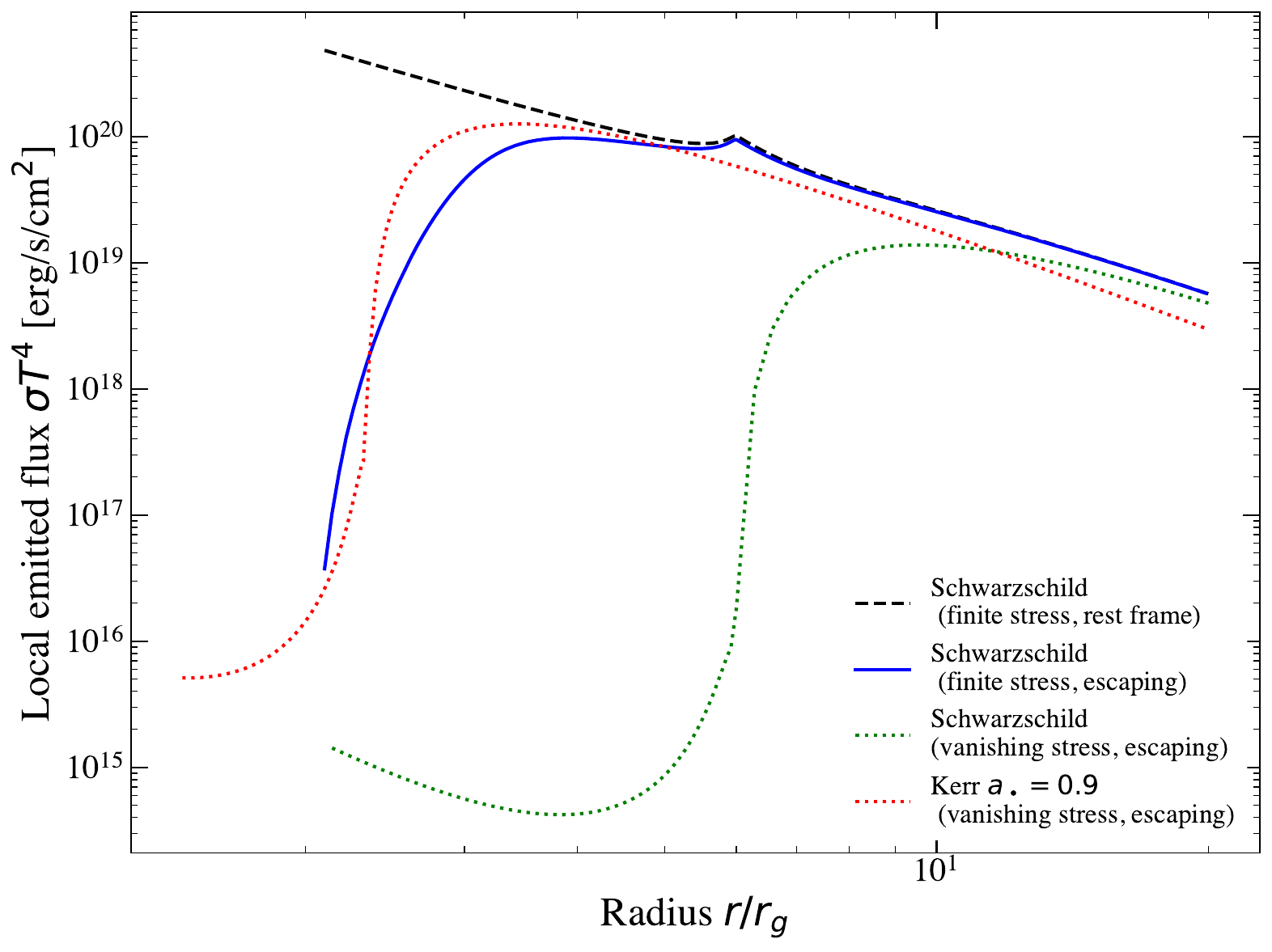}
    \caption{The locally emitted flux profiles of different black hole disks. By a green dotted  curve we show a Schwarzschild disk with a (near) vanishing ISCO stress, while by a red dotted curve we show the same flux profile for a more rapidly rotating black hole $a_\bullet = 0.9$. By a black dashed curve we show the flux, generated in the fluid rest frame, from a Schwarzschild black hole disk with ISCO stress set to the value observed in the GRMHD simulation of \citealt{Rule25}. Not all of this flux escapes to the observer, and a significant fraction is lost into the black hole. The blue solid curve shows how much of this local flux  {\it does not} end up in the black hole, and is therefore potentially observable. The moderate-stress Schwarzschild black hole looks much more similar to a rapidly spinning Kerr black hole disk than its vanishing stress counterpart. It will therefore be more difficult to distinguish these two profiles in the data.   }
    \label{fig:fluxes}
\end{figure}

\begin{figure}
    \centering
    \includegraphics[width=\linewidth]{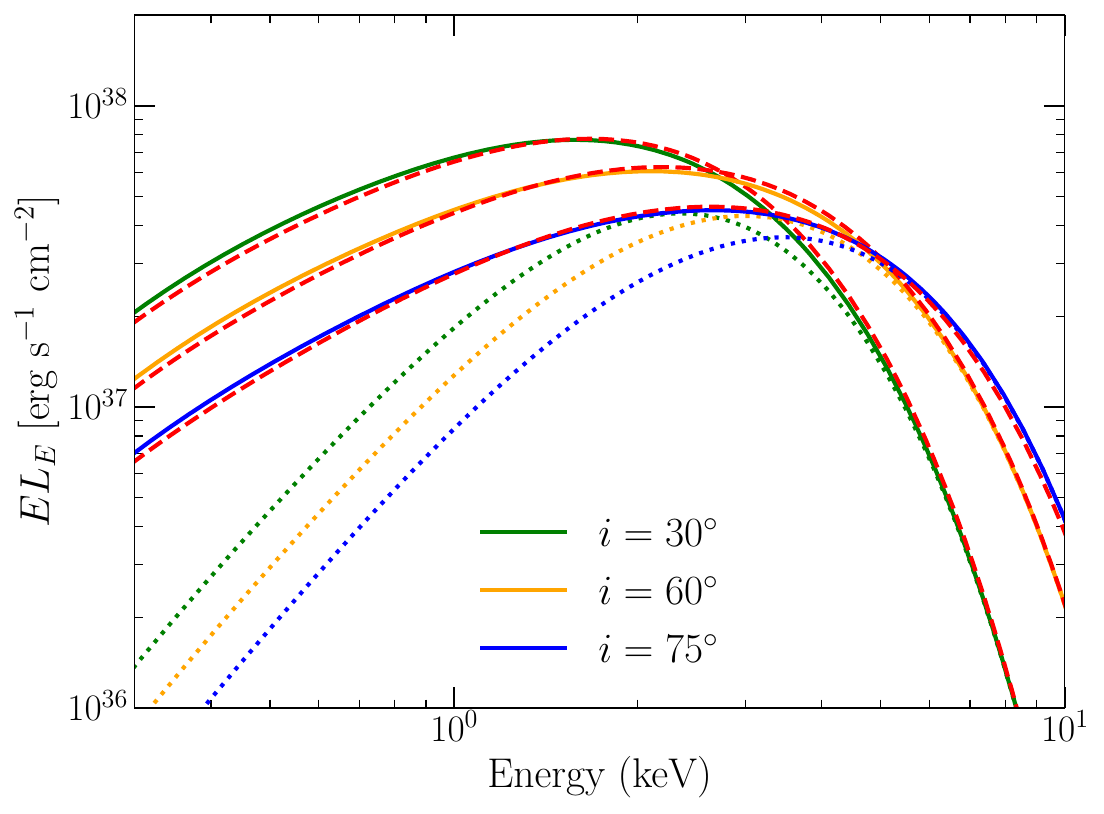}
    \caption{An example of spin-plunging region degeneracies. By red dashed curves we show three 0.3-10 keV disk spectra for vanishing ISCO stress disks around a black hole with mass $M_\bullet = 10 M_\odot$ and spin $a_\bullet = 0.85$, as observed at three different inclinations (denoted on plot). The solid curves show disks which include  emission from within the plunging region, but now for Schwarzschild black holes of the same mass observed at the same inclinations. Clearly low-spin bright-plunging region disks can mimic rapidly rotating black holes with zero emission within the plunging region. The actual emission sourced from within the plunging region is shown by dotted curves, and makes up a small fraction of the bolometric luminosity of the disk.   }
    \label{fig:theory_degeneracy}
\end{figure}

In a vanishing ISCO stress disk system, the only way in which to increase the flux emitted from small radii is to increase the black hole's spin, thereby moving the ISCO radius closer to the black hole's event horizon. This, in short, is the premise of the continuum fitting method of computing black hole spins: one places constraints on the inclination, distance and mass of the central hole from other methods, and then fits for the spin and accretion rate. As the high energy flux is very sensitive to the spin (in a vanishing ISCO stress disk), the spin is relatively straightforward to constrain from the data. 

In a model with a finite ISCO stress, but no emission from within the ISCO (a set up which is, for example, possible in {\tt XSPEC} with the {\tt kerrbb} model), it is still not possible to substantially increase the flux emitted {\it from small radii} by modifying the ISCO stress, as the disk is still curtailed at the ISCO. However, in a realistic system (like those described by {\tt fullkerr}), where the plasma remains hot and continues to emit from inside the ISCO, increasing the ISCO stress parameter both increases the bolometric luminosity of the disk system, but also substantially increases the flux emitted from small radii. This, naturally, sets up an intrinsic parameter degeneracy between ISCO stress and black hole spin, as both can modify the small-radius flux profiles.   

The simplest way in which to compare the emission profiles of  disk systems with and without ISCO stresses is to compute the locally radiated flux $F(r) = \sigma T^4$, where $T(r)$ is the local radiative temperature. This is not, however, the most relevant quantity to compare for different systems however, owing to the very different relative probabilities that a photon (emitted at a fixed radius) will escape the black hole's gravity when emitted from circularly orbiting or plunging material.  

Some photons emitted from an accretion flow,  at all radii,  will be captured by the central black hole. At radii outside of the ISCO this is typically a relatively small fraction of the total photon field, except for those disks around more rapidly spinning $(a_\bullet \gtrsim 0.9)$ black holes, were the ISCO itself becomes close to the event horizon. Typically for stable disk regions outside of $r\gtrsim 5GM_\bullet/c^2$ less than $\sim 10\%$ of all emitted photons are captured, with this fraction typically dropping below $1\%$ at radii $r \gtrsim 10GM_\bullet/c^2$ \citep[see e.g.,][for more details]{Thorne1974, Mummery25}. 

This picture is altered significantly for photons emitted from plasma within the plunging region however. Firstly, and trivially, the plunging region extends right down to the event horizon, where $100\%$ of emitted photons will be captured. In addition however, the fluid's increasing component of radial velocity will result in the relativistic beaming of its emitted radiation in the direction of the event horizon, which naturally results in an increasing capture fraction.  This means that at the same radial scale a photon emitted from plunging material is significantly more likely to be captured than one emitted from stably orbiting material. This modifies the flux which is potentially observable to a distant observer. 

It is possible to compute the fraction of photons emitted from a given disk radius which are ultimately captured by the black hole. We denote this fraction $f_\bullet(r)$. The details of this calculation for photons emitted from the main body of the disk are provided in \cite{Thorne1974}, and for material within the plunging region are discussed in \cite{Mummery25}. The more relevant quantity to compare is therefore $\widetilde F(r) = [1 - f_\bullet(r)]F(r)$. This is the {\it potentially observable} locally radiated flux. 

In Figure \ref{fig:fluxes} we show a set of canonical local disk fluxes, computed with and without a finite ISCO stress. These systems have broadly typical X-ray binary parameters, namely $M_\bullet=10M_\odot$ and $\dot M = 0.1 \dot M_{\rm edd}$\footnote{We normalize the mass accretion rate by the spin-dependent accretion efficiency $\eta(a)$, leading to slight differences in the flux in the Newtonian regime.}.    By a green dotted  curve we show a Schwarzschild disk with a (near) vanishing ISCO stress, while by a red dotted curve we show the same flux profile for a more rapidly rotating black hole $a_\bullet = 0.9$. These are clearly extremely different, and this is in essence the premise of the continuum fitting method. 

We also show, however, the flux, as measured in the fluid rest frame, from a Schwarzschild black hole disk with ISCO stress set to the value observed in the GRMHD simulation of \citealt{Rule25} (namely $\delta_{\cal J} = 0.053$). Not all of this flux escapes to the observer, and a significant fraction is lost into the black hole. The blue solid curve shows how much of this local flux escapes the black holes vicinity, and is therefore potentially observable. We stress to the reader that going from even these escaping-flux profiles to observable X-ray spectra is highly non-trivial, and the observed bolometric emission from these Schwarzschild disks remains dominated by stable disk regimes.  The moderate-stress Schwarzschild black hole do, however, look much more similar to a rapidly spinning Kerr black hole disk than its vanishing stress counterpart. Interestingly,  the turnover in the local flux at small radii in a bright plunging region disk is driven by photon capture, rather than a disk boundary condition. It will clearly be more difficult to distinguish these two profiles, when compared to two vanishing ISCO stress profiles, from X-ray data.  Once reasonable uncertainties in the disk-observer inclination angle and mass accretion rate are incorporated, this difficultly will only increase.

This disk frame similarities result in X-ray spectra in the distant observers frame which are, even in principle, extremely difficult to distinguish. This is highlighted in Figure \ref{fig:theory_degeneracy}, where we show  (by red dashed curves) three 0.3-10 keV disk spectra produced by vanishing ISCO stress disks around a black hole with mass $M_\bullet = 10 M_\odot$ and spin $a_\bullet = 0.85$, as observed at three different inclinations (denoted on plot). The accretion rate in this plot was taken to be a fiducial $0.1 \dot M_{\rm edd}$, but is largely inconsequential. The solid curves show disks which include  emission from within the plunging region, but now for Schwarzschild black holes of the same mass observed at the same inclinations. These spectra include all relativistic photon propagation effects. Clearly low-spin bright-plunging region disks can mimic rapidly rotating black holes with zero emission within the plunging region. This is in effect the main point of this paper. 

We note here that the key physical character of this potential degeneracy is necessarily limited to the smallest radii. These small radii cannot be probed by models which do not treat the ISCO stress as a free parameter, and also do not solve the disk thermodynamic equations within the ISCO.   This second point is crucial, and likely explains why \cite{Li05} found that varying that ISCO stress parameter was mainly degenerate with a change in accretion rate $\dot M$. This, we shall show, is absolutely not the case when intra-ISCO emission is included, as was also found by \cite{Davis05, Davis06} who, despite not having a model for the plunging region, also found a strong spin-ISCO stress degeneracy.

\begin{figure*}
    \centering
    \includegraphics[width=0.49\linewidth]{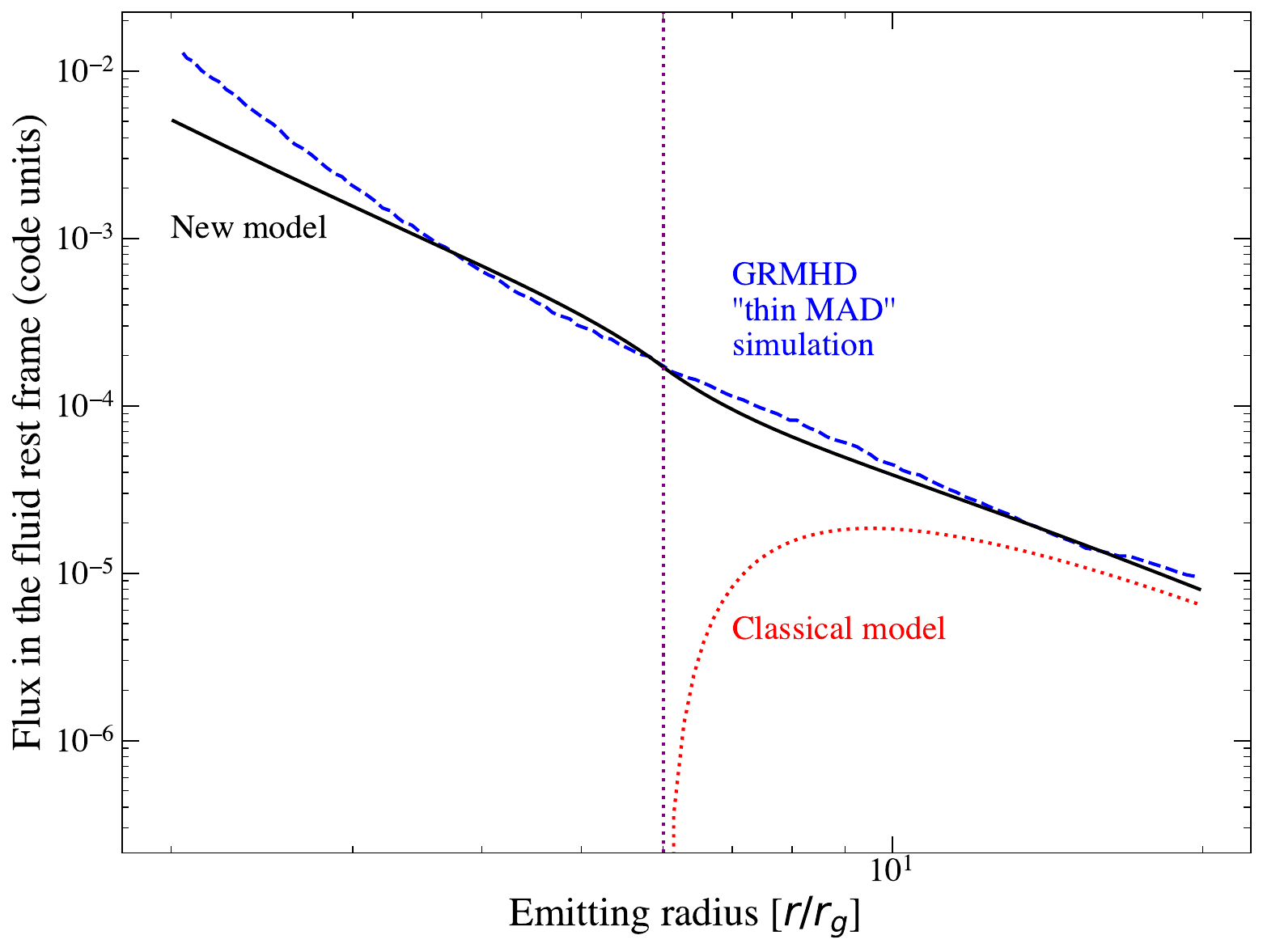}
    \includegraphics[width=0.49\linewidth]{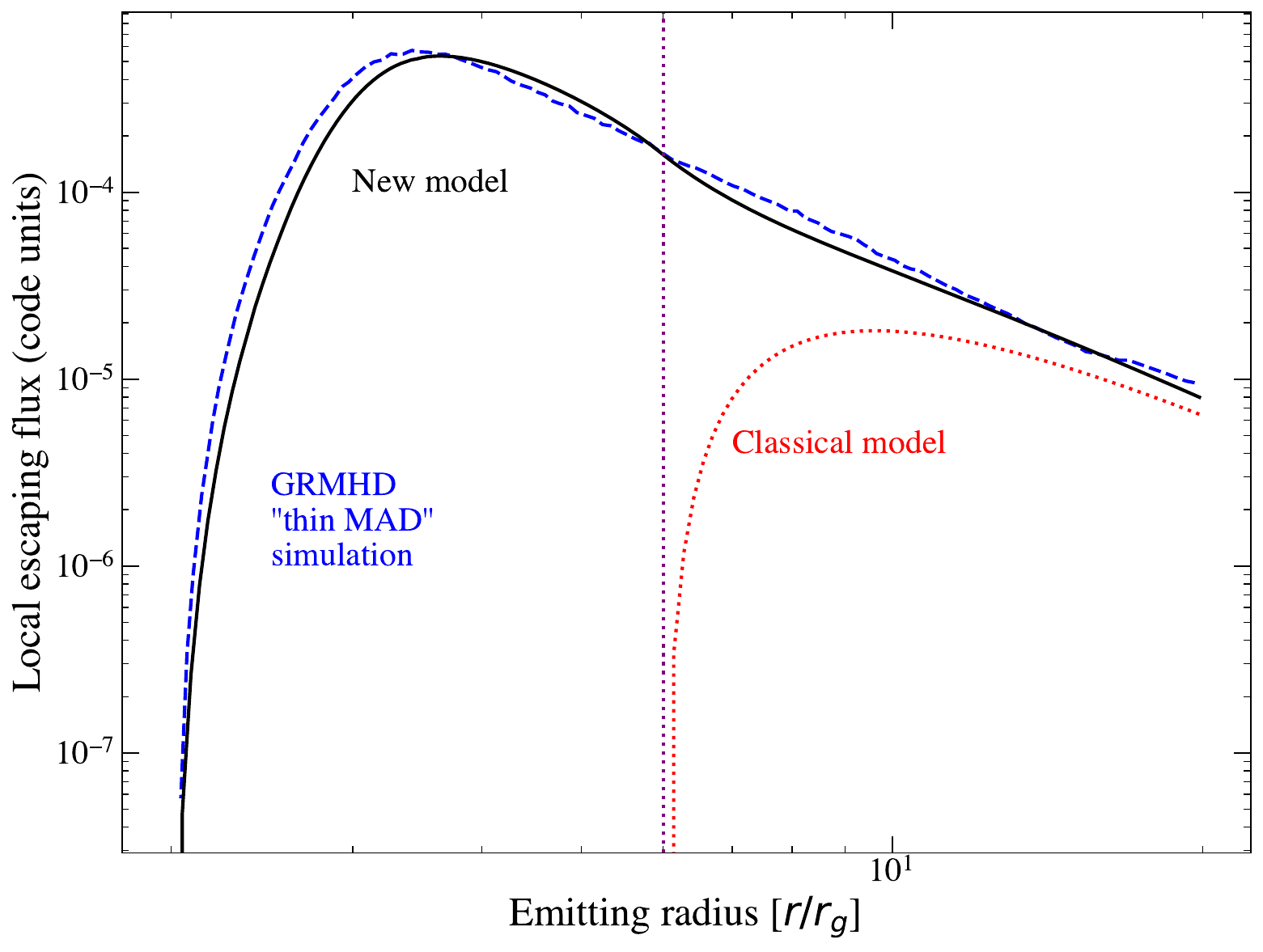}
    \caption{{\bf Left:} The locally liberated flux, as measured in the fluid rest frame, for the Schwarzschild (spin $a=0$) simulations of \citealt{Dhang25} (blue dashed curve) plotted against disk radius. This is compared to the classical \citealt{NovikovThorne73} model (red dotted line) which clearly is an extremely poor approximation to the simulation results. The new model of \citealt{MummeryBalbus2023} is shown by a black curve, which represents a much improved description. The ISCO is shown by the purple vertical dotted line. The {\bf right} hand panel shows the amount of flux which escapes the black hole (i.e., is not captured) as a function of radius. These simulations were of thin disks, and are deemed ``MAD'' (i.e., they have strong magnetic fields).  }
    \label{fig:madsim0}
\end{figure*}

\begin{figure*}
    \centering
    \includegraphics[width=0.49\linewidth]{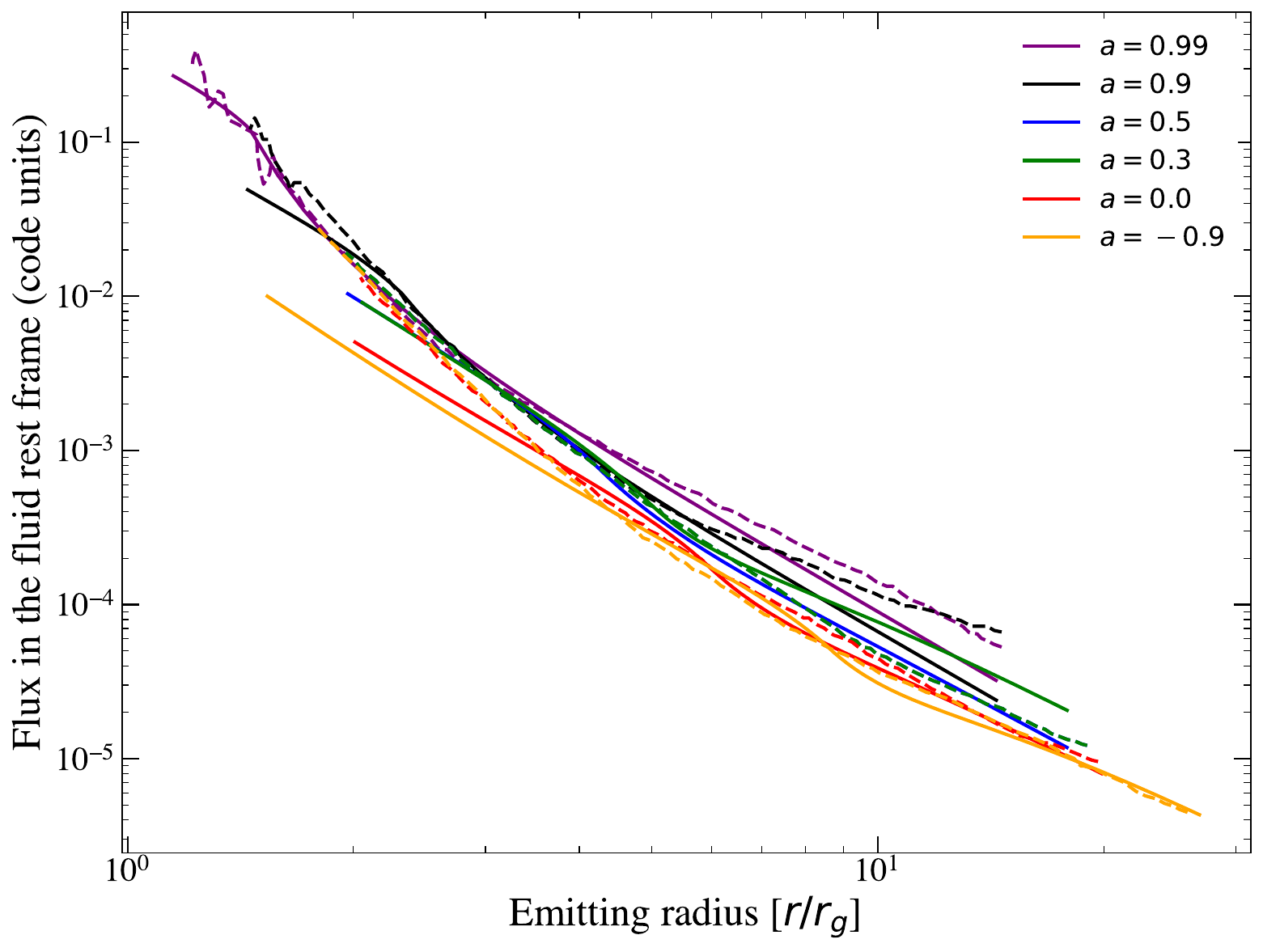}
    \includegraphics[width=0.49\linewidth]{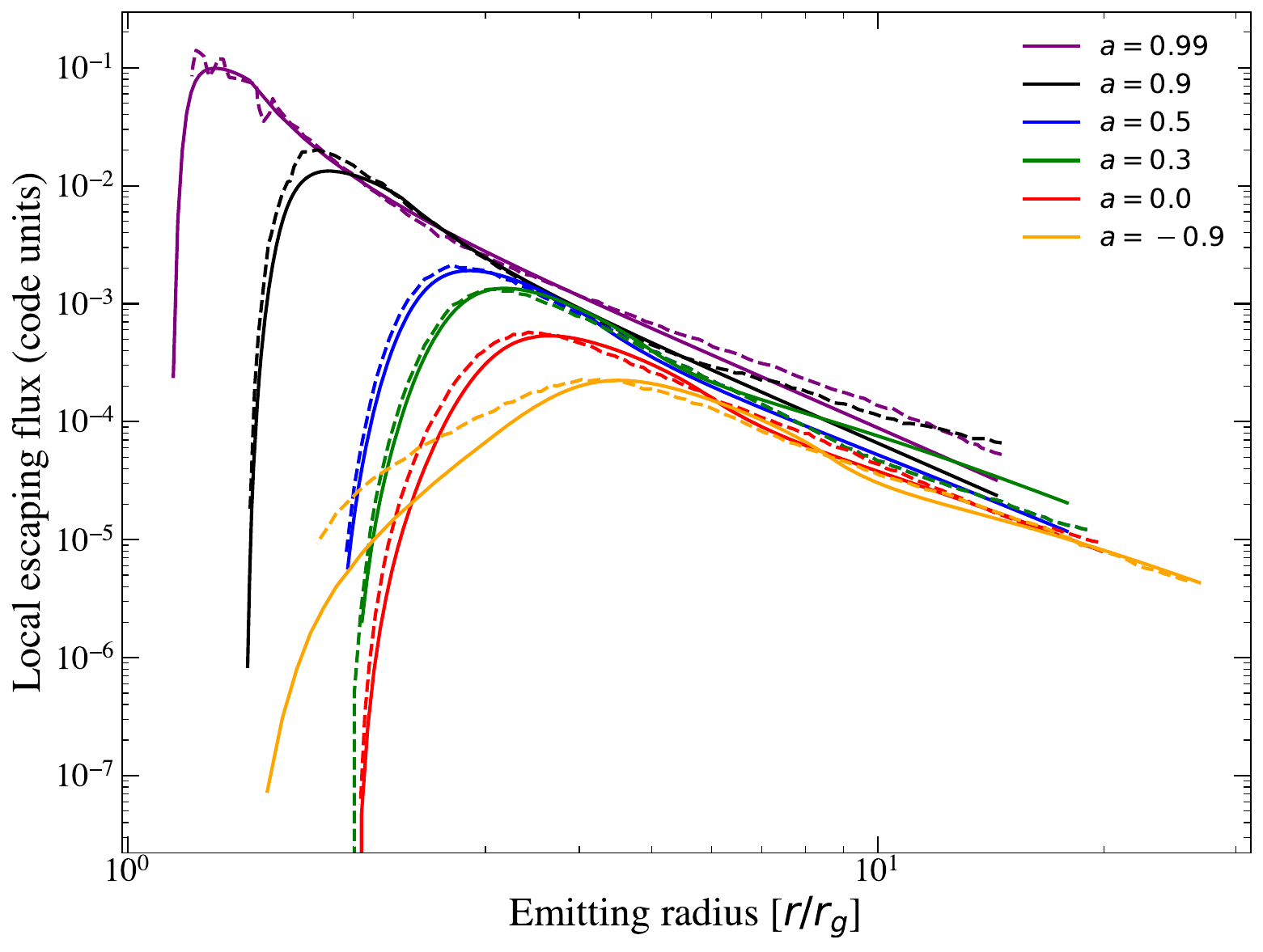}
    \caption{{\bf Left:} the flux in the fluid rest frame, as a function of radius, for the six different spin values in \citealt{Dhang25}. They show minimal difference as the spin is varied. This is the exact opposite of what would be the case in a series of vanishing ISCO temperature disks (see e.g., the red and green curves in Figure \ref{fig:fluxes}). {\bf Right}: the escaping fluxes from these disk systems as a function of radius. The different dynamical properties of plunging and circularly-orbiting material leads to some differences at radii $r \lesssim 3 r_g$. Photons emitted here are, however, extremely strongly gravitationally red-shifted (owing to the presence of a horizon $\sim 1r_g$ away), and will contribute minimally to the observed X-ray flux. The models of \citealt{MummeryBalbus2023} are shown by solid curves, and reproduce all of the simulation data (dashed curves) well.     }
    \label{fig:madsim_all}
\end{figure*}

\subsection{Comparison to recent simulation data }
It is of interest to compare directly the models of intra-ISCO emission used in this work with modern numerical predictions of the local liberated (and escaping) flux of black hole accretion disks, of a variety of spins. In Figure \ref{fig:madsim0} we show the locally liberated flux, as measured in the fluid rest frame, for the Schwarzschild (spin $a=0$) simulations of \cite{Dhang25} (blue dashed curve) plotted against disk radius. This is compared to the classical \cite{NovikovThorne73} model (red dotted line) which clearly is an extremely poor approximation to the simulation results. The new model of \cite{MummeryBalbus2023} is shown by a black curve, which represents a much improved description\footnote{In their paper \citealt{Dhang25} compared this locally liberated flux to the models of \citealt{NovikovThorne73} and \citealt{AgolKrolik00}, stating that the model of \citealt{AgolKrolik00} can be extended inside the ISCO. This is mathematically true, in the sense that the model will return some number as a function of radius, but this number is physically meaningless. The \citealt{AgolKrolik00} model assumes the fluid follows circular orbits, and this is very much not the case inside the ISCO. This is why \citealt{AgolKrolik00} curtail their models at the ISCO in their original paper. This extension is the origin of the unphysical flux-divergence seen in the ``AK'' models in \citealt{Dhang25} at the photon radius where a matter particle would have to be rotating at the speed of light, which it clearly does not in a realistic physical model.}. 

As \citealt{Dhang25} chose a gas-pressure equation of state (adiabatic index $\Gamma = 5/3$), we use equation (67) of \cite{MummeryBalbus2023} \citep[assuming electron scattering opacity dominates, a reasonable assumption for XRBs and in keeping with the ad-hoc cooling function used by][see section \ref{conc} for further discussion]{Dhang25} to compute the local flux $F=\sigma T^4$  within the ISCO radius (denoted $r_I$) 
\begin{equation}
    F(r) = F_I \left({r\over r_I}\right)^{-5}\, \left[{c\over u_I}\sqrt{2r_g \over 3r_I}\left({r_I \over r}-1\right)^{3/2} + 1\right]^{-5/4} , 
\end{equation}
where $u_I$ is the radial velocity at which the fluid crosses the ISCO \citep[typical values observed in the simulations of][were $u_I \sim 0.1c$ -- a high value resulting from the large magnetic field strengths in the simulations]{Dhang25}\footnote{The authors are extremely grateful to Prasun Dhang for sharing the simulation data.}, and $F_I$ is the flux at the ISCO radius. The flux outside of the ISCO in this case is the same as in {\tt fullkerr} (which includes radiation pressure in addition to gas pressure). 

To reproduce the simulation data we required an ISCO stress parameter $\delta_{\cal J} \approx 0.07$ (which sets $F_I$), and a radial velocity $u_I \approx 0.13c$ (although these values are not fit for, merely chosen by eye). It is interesting that this ISCO stress parameter is within a factor $\sim 2$ of those required from observations (as we shall discuss shortly), and the lower magnetic field strength simulations of \cite{Rule25}, suggesting a convergence in different modeling approaches. 

In Figure \ref{fig:madsim_all} we show the flux in the fluid rest frame (left panel), as a function of radius, for the six different spin values in \cite{Dhang25}. They show minimal difference as the spin is varied. This is the exact opposite of what would be the case in a series of vanishing ISCO temperature disks (see e.g., the red and green curves in Figure \ref{fig:fluxes}). In the right panel we show the escaping fluxes from these disk systems as a function of radius. The different dynamical properties of plunging and circularly-orbiting material leads to some differences at radii $r \lesssim 2 GM_\bullet/c^2$ (see earlier discussion). Photons emitted here are, however, extremely strongly gravitationally red-shifted (owing to the presence of a horizon $\sim 1GM_\bullet/c^2$ away), and will contribute minimally to the observed X-ray flux. The models of \cite{MummeryBalbus2023} are shown by solid curves, and reproduce all of the simulation data (dashed curves) well. 

A more detailed comparison between models and simulations with different spin values are shown in Appendix \ref{app:sims}. 

\section{Data and analysis approach}\label{obs}
In this section we introduce the observational datasets used in section \ref{sec:results} to analyse spin-stress degeneracies in X-ray binaries. 

As we are partly motivated by the (possible) discrepancies between the spin measurements inferred from X-ray and Gravitational Wave techniques, we analyse a high-mass X-ray binary in this work. We select M33 X-7 as our source of choice for a number of reasons. Firstly, M33 X-7 has a clean disk-dominated X-ray spectrum (in contrast with, say, the high mass X-ray binary Cyg X-1 which has numerous complicating additional components). Secondly, by virtue of being located in M33, M33 X-7 has a  small {\it fractional} systematic error in its distance, which aids in constraining the system parameters. Thirdly, M33 X-7 has a high inferred spin from conventional approaches and a low absorbing column density along our line of sight and, finally, it is observed to have a hot disk, putting the curvature of the spectrum in the middle of the \xmm\ band. This final consideration is important as it shows that the degeneracy between spin and stress is not limited to just those sources were one observes only the highest energy emission from the disk.

\subsection{M33 X-7 dataset}

We download the M33 X-7 data from the HEASARC repository, and selected two \xmm\ observations which had particularly useful properties for the type of analysis we wish to perform. These \xmm\ observations had high count rates and good data quality, with minimal spectral contamination from coronal (power law) components. Vanishing ISCO stress model fits to M33 X-7 provide consistent results \citep{Liu08}, and so the specific choice of these two spectra will have minimal impact on our results. The M33 X-7 data properties are summarised in Table \ref{tab_obs}.

{The \xmm\ data were reduced using the \xmm\ Science Analysis System (SAS) V22 and calibration files updated on 30 April 2024. Our focus centered on analysing the EPIC data.  To achieve this, we employed the EPPROC (EMPROC) tool to generate clean calibrated pn (MOS) event lists. Subsequently, data were filtered to remove the impact of flaring particle background, identifying intervals where the single event count rate in the 10--12\,keV band exceeds 0.4 (0.35) counts~s$^{-1}$ for pn (MOS) as background-dominated intervals.} {Source events were extracted using EVSELECT, selecting both single and double pixel events from a circular region of 25 (30) arcsec radius centered on the source for pn (MOS). For the MOS detectors, background events were extracted from a nearby source-free circular region with a radius of 120 arcsec. For the pn detector, we used a box-shaped, source-free background region of 210$\times$64 arcsec$^2$, located within the same chip section of the central circular area of the Full Frame mode field of view. This region avoids known instrumental emission lines, including Cu K$\alpha$, Zn K$\alpha$, and K~K$\alpha$.} {Response files were generated using the RMFGEN and ARFGEN tasks. The observed 0.3–10\,keV count rates for the pn (MOS) detectors on 13 and 17 July were 0.4 and 0.3 (0.13 and 0.11) counts~s$^{-1}$, respectively—well below the pile-up threshold for Full Frame mode. We verified the absence of significant pile-up by running EPATPLOT and found the observed-to-model single and double pattern fractions to be consistent with 1 within 1-$\sigma$ statistical errors, finding them consistent with no clear signs of pile-up.}

\begin{table}
    \centering
    \begin{tabular}{ccccc}
    \hline\hline
    Date & MJD & Mission & ObsID & Length (ks)  \\
    \hline
    2019-07-13 & 58677 & \xmm\ & 0831590401 & 19.6  \\ 
    2019-07-17 & 58681 & \xmm\ & 0831590201 & 15.1  \\
    \hline\hline
     \end{tabular}
    \caption{A summary of the X-ray observations of the high-flux soft state of M33 X-7 analysed in this work. }
    \label{tab_obs}
\end{table}

\subsection{MAXI J1820+070 dataset}
We are also interested in the physical properties of sources which are known to have detectable emission from within the plunging region. The X-ray binary MAXI J1820+070 is one such example, as was first suggested by \cite{Fabian20} and confirmed by \cite{Mummery24Plunge}. 

In this work we reuse the {\it NuSTAR} X-ray spectral data first presented in \cite{Fabian20} which formed the basis for the fits in \cite{Mummery24Plunge}. We refer the reader to \cite{Fabian20} for a description of the data reduction, and to \cite{Mummery24Plunge} for a description of the signatures of the plunging region emission. The data used in this paper correspond to {\it NuSTAR} observations Nu25, Nu27, Nu29 and Nu31 \citep[with the last two digits distinguishing the different ObsIDs][]{Fabian20}. The properties of the plunging region (in particular the parameter $\delta_{\cal J}$) were stable across these four epochs, and so the choice of epoch studied is not of qualitative importance. 

\subsection{MAXI J0637$-$430 dataset}
A second source with detected plunging region emission is MAXI J0637$-$430. We use the joint {\it Swift}, {\it NICER} and {\it NuSTAR} observations of MAXI J0637$-$430, as presented in \cite{Lazar21} and re-analysed in \cite{Mummery24PlungeB}. We refer the reader to these two works for a detailed discussion of the data analysis and reduction. Again, multiple epochs of MAXI J0637$-$430 data show signatures of plunging region emission, well described by broadly constant values of $\delta_{\cal J}$, meaning that the results presented in this work are qualitatively independent of the particular epoch chosen. 

\subsection{Spectral models and system parameters}
We will be using the relativistic thin disk models {\tt kerrbb} \citep{Li05} and {\tt fullkerr} \cite{Mummery24Plunge} in this work. At radii significantly larger than the ISCO, these models are identical, and are both described by the classical \cite{NovikovThorne73} and \cite{PageThorne74} time-independent relativistic thin disk solutions. Closer to the ISCO, however, the two models start to deviate, and the scale of this deviation is dictated by an ISCO stress parameter $\delta_{\cal J}$. Physically, $\delta_{\cal J}$ corresponds, as we have discussed, to the angular momentum passed back to the disk by plunging fluid. This transport is mediated by growing magnetic stresses in the plunging region (Figure \ref{fig:physics}), driven by flux-freezing \citep{Krolik99}.  Typical values seen in GRMHD simulations put this parameter broadly in the range $\delta_{\cal J} \sim 0.01-0.1$. 

Inside of the ISCO {\tt kerrbb} assumes zero photon emission (even if a finite ISCO stress is imposed), whereas {\tt fullkerr} uses the solutions of the intra-ISCO thermodynamic equations derived by \cite{MummeryBalbus2023}.  

We also include {\tt tbabs} \citep{Wilms00}, to model the absorption of X-ray photons by neutral Hydrogen along their path to the observer. This introduces the additional free parameter $n_H$ (the integrated effective Hydrogen column density, with dimensions of 1/area), which is fit to each epoch individually. For M33 X-7 we fit data taken from three different instruments aboard the \xmm\ satellite, EMOS1, EMOS2 and EPN. To allow for slight systematic differences between these instruments we include a simple multiplicative factor (named {\tt constant} in {\tt XSPEC} notation). Our full model is therefore, in {\tt XSPEC} notation 
\begin{equation}
{\tt model} = {\tt constant} \times {\tt tbabs}({\tt disk}
), \nonumber
\end{equation}
where {\tt disk} is either {\tt kerrbb} or {\tt fullkerr} depending on the model used.

For our analysis of M33 X-7 we fix the following system parameters to known values: $M_\bullet = 15 M_\odot, i = 75^\circ, D=845$ kpc, and do not let them vary throughout the analysis. These particular values correspond to the canonical choices in the literature, see e.g., \citep{Liu08}.

As black hole spin parameters are also somewhat degenerate with the colour-correction factor $f_c$ \citep{Salvesen21}, we fix $f_c = 1.7$ in the main body of the disk for both models. For {\tt fullkerr} we also fix $f_c = 1.7$ within the ISCO.  (This value is in some sense natural \cite{Davis06}, and similar values have been inferred in the literature using radiative transfer simulations of photons through disk atmospheres).  Formally for much of the natural parameter space of X-ray binary disks, the intra-ISCO region of the disk rapidly becomes ``photon-starved'' \citep{Mummery24Plunge}, meaning that the emission from within the ISCO has $f_c > 1.7$,  potentially reaching substantial values \cite{Davis19}. This is usually modeled within {\tt fullkerr} with a tuneable parameter $\xi$, but we found that letting $\xi$ vary produced qualitatively identical results, at the expense of introducing an additional free parameter (if a degeneracy exists with a parameter frozen, adding more free variables can only make the potential degeneracy worse). 

All model fitting is done in the X-ray spectral fitting package {\tt XSPEC} \cite{Arnaud96}. We shall use the $\chi^2$ statistic when comparing the ability of different models to describe the data, while the usual reduced chi-squared is denoted  $\chi^2_r$.

\section{Observational results}\label{sec:results}

\subsection{Fitting the M33 X-7 X-ray spectrum}

The vanishing ISCO stress {\tt kerrbb} model has only a handful of free parameters $n_H, \dot M, a_\bullet$. This does a perfectly acceptable job at fitting the data. We recover the well-known result in the literature that $a_\bullet = 0.84 \pm 0.05$ when $\delta_{\cal J} = 0$. The reduced $\chi^2$ statistic for this fit is $\chi^2_r = 204.9/212$. 
\begin{figure}
    \centering
    \includegraphics[width=\linewidth]{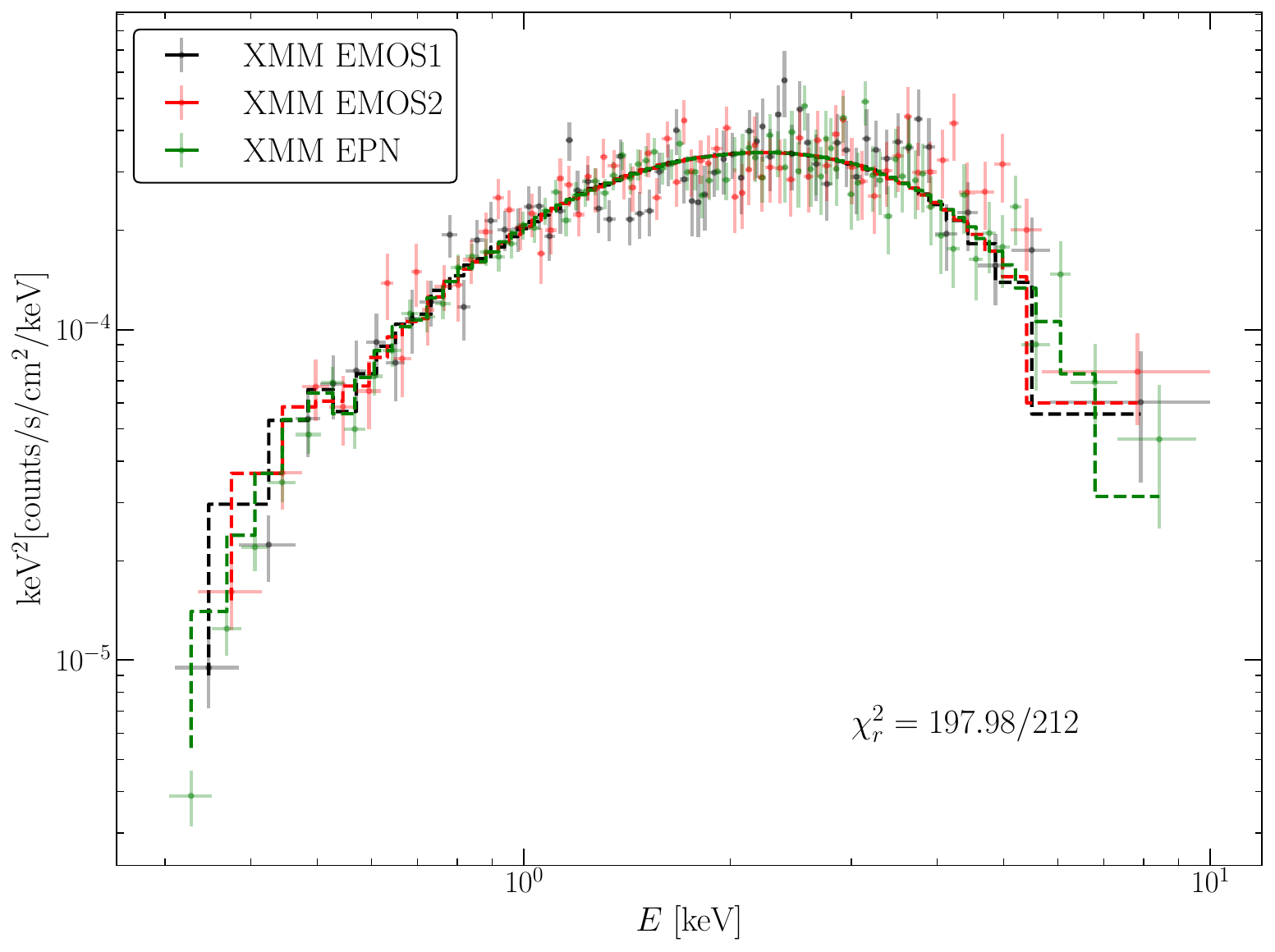}
    \includegraphics[width=\linewidth]{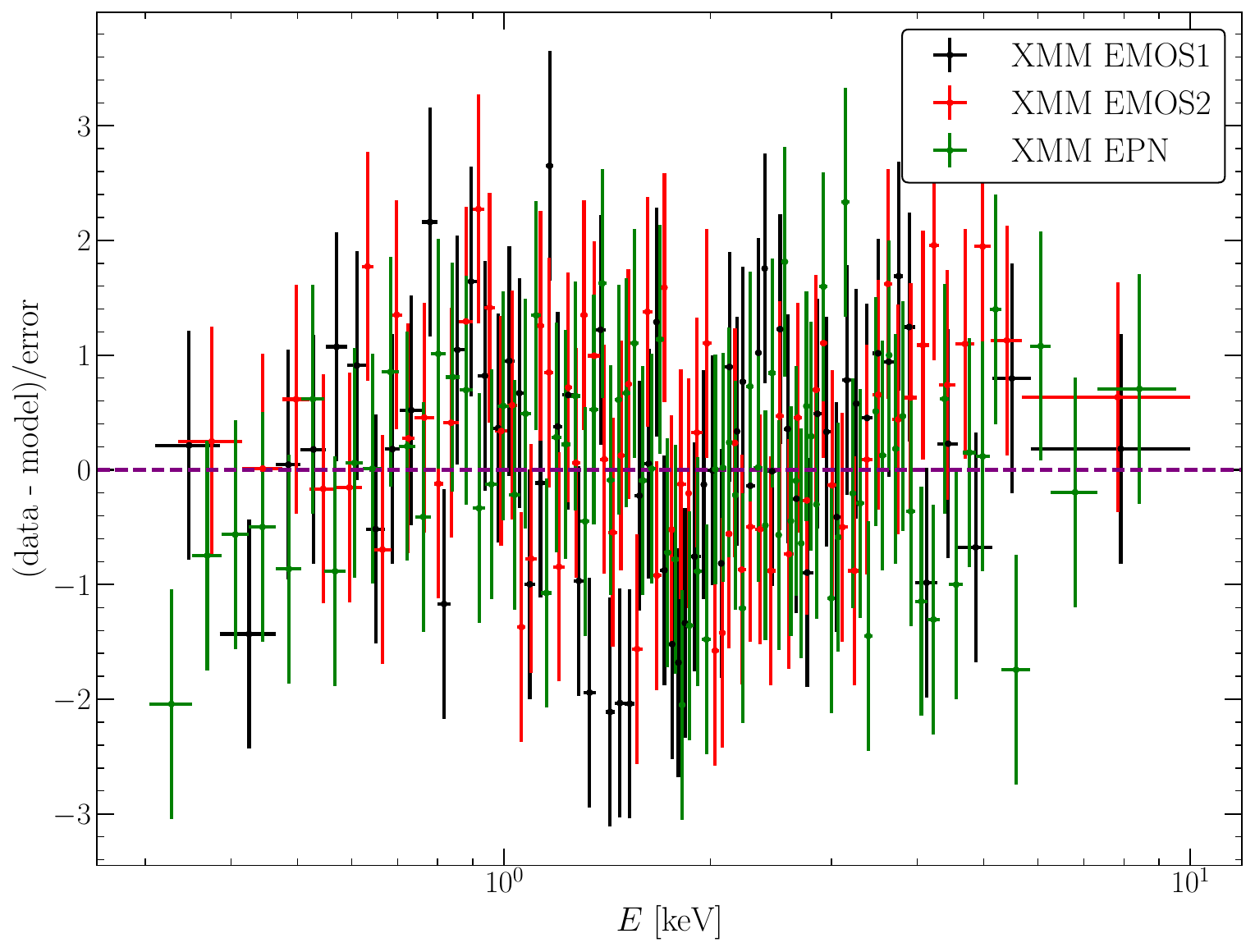}
    \caption{Upper: The X-ray spectrum (with {\tt fullkerr} fit) of M33 X-7 (observation taken on 2019/07/13), showing a soft thermal spectrum which peaks at $E \sim 2-3$ keV and is well described by emission from an accretion flow. The different colours represent different instruments aboard the \xmm\ satellite. Model fits are shown by dashed curves with matching colours.  Lower: the normalised model-data discrepancies, showing that there are no systematic model-data deviations. This fit assumed a Schwarzschild black hole with a finite ISCO stress $(\delta_{\cal J} = 0.078)$ and emission from within the plunging region. If a vanishing ISCO stress were assumed then a much higher spin would be required $a_\bullet = 0.84 \pm 0.05$.  }
    \label{fig:spectrum}
\end{figure}
\begin{figure*}
    \centering
    \includegraphics[width=.48\linewidth]{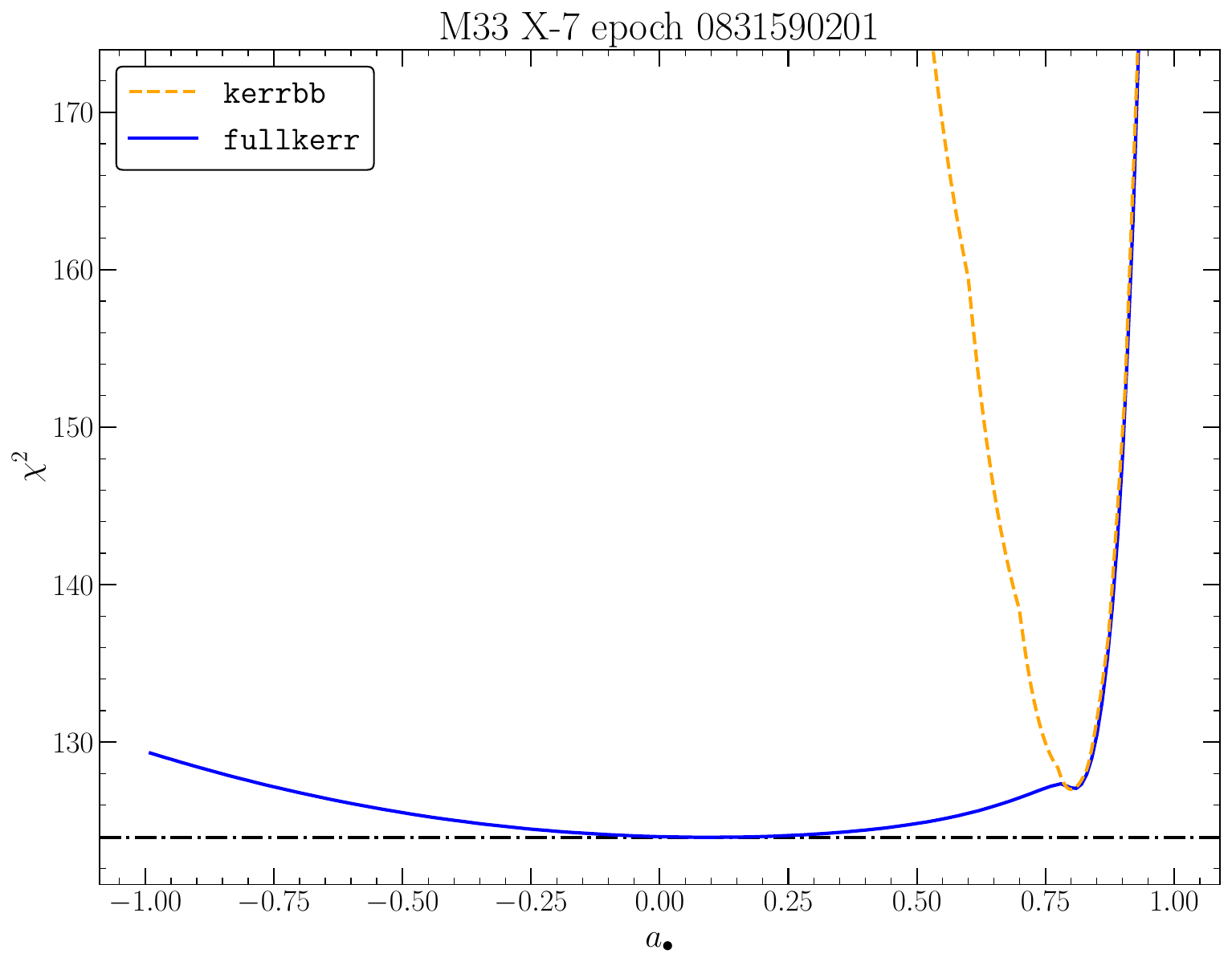}
    \includegraphics[width=.48\linewidth]{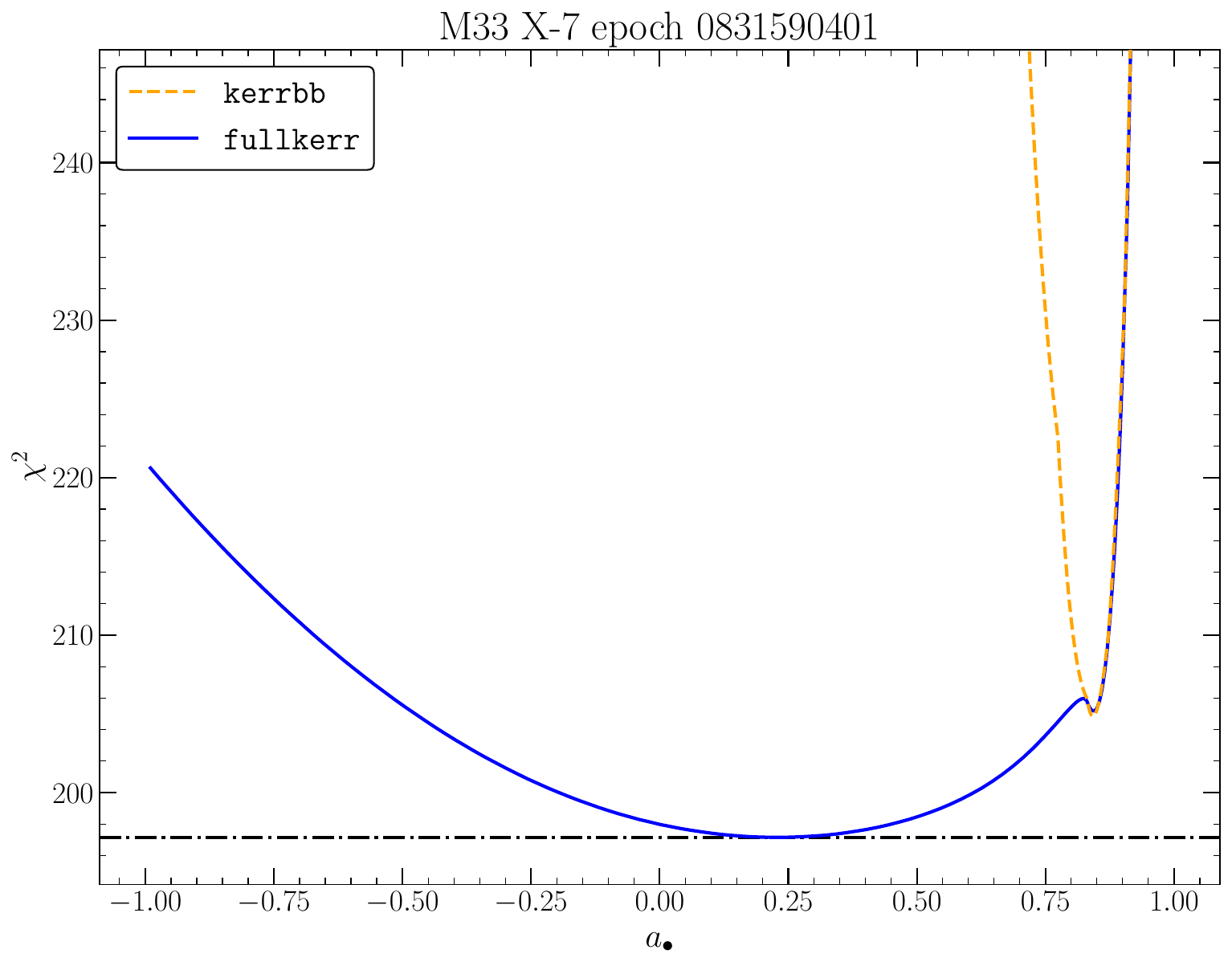}
    \includegraphics[width=.48\linewidth]{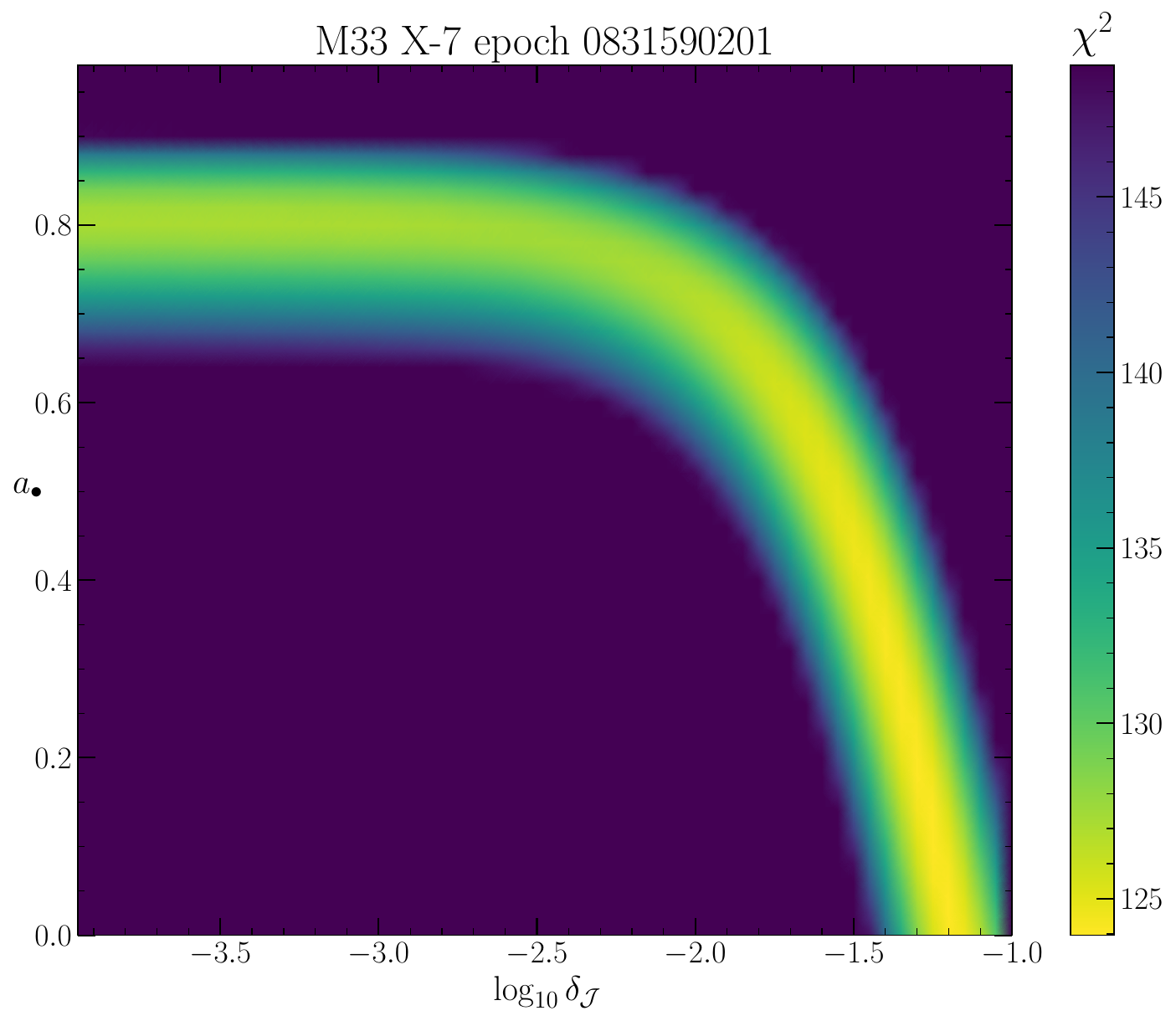}
    \includegraphics[width=.48\linewidth]{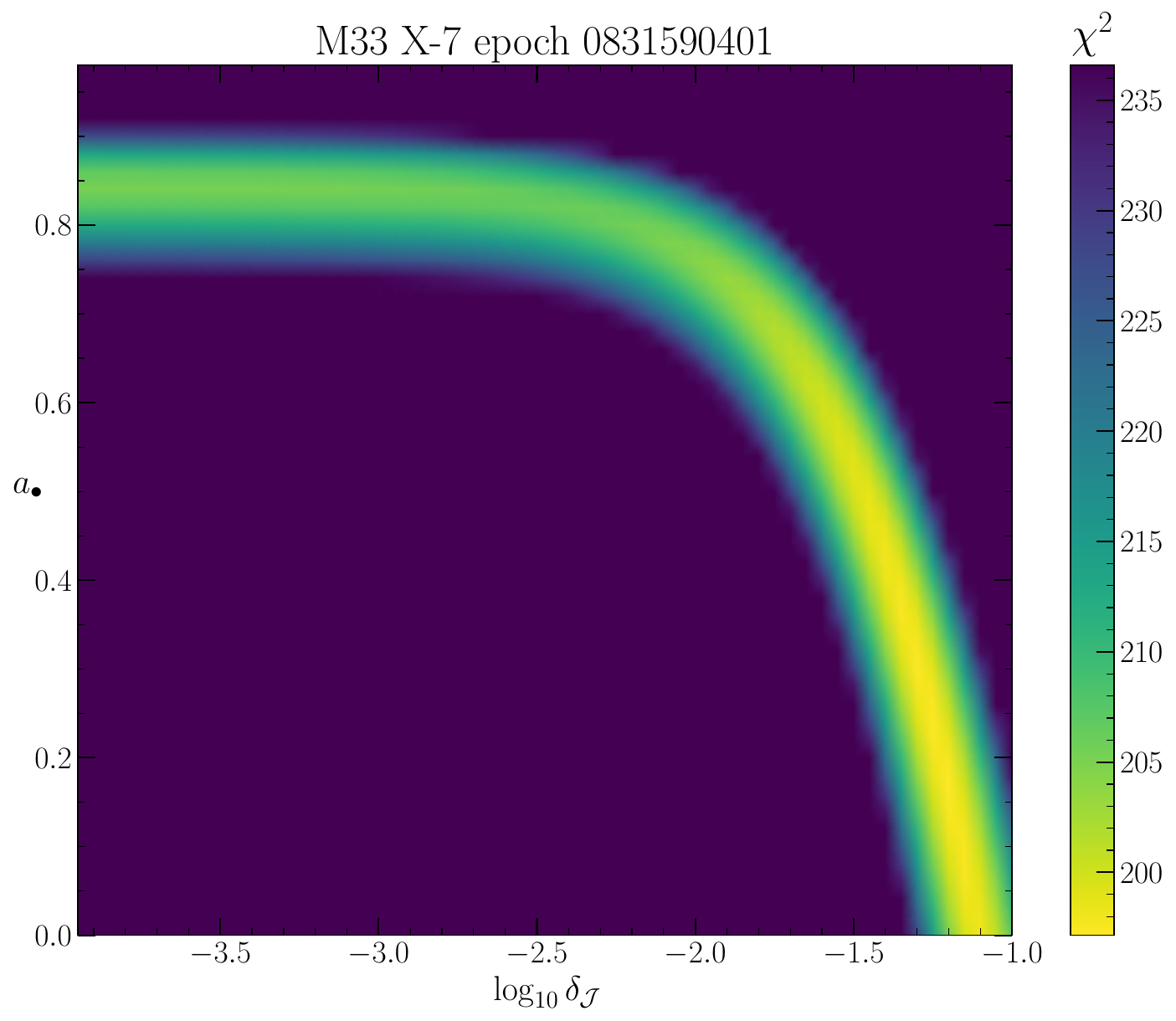}
    \caption{The formal spin-stress degeneracy for both epochs of our M33 X-7 analysis. The upper row shows the one-dimensional $\chi^2$-statistic as a function of black hole spin for both epochs, while the lower row shows the two dimensional $a_\bullet-\delta_{\cal J}$ degeneracy, with $\chi^2$-statistic shown by the colour bar. In the upper row we see that the spin posteriors for finite ISCO stress disks are significantly broader than their vanishing ISCO stress counterparts, a result of the additional modelling flexibility available to a model which can include extra small-scale emission. At high spins the best-fitting  {\tt kerrbb} and {\tt fullkerr} models are identical, as they describe identical physical systems in the limit $\delta_{\cal J} \to 0$. This can be seen reflected in the fit statistic as a function of black hole spin for both models at spins $a_\bullet > 0.8$.  For both epochs formally better fits are found for significantly lower spins $(a_\bullet \simeq 0.2)$ and a finite ISCO stress, when compared to a vanishing ISCO stress disk with $a_\bullet \simeq 0.8$, although the improvement is minor (at the $\sim 2-3\sigma$ level).    The lower row shows that the degeneracy is rather sharp in the spin-stress plane. If the stress parameter $\delta_{\cal J} \lesssim 0.01$, then the spin inference is unaffected. However, over the range  $0.01 \lesssim \delta_{\cal J} \lesssim 0.1$ the best-fitting spin parameter varies hugely (from $a_\bullet \simeq 0.8$ to $a_\bullet < 0$). This exact regime of $\delta_{\cal J}$ is what is required to reproduce the observed X-ray spectra of MAXI J1820 and MAXI J0637, meaning that many astronomical systems may inhabit this particularly sensitive region of parameter space.  }
    \label{fig:degeneracy}
\end{figure*}

Moving to {\tt fullkerr}, we have one more free parameter, $\delta_{\cal J}$. Freeing this parameter we find that it is very difficult to measure the spin. Freezing the spin to that of a Schwarzschild black hole, we find another formally acceptable  (in fact slightly better)  fit, with $\delta_{\cal J} = 0.078$. The reduced $\chi^2$ parameter for this fit is $\chi^2_r = 198.0/212$, a formal improvement of $\Delta \chi^2 = - 6.9$ for no additional free parameters (as the black hole spin is now not allowed to vary).   This value of $\delta_{\cal J}$ is consistent with the simulations of \cite{Dhang25} discussed earlier.  We show this fit to the data in Figure (\ref{fig:spectrum}). In the upper panel we show the X-ray spectrum (with {\tt fullkerr} fit) of M33 X-7 (for the observation taken on 2019/07/13), showing a soft thermal spectrum which peaks at $\sim 2-3$ keV and is well described by emission from the {\tt fullkerr} model. In the lower panel we show the normalised model-data discrepancies, showing that there are no systematic model-data deviations. Qualitatively identical results are found for the observation taken on 2019/07/17 (the {\tt kerrbb} model finds $a_\bullet = 0.81 \pm 0.05$, with $\chi_r^2 = 127.1$, while the {\tt fullkerr} model in a fixed Schwarzschild spacetime finds $\delta_{\cal J} = 0.065$ with $\chi_r^2 = 124.0/118$, i.e., $\Delta \chi^2 = -3.1$). 

Clearly, equally good fits to observational data by a finite ISCO stress Schwarzschild black hole and a vanishing ISCO stress Kerr black hole with rapid rotation ($a_\bullet \approx 0.84$) confirm, as suspected, that a parameter degeneracy exists.

\subsection{Formal spin-stress degeneracy }

\begin{figure}
\centering
\includegraphics[width=\linewidth]{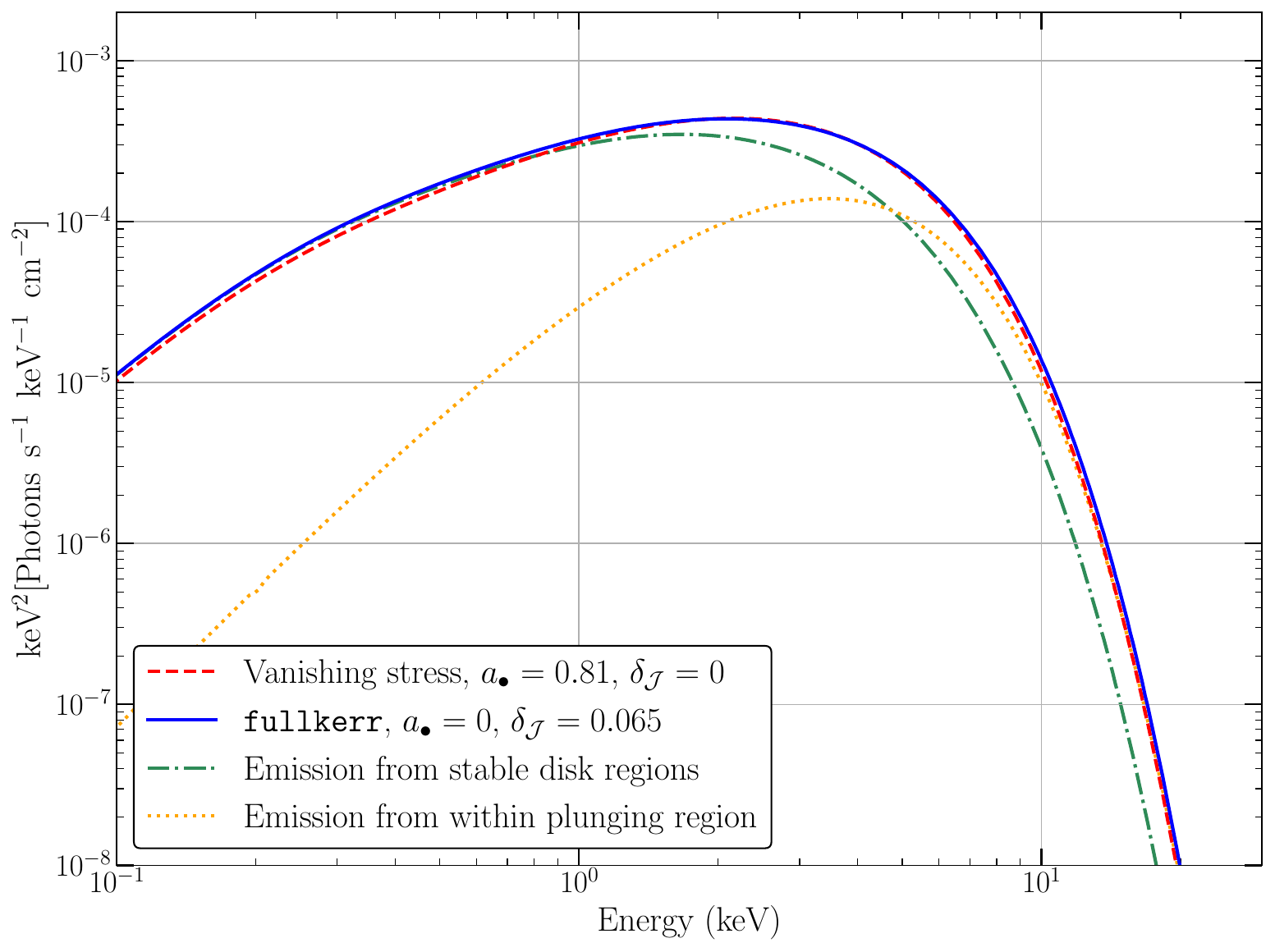}
\includegraphics[width=\linewidth]{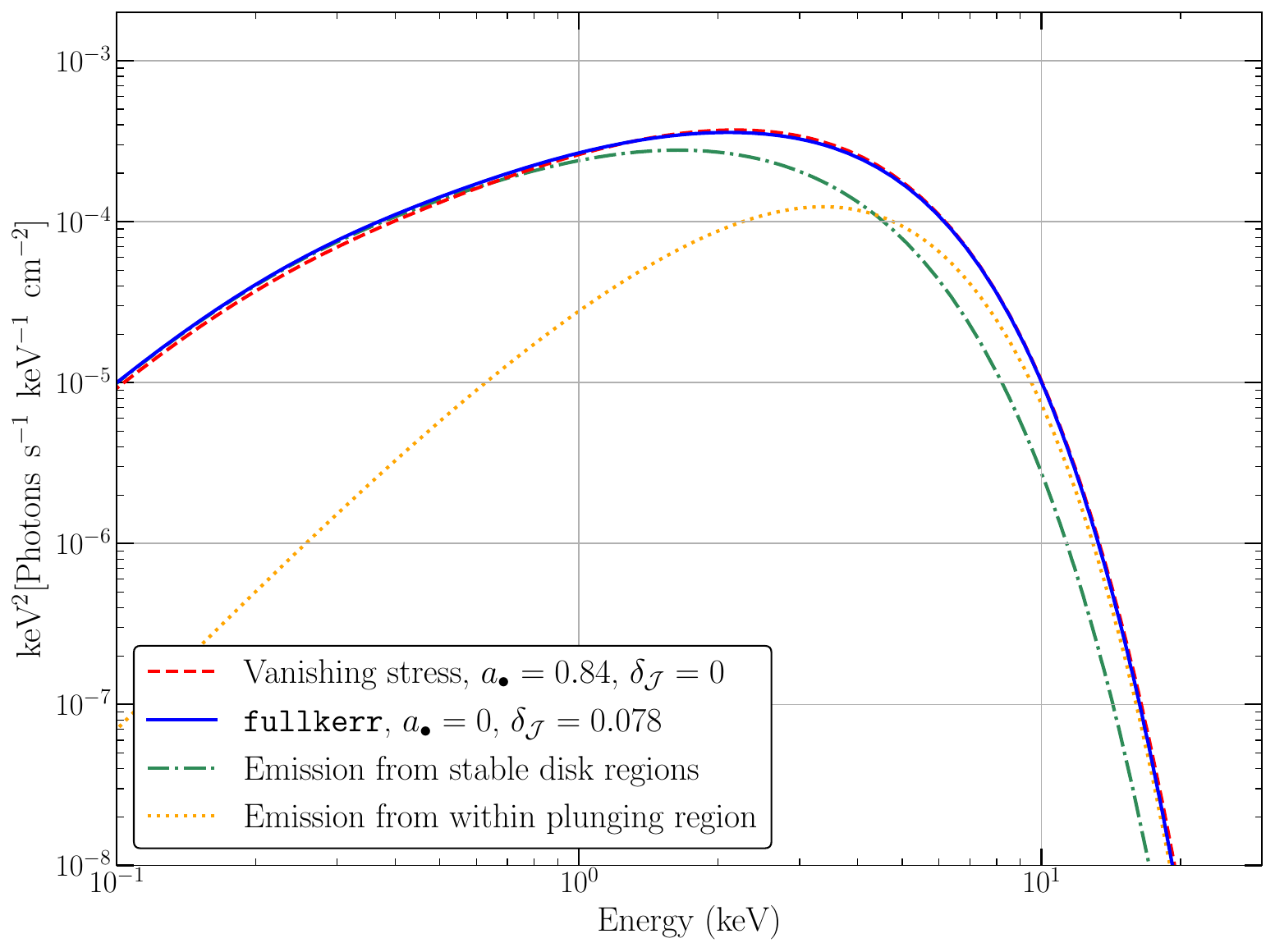}
\caption{{\bf Upper:} The unfolded X-ray spectra of the best-fitting models to the M33 X-7 observations taken on 2019-07-17. By a red dashed curve we show the best fitting disk model with zero ISCO temperature, and a high black hole spin $a_\bullet = 0.81$. By a blue solid curve we show the best fitting Schwarzschild model ($a_\bullet = 0$), which includes emission from within the plunging region with ISCO angular momentum parameter $\delta_{\cal J} = 0.065$.  We also split off the contribution to the total observed {\tt fullkerr} emission from disk regions outside (green dot-dashed) and inside (orange dashed) the ISCO. The plunging region contributes only a relatively small fraction of the {\it bolometric} luminosity of the disk system, but its contribution is vitally important in X-ray bands. The blue solid and red dashed curves are near-indistinguishably good fits to the data (see Figure \ref{fig:degeneracy}). {\bf Lower:} the same plot, but now for the observation taken on 2019-07-13.   }
\label{fig:components}
\end{figure}

\begin{figure}
\centering
\includegraphics[width=\linewidth]{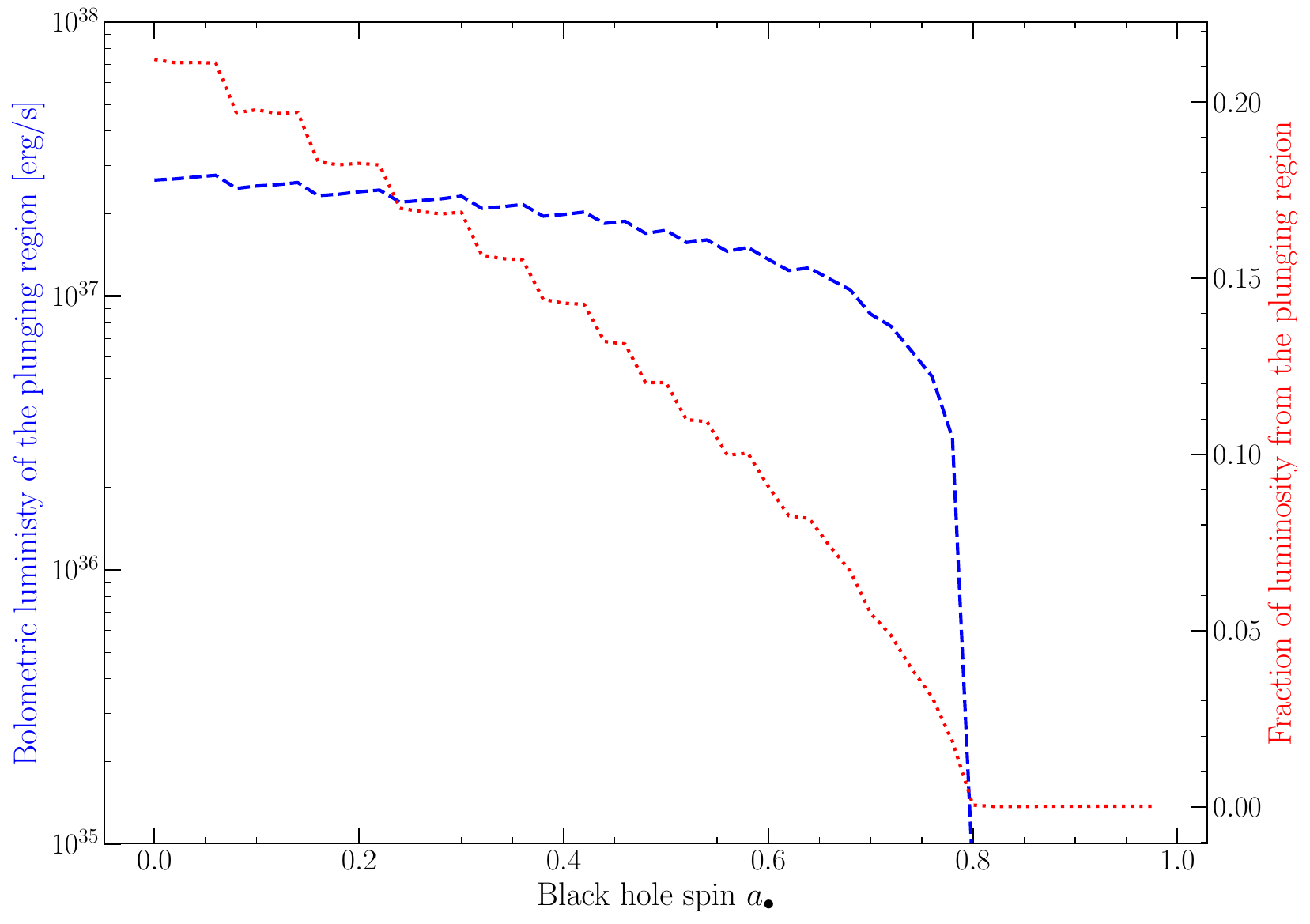}
\caption{ The properties of the plunging region emission of the best-fitting disk models fit to the M33 X-7 observations taken on 2019-07-17, as a function of the assumed black hole spin.  The left axis (blue dashed curve) shows the bolometric luminosity of the emission from the plunging region which is relatively insensitive to the assumed black hole spin, except for when the spin is assumed to be high $a_\bullet \gtrsim 0.75$, when a vanishing ISCO temperature disk becomes the best fitting solution. The right axis (red dotted curve) shows the fraction of the bolometric luminosity of the entire disk system which originates from within the plunging region. Both of these curves are made from the perspective of a distant observer orientated at $i = 75^\circ$ (the assumed value of our observations), and these results would be quantitatively (though not qualitatively) different for different inclinations. Note that the left axis  is plotted on a logarithmic scale while the right axis is plotted on a linear scale.   }
\label{fig:plunge}
\end{figure}

\begin{figure*}
    \centering
    \includegraphics[width=0.49\linewidth]{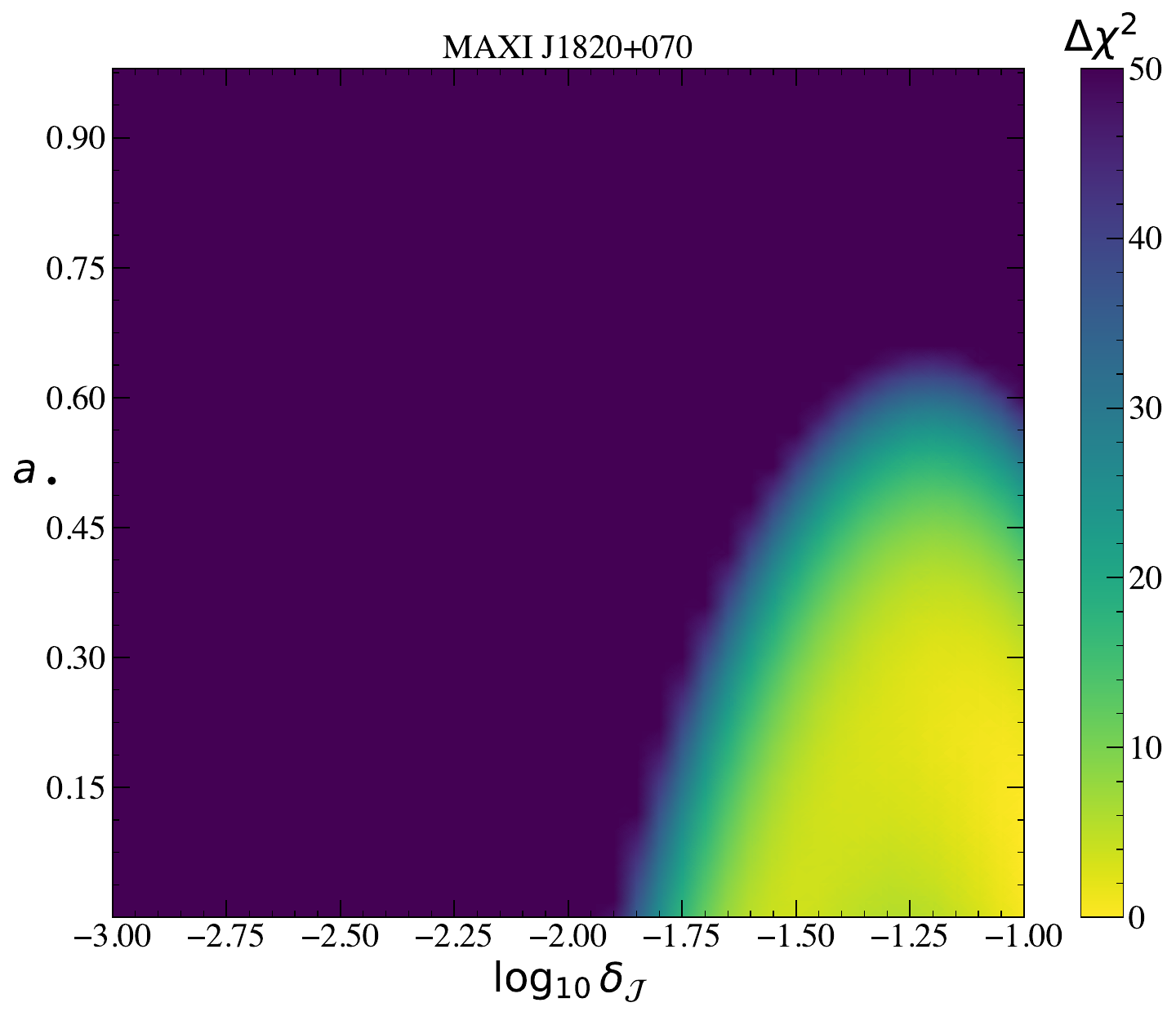}
    \includegraphics[width=0.49\linewidth]{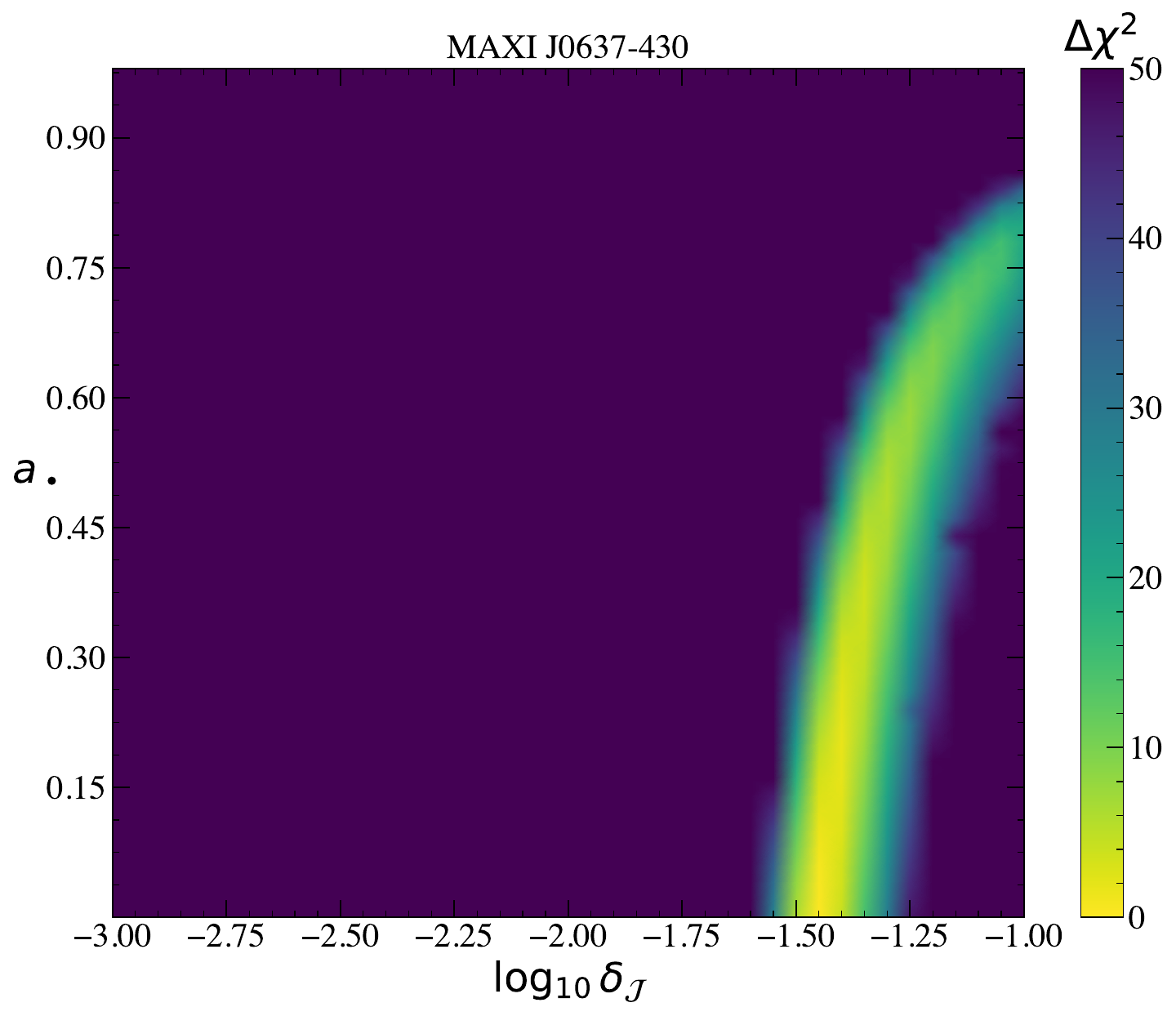}
    \caption{The spin-stress degeneracies for 
    MAXI J1820+070 (right panel) and MAXI J0637$-$430 (left panel). The best fitting parameters for all three systems favour low spins $a_\bullet \sim 0.2$ and moderate stresses $-1.5 \lesssim \log_{10}\delta_{\cal J} \lesssim -1.0$. Interestingly, when high quality data is available, the two MAXI sources no longer display a high-spin low-stress section of the degeneracy, as this is no longer an acceptable fit. The two dimensional spin-stress plane instead collapses down to the moderate-stress low-spin section of the M33 X-7 parameter degeneracy (see Figure \ref{fig:degeneracy}).  }
    \label{fig:3xrbs}
\end{figure*}

Having identified that a spin-stress degeneracy exists for the continuum X-ray spectrum of M33 X-7, in this section we examine this in more detail. Our key results are displayed in Figure (\ref{fig:degeneracy}).

The upper row of Fig. \ref{fig:degeneracy} shows the one-dimensional $\chi^2$-statistic as a function of black hole spin for both epochs considered in this work. In these two upper panels we see that the spin posteriors for finite ISCO stress disks (blue solid curves) are significantly broader than their vanishing ISCO stress counterparts (orange dashed curves), a result of the additional modeling flexibility available to a model which can include extra small-scale emission. At high spins the best-fitting  {\tt kerrbb} and {\tt fullkerr} models are identical, as they describe identical physical systems in the limit $\delta_{\cal J} \to 0$. This model identity can be seen reflected in the fit statistic as a function of black hole spin for both models at spins $a_\bullet > 0.8$.  

We note that for both epochs formally better fits are found for significantly lower spins $(a_\bullet \simeq 0.2)$ and a finite ISCO stress, when compared to a vanishing ISCO stress disk with $a_\bullet \simeq 0.8$, although the improvement is relatively minor ($\Delta \chi^2 = -6.9$ and $-3.1$ respectively, for the case when the spin is frozen to 0 rather than the stress), and the vanishing ISCO stress best fit is formally acceptable in a statistical sense.  If we were to combine multiple epochs in our M33 X-7 fitting, a spin value $a_\bullet \sim 0.2$ would become increasingly required in a statistical sense. It is our intention to discuss {\it degeneracies} in this work however, so we simply analyse the two epochs independently. 

These results imply that black hole spin constraints found with vanishing ISCO stress models can, in a statistical sense, be formally considered black hole spin {\it upper limits}. In a physical sense the value of the spin parameter inferred from a continuum X-ray spectrum is very sensitive to the chosen value of the ISCO stress parameter $\delta_{\cal J}$. 

This can be clearly seen in the two panels in the lower row of Fig. \ref{fig:degeneracy}, which show the two dimensional $a_\bullet-\delta_{\cal J}$ degeneracy, with $\chi^2$-statistic denoted by a colour bar (the colour-bar saturates at a level $\Delta \chi^2 = 30$ above the minimum, any model in this saturated region is therefore rejected at high confidence). These panels show that the degeneracy is rather sharp in the spin-stress plane. If the stress parameter is $\delta_{\cal J} \lesssim 0.01$, then the black hole spin inference is largely unaffected. However, over the range  $0.01 \lesssim \delta_{\cal J} \lesssim 0.1$ the best-fitting spin parameter varies hugely (from $a_\bullet \simeq 0.8$ to $a_\bullet < 0$). This exact parameter regime of $\delta_{\cal J}$ is what is required to reproduce the observed X-ray spectra of MAXI J1820 \citep[$\delta_{\cal J} \simeq 0.04$;][]{Mummery24Plunge} and MAXI J0637 \citep[$\delta_{\cal J} \simeq 0.03$;][]{Mummery24PlungeB}, which cannot be described adequately with a vanishing ISCO stress model. This is also the range spanned by all the GRMHD simulations the authors are aware of \citep[e.g.,][]{Shafee08, Noble10, Penna10, Schnittman16, Rule25}. 

We wish to point out that while the inclusion of emission sourced from within the plunging region is clearly very important for attempts to measure black hole spins, the majority of the {\it bolometric} luminosity of these disk systems is still sourced from the stable disk regions at radii $r > r_I$. This was discussed in detail in \cite{Mummery24Plunge} (see their Figures 7 \& 8), but we recap the key 
results here for the specific case of M33 X-7.  

In Figure \ref{fig:components} we show the unfolded X-ray spectra of the best-fitting models to the M33 X-7 observations taken on 2019-07-17. By a red dashed curve we show the best fitting disk model with zero ISCO temperature, and a high black hole spin $a_\bullet = 0.81$ (the formally best fitting spin for that epoch). By a blue solid curve we show the best fitting Schwarzschild model ($a_\bullet = 0$), which includes emission from within the plunging region with ISCO angular momentum parameter $\delta_{\cal J} = 0.065$.  We also split off the contribution to the total observed {\tt fullkerr} emission from disk regions outside (green dot-dashed) and inside (orange dashed) the ISCO. As can be seen visually, the plunging region contributes only a relatively small fraction of the {\it bolometric} luminosity of the M33 X-7 disk system, but its contribution is vitally important in X-ray bands. A large fraction of the extra emission (for moderately stressed disks) arises from the near- (but extra-)ISCO region, where additional dissipation occurs owing to the angular momentum flux from the intra-ISCO flow.  The blue solid and red dashed curves are near-indistinguishably good fits to the data (see Figure \ref{fig:degeneracy}). In the lower panel we repeat this experiment for the data taken on 2019-07-13. 

In Figure \ref{fig:plunge} we elucidate more quantitatively the luminosity emergent from the plunging region of M33 X-7, for different assumed black hole spins of the central hole.  The left axis (blue dashed curve; logarithmic scale) shows the bolometric luminosity of the emission from the plunging region which is relatively insensitive to the assumed black hole spin, except for when the spin is assumed to be high $a_\bullet \gtrsim 0.75$, when a  disk with a vanishing ISCO flux becomes the best fitting solution. The right axis (red dotted curve; linear scale) shows the fraction of the bolometric luminosity of the entire disk system which originates from within the plunging region. Both of these curves are made from the perspective of a distant observer orientated at $i = 75^\circ$ (the assumed value of our observations), and these results would be quantitatively (though not qualitatively) different for different inclinations.

\subsection{The spin-plunge degeneracy for MAXI J1820+070 and MAXI J0627-430}
We have identified a two-branch degeneracy in the spin-stress plane for M33 X-7, which can either be described by $a_\bullet \approx 0.85$ with $\delta_{\cal J} \to 0$, or by $-1<a_\bullet < 0.85$ for $0.01 < \delta_{\cal J} < 0.1$. The question then becomes where do sources with higher quality data (i.e., brighter, Galactic, sources) lie.

In Figure \ref{fig:3xrbs} we repeat the experiment\footnote{Note that this means that $\xi$ is fixed to zero, a parameter which was allowed to vary in both \citealt{Mummery24Plunge} and \citealt{Mummery24PlungeB}, leading to slightly changed best-fit parameters in this plot when compared to those papers. Letting $\xi$ vary (for any of the sources in this paper) does not lead to a changed degeneracy profile, merely better fits to the data for moderately stressed disks (this is unsurprising as there is another free parameter).} of the previous section, but now for MAXI J1820+070 and MAXI J0637$-$430.  These two sources require plunging region emission at high statistical significance \citep{Mummery24Plunge, Mummery24PlungeB}. It is important to test whether these sources also lie in the same low-spin moderate-stress region of parameter space of the low-spin branch of M33 X-7. Figure \ref{fig:3xrbs} shows that they do.  MAXI J1820+070 has four epochs where significant emission from the plunging region is detected \citep{Fabian20, Mummery24Plunge}, while MAXI J0637$-$430 has three \citep{Mummery24PlungeB}. For these two sources we simply use the brightest epoch in both cases (i.e., the spectra with the best data), and we have verified that our results are unchanged if we choose any of the other epochs of data for either source.  

It is clear from a visual inspection of Figure \ref{fig:3xrbs} that when plunging region emission is detectable the high spin low-stress solution drops away, leaving only the low-spin moderate-stress region of the M33 X-7 degeneracy. The best fitting parameters for all three systems favour low spins $a_\bullet \sim 0.2$ and moderate stresses $-1.5 \lesssim \log_{10}\delta_{\cal J} \lesssim -1.0$.  This suggests that when high quality data is available the high-spin low-stress region of parameter space is ruled out.  

\section{Discussion and conclusions}\label{conc}
In this work we have examined the ability of thermal continuum methods to constrain black hole spins in the presence of emission from the plunging region, ubiquitously seen in full numerical simulations of the accretion process. 

The formal statement we can make is that black hole spin constraints found with classical vanishing ISCO stress models can, in a statistical sense, be formally considered black hole spin {\it upper limits}. This of course does not imply that these existing constraints are wrong (this is not a general claim one can make), simply that they could be biased, and if they are biased they are biased towards high spins. There are generally two branches available to disk continuum spectrum fits, a high-spin low-stress branch and a low-spin moderate-stress branch.

When data are available that shows signatures of the plunging region this is associated with the exact same regions of parameter space inhabited by the low-spin moderate-stress branch, suggesting that when high quality data are available the high-spin low-stress region of parameter space is ruled out.   

Clearly therefore it is essential that the physics of the plunging region is understood in detail, and for a data-modeling perspective this means that robust properties of the stress parameter $\delta_{\cal J}$ are determined. The spin-stress degeneracy is rather sharp, by which we mean it is a strong function of $\delta_{\cal J}$ in the relevant parameter regime. If the stress parameter is $\delta_{\cal J} \lesssim 0.01$, then continuum fitting black hole spin inference is largely unaffected. However, over the range  $0.01 \lesssim \delta_{\cal J} \lesssim 0.1$ the best-fitting spin parameter varies hugely (from $a_\bullet \simeq 0.8$ to $a_\bullet < 0$ for M33 X-7). This exact parameter regime of $\delta_{\cal J}$ is what is required to reproduce the observed X-ray spectra of MAXI J1820 \citep{Mummery24Plunge} and MAXI J0637 \citep{Mummery24PlungeB}, which cannot be described adequately with a vanishing ISCO stress model. This possibly implies that all astronomical disk systems may inhabit this particularly sensitive region of parameter space, where spin inference is at its most delicate. 

\subsection{The importance of simulations with full radiation physics}
The parameter regime of interesting plunging region physics ($0.01 \lesssim \delta_{\cal J} \lesssim 0.1$) is motivated by over a decade of GRMHD simulations of the accretion process. While these simulations have become increasingly sophisticated (in terms of e.g., the number of simulations, their resolution, the disk thicknesses probed, the magnetic field strengths and geometry, the range of black hole spins, etc.) the inclusion of full radiation physics (i.e., self consistently evolving the stress-energy tensor of the radiation field) has only just become possible \citep{Stone24}. 

The existing simulations, including those discussed in section \ref{physics}, utilised {\it ad-hoc} cooling functions in their evolution. What this means on a technical level is that, rather than evolving the radiation field explicitly, the equation of energy-momentum conservation of the flow is modified to have a sink function $S^\nu$ included
\begin{equation}
	\nabla_\mu T^{\mu \nu} = - S^\nu, 
\end{equation}
which has some functional form which aims (typically) to remove energy and momentum in the system in the manner in which (it is hoped) approximates what a photon field would do. Various parameterisations of this sink function have been used in the literature \citep[e.g.][who popularised two of the most commonly used forms]{Noble10, Penna10}, but this approach no doubt misses important physics.  

It is essential that the experiments discussed in section \ref{physics} are repeated with the upgraded {\tt AthenaK} code, and that realistic flux profiles are computed down to horizon scales. If these full-radiation physics experiments confirm the picture formed from a decade of GRMHD simulations then it appears clear that currently published spin measurements may be in substantial error, and a broad re-evaluation will likely have to be performed. 

\subsection{An agreement between all methods?}
While we have focussed on the existing discrepancy between GW and EM spin inferences in this work, the results of this paper can in fact be interpreted as a coming together of all approaches to inferring  black hole properties. If gravitational wave sources are in fact telling us that astronomical (stellar mass) black holes are slowly rotating, then our work suggests that this means that black hole disk systems are all moderately stressed. 

This moderately stressed result is, however, exactly what every GRMHD simulations has been finding for over a decade now. Taken literally, GRMHD simulations tell us that we should be fitting soft state black hole X-ray spectra with ISCO stress parameters in the range $\delta_{\cal J} \sim 0.01-0.1$ which, we have shown here, imply lower spins. It seems possible that  the physical characteristics of the X-ray binary and LIGO-merger populations are in much stronger accord than previously appreciated. This should be checked as a matter of urgency through population-level studies of soft-state XRB spectra. 

\subsection{Future routes to breaking the degeneracy} 

The results presented in this paper suggest that inferring the black hole spin from the X-ray continuum emission is significantly more difficult than has previously been suggested in the literature, owing to the fact that the typical additional dissipation observed in the inner disk regions in GRMHD simulations puts black hole disks (of all spins) in highly degenerate regions of parameter space. 

We stress that we do not believe that it will be impossible, with future modeling advances, to constrain black hole spins from their continuum X-ray emission. The reason for this is that while the locally emitted flux from within the plunging region of low-spin black holes and the main body of the disk of high-spin black holes are similar (see simulations Fig. \ref{fig:madsim_all}, or theoretical calculations Fig. \ref{fig:fluxes}), the plunging region is physically distinct from the main body of the disk in a number of observationally meaningful ways. These differences will, likely, imprint additional signatures into the data, which may well be able to be disentangled once more sophisticated models are developed. We discuss possible avenues for future study here. 

{\it Energy dependent variability --} while the absolute scale of variability of black hole accretion flows in the soft (thermal-dominated) state is low (significantly lower than in the hard state), the energy-dependence of this variability may well offer insight into the importance (or not) of the plunging region in a given observation. \cite{Fabian20} demonstrated that the emission in MAXI J1820+070 at $E\sim 8$ keV \citep[the region of the spectrum dominated by photons emitted from the plunging region][]{Mummery24Plunge} was significantly more variable than emission at (e.g.,) $E\sim 5$ keV (dominated by the main body of the disk). It seems likely that this is physical -- emission from within the plunging region is sensitive to the turbulent transport of angular momentum in highly relativistic regimes (i.e., to $\delta_{\cal J}$). As this parameter is intricately tied to the properties MHD turbulence, it seems likely that it will vary stochastically. It is also known from GRMHD simulations that accretion within the plunging region is often mediated through spiral structures \citep{MummeryStone24}, which also seem likely to act to amplify variability of the emission form this region.  Future theoretical studies, along with comparisons to systems with known plunging region emission, may well offer valuable insight into whether spectral-variability studies can break the spin-stress degeneracy discussed here. 

{\it Polarization --} the recent launching of the Imaging X-ray Polarimetry Explorer (IXPE) satellite \citep[e.g.][]{Weisskopf22} now allows polarisation information (the degree and angle of the polarisation vector) for sources accreting in the soft state to be measured \citep[e.g.,][]{Steiner24}. As we have discussed throughout this paper, photons emitted from the same absolute radial scale have differing probabilities of being captured by the black hole depending on whether they were emitted from plunging or stably-orbiting disk material. They also, for the exact same physical reasons, have differing probabilities of illuminating the far side of the disk \citep[something also discussed in][]{AgolKrolik00}, which may well imprint clear polarisation signatures in the emission \citep[as scattering off a disk atmosphere generally leads to polarisation signatures][]{Schnittman09}. While the number of sources with detailed X-ray spectral and polarisation information available is naturally smaller than those with just spectral information, these sources may well be prime targets for future spin constraints, and it seems possible this may well break the degeneracy discussed here. Detailed models of polarisation signatures including emission from within the plunging region \citep[i.e., an extension of][to include the models for intra-ISCO emission used here]{Schnittman09} will be required for this science case,  though this should not be particularly difficult. 

{\it Reflection --} in this paper we have focused exclusively  on the properties of the continuum emission from black hole disks, leaving an analysis of reflection features to a future work. It is worth mentioning however that owing to the very different properties of the flow which are probed by reflection and continuum fitting techniques (broadly speaking, the density, velocity and illuminating flux via reflection versus the liberated flux via continuum fitting), reflection techniques do offer a promising route to breaking spin-plunging region degeneracies. This, of course, will only be the case if reflection models include the plunging region (so that the degeneracy can be self consistently tested). Classical calculations \citep{Reynolds97} and modern numerical simulations \citep{Kinch21} suggest that the plunging region may well play an important role in reflection studies, particularly at larger accretion rates (when the density of the flow is higher, preventing over ionization of the plunging flow, see e.g., \citealt{Wilkins20}). 

{\it Trends with accretion rate --} the analysis of M33 X-7 presented here included two epochs where the luminosity of the source was high. Similarly, MAXI J1820+070 and MAXI J0637$-$430 were examined in bright soft states. An important question, one to which simulations will offer valuable insight, is how the physics of the plunging region emission (in effect the value of $\delta_{\cal J}$) varies with luminosity (or effective $\dot M$). The spin of the black hole will of course not vary epoch-to-epoch, but if $\delta_{\cal J}$ were to systematically varies with $\dot M$ (something for which we stress there is, as far as we are aware, little current numerical evidence either for or against), then there may well be signatures of the plunging region which reveal themselves in sources with luminosities which vary over orders of magnitude. We note that across four epochs MAXI J1820+070 showed a broadly constant value of $\delta_{\cal J}$ as $\dot M$ fell by 50\% and similarly MAXI J0637$-$430 also showed a broadly consistent $\delta_{\cal J}$ across a factor 2 change in $\dot M$. Further observational work is required to determine if this is typical behavior. 

{\it High quality X-ray data and detecting the plunging region --} while, as we have argued here, disk systems with a vanishing ISCO flux and a high black hole spin are likely to be highly degenerate with more slowly (or not at all) spinning black holes with substantial dissipation near and within the plunging region, the same is not true in reverse. Some disk systems, for example the two sources MAXI J1820+070 and MAXI J0637-430, cannot be described by simple vanishing ISCO flux disk systems for any black hole spin, even with the added complication of detectable coronal components \citep{Mummery24Plunge, Mummery24PlungeB}. When plunging region emission is detected, constraints on the black hole spin can be placed which are much stronger than those discussed here in the case of M33 X-7. The physics of this is simple, for photons from the plunging region to be detected, the plunging region must have a substantial radial extent. As the radial extent of the plunging region drops to zero in the $a_\bullet \to 1$ limit, high spins are then strongly disfavored by the data. Ultimately, the best route to breaking the spin-plunge degeneracies may well be in collecting sufficiently high quality soft-state data so as to seek to break the degeneracy between the plunging region and the black hole spin from the plunging region direction. As argued in \cite{Mummery24PlungeB} this will likely require {\it NuSTAR} data, owing to the necessity of including high photon energy ($E \gtrsim 20$ keV) data in continuum fitting exercises. These high energy observations are included so that the signatures of the
plunging region are not modelled out by coronal components with the wrong power-law indices which can, over a restricted interval, successfully mimic the additional photon flux sourced from within the ISCO. 


\subsection{Summary and conclusions}

In this paper we have demonstrated, using a combination of  theoretical calculations, numerical simulations and observational data, that the inclusion of emission from within the innermost stable circular orbit (ISCO) results in a black hole with a low spin producing a thermal continuum X-ray spectrum that mimics that produced by a much more rapidly rotating black hole surrounded by a disk with no emission from within the ISCO. We were motivated partly by a claimed discrepancy between the spins inferred from GW techniques (GW binaries), and those from EM techniques (HMXRBs). 

A similar argument (that EM spin measurements may be in substantial error) was recently made by \cite{Belczynski24}. This argument, however, was premised on the potential contamination of the X-ray spectrum of XRB systems by a ``warm corona''. Such a warm corona is not modelled physically (it is simply introduced as an {\it ad hoc} modelling component not coupled to the state of the system at all), and must be very hot to prevent atomic lines (which are not observed) appearing in the X-ray spectrum while only producing ``warm'' emission \citep[see discussion in][]{Kara25}. It is not clear if such a hot-yet-warm corona can exist, or if it would be stable, and we are not aware of any claims that such a corona has been observed in numerical simulations of the accretion process. 

The argument put forward here is much simpler, and is based on better understood physics. In effect we claim that missing small-radial-scale physics in traditional models fit to continuum X-ray spectra leads to a natural tendency of these classical models to return high black hole spins when fit to data. This missing physics is simply hot plasma which continues to radiate down to the horizon of a black holes spacetime. This plasma is kept hot as it approaches and crosses the ISCO by magnetohydrodynamical stresses, with magnetic field strengths growing throughout the plunging region due to flux freezing (Alfv\'ens theorem) and an accelerating fluid dragging field lines into a shrinking volume. 

The key result  of this paper is that the exact region of parameter space implied by GRMHD simulations, and observational analyses of XRB systems with detectable plunging region emission, is precisely the parameter regime which would result in a systematic lowering of continuum fitting spin measurements more broadly.   It is clear that the definitive answer to this question is close to hand, and future dedicated simulations with full radiation physics should shortly provide confirmation of where on the spin-plunging region degeneracy a typical disk system lies. 

\section*{Acknowledgments} 
This work was supported by a Leverhulme Trust International Professorship grant [number LIP-202-014]. J.J. acknowledges support from the Leverhulme Trust, Isaac Newton Trust and St Edmund's College. AI acknowledges support from the Royal Society. For the purpose of Open Access, AM has applied a CC BY public copyright licence to any Author Accepted Manuscript version arising from this submission. 
 
\section*{Data accessibility statement}
The {\tt XSPEC} model {\tt fullkerr} is available at the following GitHub repository:  \url{https://github.com/andymummeryastro/fullkerr}. The X-ray spectra of M33 X-7, MAXI J1820+070 and MAXI J0637$-$430 are all publicly available. All of the data used in this work can be downloaded from the HEASARC website \url{https://heasarc.gsfc.nasa.gov}.

\label{lastpage}

\bibliographystyle{mnras}
\bibliography{andy}

\begin{thebibliography}{}
\makeatletter
\relax
\def\mn@urlcharsother{\let\do\@makeother \do\$\do\&\do\#\do\^\do\_\do\%\do\~}
\def\mn@doi{\begingroup\mn@urlcharsother \@ifnextchar [ {\mn@doi@} {\mn@doi@[]}}
\def\mn@doi@[#1]#2{\def\@tempa{#1}\ifx\@tempa\@empty \href {http://dx.doi.org/#2} {doi:#2}\else \href {http://dx.doi.org/#2} {#1}\fi \endgroup}
\def\mn@eprint#1#2{\mn@eprint@#1:#2::\@nil}
\def\mn@eprint@arXiv#1{\href {http://arxiv.org/abs/#1} {{\tt arXiv:#1}}}
\def\mn@eprint@dblp#1{\href {http://dblp.uni-trier.de/rec/bibtex/#1.xml} {dblp:#1}}
\def\mn@eprint@#1:#2:#3:#4\@nil{\def\@tempa {#1}\def\@tempb {#2}\def\@tempc {#3}\ifx \@tempc \@empty \let \@tempc \@tempb \let \@tempb \@tempa \fi \ifx \@tempb \@empty \def\@tempb {arXiv}\fi \@ifundefined {mn@eprint@\@tempb}{\@tempb:\@tempc}{\expandafter \expandafter \csname mn@eprint@\@tempb\endcsname \expandafter{\@tempc}}}

\bibitem[\protect\citeauthoryear{{Abbott} et~al.}{{Abbott} et~al.}{2023a}]{Abbott23}
{Abbott} R.,  et~al., 2023a, \mn@doi [Physical Review X] {10.1103/PhysRevX.13.011048}, \href {https://ui.adsabs.harvard.edu/abs/2023PhRvX..13a1048A} {13, 011048}

\bibitem[\protect\citeauthoryear{Abbott et~al.}{Abbott et~al.}{2023b}]{KAGRA:2021vkt}
Abbott R.,  et~al., 2023b, \mn@doi [Phys. Rev. X] {10.1103/PhysRevX.13.041039}, 13, 041039

\bibitem[\protect\citeauthoryear{{Abramowicz}, {Lanza}  \& {Percival}}{{Abramowicz} et~al.}{1997}]{Abramowicz97}
{Abramowicz} M.~A.,  {Lanza} A.,   {Percival} M.~J.,  1997, \mn@doi [\apj] {10.1086/303869}, \href {https://ui.adsabs.harvard.edu/abs/1997ApJ...479..179A} {479, 179}

\bibitem[\protect\citeauthoryear{{Agol} \& {Krolik}}{{Agol} \& {Krolik}}{2000}]{AgolKrolik00}
{Agol} E.,  {Krolik} J.~H.,  2000, \mn@doi [\apj] {10.1086/308177}, \href {https://ui.adsabs.harvard.edu/abs/2000ApJ...528..161A} {528, 161}

\bibitem[\protect\citeauthoryear{{Arnaud}}{{Arnaud}}{1996}]{Arnaud96}
{Arnaud} K.~A.,  1996, in {Jacoby} G.~H.,  {Barnes} J.,  eds,  Astronomical Society of the Pacific Conference Series Vol. 101, Astronomical Data Analysis Software and Systems V. p.~17

\bibitem[\protect\citeauthoryear{{Balbus} \& {Hawley}}{{Balbus} \& {Hawley}}{1991}]{BalbusHawley91}
{Balbus} S.~A.,  {Hawley} J.~F.,  1991, \mn@doi [\apj] {10.1086/170270}, \href {https://ui.adsabs.harvard.edu/abs/1991ApJ...376..214B} {376, 214}

\bibitem[\protect\citeauthoryear{{Balbus} \& {Hawley}}{{Balbus} \& {Hawley}}{1998}]{BalbusHawley98}
{Balbus} S.~A.,  {Hawley} J.~F.,  1998, \mn@doi [Reviews of Modern Physics] {10.1103/RevModPhys.70.1}, \href {https://ui.adsabs.harvard.edu/abs/1998RvMP...70....1B} {70, 1}

\bibitem[\protect\citeauthoryear{{Balbus} \& {Papaloizou}}{{Balbus} \& {Papaloizou}}{1999}]{BalbPap99}
{Balbus} S.~A.,  {Papaloizou} J. C.~B.,  1999, \mn@doi [\apj] {10.1086/307594}, \href {https://ui.adsabs.harvard.edu/abs/1999ApJ...521..650B} {521, 650}

\bibitem[\protect\citeauthoryear{{Bardeen}}{{Bardeen}}{1970}]{Bardeen1970}
{Bardeen} J.~M.,  1970, \mn@doi [\nat] {10.1038/226064a0}, \href {https://ui.adsabs.harvard.edu/abs/1970Natur.226...64B} {226, 64}

\bibitem[\protect\citeauthoryear{{Bardeen}, {Press}  \& {Teukolsky}}{{Bardeen} et~al.}{1972}]{Bardeen72}
{Bardeen} J.~M.,  {Press} W.~H.,   {Teukolsky} S.~A.,  1972, \mn@doi [\apj] {10.1086/151796}, \href {https://ui.adsabs.harvard.edu/abs/1972ApJ...178..347B} {178, 347}

\bibitem[\protect\citeauthoryear{{Belczynski} et~al.,}{{Belczynski} et~al.}{2020}]{belczynski20}
{Belczynski} K.,  et~al., 2020, \mn@doi [\aap] {10.1051/0004-6361/201936528}, \href {https://ui.adsabs.harvard.edu/abs/2020A&A...636A.104B} {636, A104}

\bibitem[\protect\citeauthoryear{{Belczynski}, {Done}, {Hagen}, {Lasota}  \& {Sen}}{{Belczynski} et~al.}{2024}]{Belczynski24}
{Belczynski} K.,  {Done} C.,  {Hagen} S.,  {Lasota} J.-P.,   {Sen} K.,  2024, \mn@doi [\aap] {10.1051/0004-6361/202450229}, \href {https://ui.adsabs.harvard.edu/abs/2024A&A...690A..21B} {690, A21}

\bibitem[\protect\citeauthoryear{{Davis} \& {El-Abd}}{{Davis} \& {El-Abd}}{2019}]{Davis19}
{Davis} S.~W.,  {El-Abd} S.,  2019, \mn@doi [\apj] {10.3847/1538-4357/ab05c5}, \href {https://ui.adsabs.harvard.edu/abs/2019ApJ...874...23D} {874, 23}

\bibitem[\protect\citeauthoryear{{Davis}, {Blaes}, {Hubeny}  \& {Turner}}{{Davis} et~al.}{2005}]{Davis05}
{Davis} S.~W.,  {Blaes} O.~M.,  {Hubeny} I.,   {Turner} N.~J.,  2005, \mn@doi [\apj] {10.1086/427278}, \href {https://ui.adsabs.harvard.edu/abs/2005ApJ...621..372D} {621, 372}

\bibitem[\protect\citeauthoryear{{Davis}, {Done}  \& {Blaes}}{{Davis} et~al.}{2006}]{Davis06}
{Davis} S.~W.,  {Done} C.,   {Blaes} O.~M.,  2006, \mn@doi [\apj] {10.1086/505386}, \href {https://ui.adsabs.harvard.edu/abs/2006ApJ...647..525D} {647, 525}

\bibitem[\protect\citeauthoryear{{Dhang}, {Dexter}  \& {Begelman}}{{Dhang} et~al.}{2025}]{Dhang25}
{Dhang} P.,  {Dexter} J.,   {Begelman} M.~C.,  2025, \mn@doi [\apj] {10.3847/1538-4357/ada76e}, \href {https://ui.adsabs.harvard.edu/abs/2025ApJ...980..203D} {980, 203}

\bibitem[\protect\citeauthoryear{{Fabian}, {Rees}, {Stella}  \& {White}}{{Fabian} et~al.}{1989}]{Fabian1989}
{Fabian} A.~C.,  {Rees} M.~J.,  {Stella} L.,   {White} N.~E.,  1989, \mn@doi [\mnras] {10.1093/mnras/238.3.729}, \href {https://ui.adsabs.harvard.edu/abs/1989MNRAS.238..729F} {238, 729}

\bibitem[\protect\citeauthoryear{{Fabian} et~al.,}{{Fabian} et~al.}{2020}]{Fabian20}
{Fabian} A.~C.,  et~al., 2020, \mn@doi [\mnras] {10.1093/mnras/staa564}, \href {https://ui.adsabs.harvard.edu/abs/2020MNRAS.493.5389F} {493, 5389}

\bibitem[\protect\citeauthoryear{{Fishbach} \& {Kalogera}}{{Fishbach} \& {Kalogera}}{2022}]{Fishbach22}
{Fishbach} M.,  {Kalogera} V.,  2022, \mn@doi [\apjl] {10.3847/2041-8213/ac64a5}, \href {https://ui.adsabs.harvard.edu/abs/2022ApJ...929L..26F} {929, L26}

\bibitem[\protect\citeauthoryear{{Gammie}}{{Gammie}}{1999}]{Gammie99}
{Gammie} C.~F.,  1999, \mn@doi [\apjl] {10.1086/312207}, \href {https://ui.adsabs.harvard.edu/abs/1999ApJ...522L..57G} {522, L57}

\bibitem[\protect\citeauthoryear{{Hawley} \& {Krolik}}{{Hawley} \& {Krolik}}{2001}]{HawleyKrolik01}
{Hawley} J.~F.,  {Krolik} J.~H.,  2001, \mn@doi [\apj] {10.1086/318678}, \href {https://ui.adsabs.harvard.edu/abs/2001ApJ...548..348H} {548, 348}

\bibitem[\protect\citeauthoryear{{Kara} \& {Garc{\'\i}a}}{{Kara} \& {Garc{\'\i}a}}{2025}]{Kara25}
{Kara} E.,  {Garc{\'\i}a} J.,  2025, \mn@doi [arXiv e-prints] {10.48550/arXiv.2503.22791}, \href {https://ui.adsabs.harvard.edu/abs/2025arXiv250322791K} {p. arXiv:2503.22791}

\bibitem[\protect\citeauthoryear{{Kinch}, {Schnittman}, {Noble}, {Kallman}  \& {Krolik}}{{Kinch} et~al.}{2021}]{Kinch21}
{Kinch} B.~E.,  {Schnittman} J.~D.,  {Noble} S.~C.,  {Kallman} T.~R.,   {Krolik} J.~H.,  2021, \mn@doi [\apj] {10.3847/1538-4357/ac2b9a}, \href {https://ui.adsabs.harvard.edu/abs/2021ApJ...922..270K} {922, 270}

\bibitem[\protect\citeauthoryear{{Krolik }}{{Krolik }}{1999}]{Krolik99}
{Krolik } J.~H.,  1999, \mn@doi [\apjl] {10.1086/311979}, \href {https://ui.adsabs.harvard.edu/abs/1999ApJ...515L..73K} {515, L73}

\bibitem[\protect\citeauthoryear{{Krolik} \& {Hawley}}{{Krolik} \& {Hawley}}{2002}]{KrolikHawley02}
{Krolik} J.~H.,  {Hawley} J.~F.,  2002, \mn@doi [\apj] {10.1086/340760}, \href {https://ui.adsabs.harvard.edu/abs/2002ApJ...573..754K} {573, 754}

\bibitem[\protect\citeauthoryear{{Lan{\v{c}}ov{\'a}} et~al.,}{{Lan{\v{c}}ov{\'a}} et~al.}{2019}]{Lancova19}
{Lan{\v{c}}ov{\'a}} D.,  et~al., 2019, \mn@doi [\apjl] {10.3847/2041-8213/ab48f5}, \href {https://ui.adsabs.harvard.edu/abs/2019ApJ...884L..37L} {884, L37}

\bibitem[\protect\citeauthoryear{{Lasota} \& {Abramowicz}}{{Lasota} \& {Abramowicz}}{2024}]{Lasota24}
{Lasota} J.-P.,  {Abramowicz} M.,  2024, \mn@doi [arXiv e-prints] {10.48550/arXiv.2410.06200}, \href {https://ui.adsabs.harvard.edu/abs/2024arXiv241006200L} {p. arXiv:2410.06200}

\bibitem[\protect\citeauthoryear{{Lazar} et~al.,}{{Lazar} et~al.}{2021}]{Lazar21}
{Lazar} H.,  et~al., 2021, \mn@doi [\apj] {10.3847/1538-4357/ac1bab}, \href {https://ui.adsabs.harvard.edu/abs/2021ApJ...921..155L} {921, 155}

\bibitem[\protect\citeauthoryear{{Li}, {Zimmerman}, {Narayan}  \& {McClintock}}{{Li} et~al.}{2005}]{Li05}
{Li} L.-X.,  {Zimmerman} E.~R.,  {Narayan} R.,   {McClintock} J.~E.,  2005, \mn@doi [\apjs] {10.1086/428089}, \href {https://ui.adsabs.harvard.edu/abs/2005ApJS..157..335L} {157, 335}

\bibitem[\protect\citeauthoryear{{Liotine}, {Zevin}, {Berry}, {Doctor}  \& {Kalogera}}{{Liotine} et~al.}{2023}]{liotine23}
{Liotine} C.,  {Zevin} M.,  {Berry} C. P.~L.,  {Doctor} Z.,   {Kalogera} V.,  2023, \mn@doi [\apj] {10.3847/1538-4357/acb8b2}, \href {https://ui.adsabs.harvard.edu/abs/2023ApJ...946....4L} {946, 4}

\bibitem[\protect\citeauthoryear{{Liu}, {McClintock}, {Narayan}, {Davis}  \& {Orosz}}{{Liu} et~al.}{2008}]{Liu08}
{Liu} J.,  {McClintock} J.~E.,  {Narayan} R.,  {Davis} S.~W.,   {Orosz} J.~A.,  2008, \mn@doi [\apjl] {10.1086/588840}, \href {https://ui.adsabs.harvard.edu/abs/2008ApJ...679L..37L} {679, L37}

\bibitem[\protect\citeauthoryear{{Lynden-Bell}}{{Lynden-Bell}}{1969}]{LyndenBell69}
{Lynden-Bell} D.,  1969, \mn@doi [\nat] {10.1038/223690a0}, \href {https://ui.adsabs.harvard.edu/abs/1969Natur.223..690L} {223, 690}

\bibitem[\protect\citeauthoryear{{Lynden-Bell} \& {Pringle}}{{Lynden-Bell} \& {Pringle}}{1974}]{LBP74}
{Lynden-Bell} D.,  {Pringle} J.~E.,  1974, \mn@doi [MNRAS] {10.1093/mnras/168.3.603}, \href {https://ui.adsabs.harvard.edu/abs/1974MNRAS.168..603L} {168, 603}

\bibitem[\protect\citeauthoryear{{McClintock}, {Narayan}  \& {Steiner}}{{McClintock} et~al.}{2014}]{McClintock14}
{McClintock} J.~E.,  {Narayan} R.,   {Steiner} J.~F.,  2014, \mn@doi [Space Science Series] {10.1007/s11214-013-0003-9}, \href {https://ui.adsabs.harvard.edu/abs/2014SSRv..183..295M} {183, 295}

\bibitem[\protect\citeauthoryear{{Mummery}}{{Mummery}}{2025}]{Mummery25}
{Mummery} A.,  2025, \mn@doi [\mnras] {10.1093/mnras/staf060}, \href {https://ui.adsabs.harvard.edu/abs/2025MNRAS.537.1963M} {537, 1963}

\bibitem[\protect\citeauthoryear{{Mummery} \& {Balbus}}{{Mummery} \& {Balbus}}{2023}]{MummeryBalbus2023}
{Mummery} A.,  {Balbus} S.,  2023, \mn@doi [\mnras] {10.1093/mnras/stad641}, \href {https://ui.adsabs.harvard.edu/abs/2023MNRAS.521.2439M} {521, 2439}

\bibitem[\protect\citeauthoryear{{Mummery} \& {Stone}}{{Mummery} \& {Stone}}{2024}]{MummeryStone24}
{Mummery} A.,  {Stone} J.~M.,  2024, \mn@doi [\mnras] {10.1093/mnras/stae1643}, \href {https://ui.adsabs.harvard.edu/abs/2024MNRAS.532.3395M} {532, 3395}

\bibitem[\protect\citeauthoryear{{Mummery}, {Mori}  \& {Balbus}}{{Mummery} et~al.}{2024a}]{MummeryMori24}
{Mummery} A.,  {Mori} F.,   {Balbus} S.,  2024a, \mn@doi [\mnras] {10.1093/mnras/stae701}, \href {https://ui.adsabs.harvard.edu/abs/2024MNRAS.529.1900M} {529, 1900}

\bibitem[\protect\citeauthoryear{Mummery, Ingram, Davis  \& Fabian}{Mummery et~al.}{2024b}]{Mummery24Plunge}
Mummery A.,  Ingram A.,  Davis S.,   Fabian A.,  2024b, \mn@doi [Monthly Notices of the Royal Astronomical Society] {10.1093/mnras/stae1160}, 531, 366

\bibitem[\protect\citeauthoryear{{Mummery}, {Jiang}  \& {Fabian}}{{Mummery} et~al.}{2024c}]{Mummery24PlungeB}
{Mummery} A.,  {Jiang} J.,   {Fabian} A.,  2024c, \mn@doi [\mnras] {10.1093/mnrasl/slae056}, \href {https://ui.adsabs.harvard.edu/abs/2024MNRAS.533L..83M} {533, L83}

\bibitem[\protect\citeauthoryear{{Noble}, {Krolik}  \& {Hawley}}{{Noble} et~al.}{2010}]{Noble10}
{Noble} S.~C.,  {Krolik} J.~H.,   {Hawley} J.~F.,  2010, \mn@doi [\apj] {10.1088/0004-637X/711/2/959}, \href {https://ui.adsabs.harvard.edu/abs/2010ApJ...711..959N} {711, 959}

\bibitem[\protect\citeauthoryear{{Noble}, {Krolik}, {Schnittman}  \& {Hawley}}{{Noble} et~al.}{2011}]{Noble11}
{Noble} S.~C.,  {Krolik} J.~H.,  {Schnittman} J.~D.,   {Hawley} J.~F.,  2011, \mn@doi [\apj] {10.1088/0004-637X/743/2/115}, \href {https://ui.adsabs.harvard.edu/abs/2011ApJ...743..115N} {743, 115}

\bibitem[\protect\citeauthoryear{{Novikov} \& {Thorne}}{{Novikov} \& {Thorne}}{1973}]{NovikovThorne73}
{Novikov} I.~D.,  {Thorne} 1973, in Black Holes (Les Astres Occlus). pp 343--450

\bibitem[\protect\citeauthoryear{{Page} \& {Thorne}}{{Page} \& {Thorne}}{1974}]{PageThorne74}
{Page} D.~N.,  {Thorne} 1974, \mn@doi [\apj] {10.1086/152990}, \href {https://ui.adsabs.harvard.edu/abs/1974ApJ...191..499P} {191, 499}

\bibitem[\protect\citeauthoryear{{Penna}, {McKinney}, {Narayan}, {Tchekhovskoy}, {Shafee}  \& {McClintock}}{{Penna} et~al.}{2010}]{Penna10}
{Penna} R.~F.,  {McKinney} J.~C.,  {Narayan} R.,  {Tchekhovskoy} A.,  {Shafee} R.,   {McClintock} J.~E.,  2010, \mn@doi [\mnras] {10.1111/j.1365-2966.2010.17170.x}, \href {https://ui.adsabs.harvard.edu/abs/2010MNRAS.408..752P} {408, 752}

\bibitem[\protect\citeauthoryear{{Potter}}{{Potter}}{2021}]{Potter21}
{Potter} W.~J.,  2021, \mn@doi [\mnras] {10.1093/mnras/stab636}, \href {https://ui.adsabs.harvard.edu/abs/2021MNRAS.503.5025P} {503, 5025}

\bibitem[\protect\citeauthoryear{{Ramachandran} et~al.,}{{Ramachandran} et~al.}{2025}]{Ramachandran25}
{Ramachandran} V.,  et~al., 2025, \mn@doi [arXiv e-prints] {10.48550/arXiv.2504.05885}, \href {https://ui.adsabs.harvard.edu/abs/2025arXiv250405885R} {p. arXiv:2504.05885}

\bibitem[\protect\citeauthoryear{{Reynolds}}{{Reynolds}}{2019}]{Reynolds19}
{Reynolds} C.~S.,  2019, \mn@doi [Nature Astronomy] {10.1038/s41550-018-0665-z}, \href {https://ui.adsabs.harvard.edu/abs/2019NatAs...3...41R} {3, 41}

\bibitem[\protect\citeauthoryear{{Reynolds} \& {Begelman}}{{Reynolds} \& {Begelman}}{1997}]{Reynolds97}
{Reynolds} C.~S.,  {Begelman} M.~C.,  1997, \mn@doi [\apj] {10.1086/304703}, \href {https://ui.adsabs.harvard.edu/abs/1997ApJ...488..109R} {488, 109}

\bibitem[\protect\citeauthoryear{{Rule}, {Mummery}, {Balbus}, {Stone}  \& {Zhang}}{{Rule} et~al.}{2025}]{Rule25}
{Rule} J.,  {Mummery} A.,  {Balbus} S.,  {Stone} J.,   {Zhang} L.,  2025, arXiv e-prints, p. arXiv:

\bibitem[\protect\citeauthoryear{{Salpeter}}{{Salpeter}}{1964}]{Salpeter64}
{Salpeter} E.~E.,  1964, \mn@doi [\apj] {10.1086/147973}, \href {https://ui.adsabs.harvard.edu/abs/1964ApJ...140..796S} {140, 796}

\bibitem[\protect\citeauthoryear{{Salvesen} \& {Miller}}{{Salvesen} \& {Miller}}{2021}]{Salvesen21}
{Salvesen} G.,  {Miller} J.~M.,  2021, \mn@doi [\mnras] {10.1093/mnras/staa3325}, \href {https://ui.adsabs.harvard.edu/abs/2021MNRAS.500.3640S} {500, 3640}

\bibitem[\protect\citeauthoryear{{Schnittman} \& {Krolik}}{{Schnittman} \& {Krolik}}{2009}]{Schnittman09}
{Schnittman} J.~D.,  {Krolik} J.~H.,  2009, \mn@doi [\apj] {10.1088/0004-637X/701/2/1175}, \href {https://ui.adsabs.harvard.edu/abs/2009ApJ...701.1175S} {701, 1175}

\bibitem[\protect\citeauthoryear{{Schnittman}, {Krolik}  \& {Noble}}{{Schnittman} et~al.}{2016}]{Schnittman16}
{Schnittman} J.~D.,  {Krolik} J.~H.,   {Noble} S.~C.,  2016, \mn@doi [\apj] {10.3847/0004-637X/819/1/48}, \href {https://ui.adsabs.harvard.edu/abs/2016ApJ...819...48S} {819, 48}

\bibitem[\protect\citeauthoryear{{Shafee}, {McKinney}, {Narayan}, {Tchekhovskoy}, {Gammie}  \& {McClintock}}{{Shafee} et~al.}{2008}]{Shafee08}
{Shafee} R.,  {McKinney} J.~C.,  {Narayan} R.,  {Tchekhovskoy} A.,  {Gammie} C.~F.,   {McClintock} J.~E.,  2008, \mn@doi [\apjl] {10.1086/593148}, \href {https://ui.adsabs.harvard.edu/abs/2008ApJ...687L..25S} {687, L25}

\bibitem[\protect\citeauthoryear{{Shakura} \& {Sunyaev}}{{Shakura} \& {Sunyaev}}{1973}]{SS73}
{Shakura} N.~I.,  {Sunyaev} R.~A.,  1973, \aap, \href {https://ui.adsabs.harvard.edu/abs/1973A&A....24..337S} {24, 337}

\bibitem[\protect\citeauthoryear{{Steiner} et~al.,}{{Steiner} et~al.}{2024}]{Steiner24}
{Steiner} J.~F.,  et~al., 2024, \mn@doi [\apjl] {10.3847/2041-8213/ad58e4}, \href {https://ui.adsabs.harvard.edu/abs/2024ApJ...969L..30S} {969, L30}

\bibitem[\protect\citeauthoryear{{Stone} et~al.,}{{Stone} et~al.}{2024}]{Stone24}
{Stone} J.~M.,  et~al., 2024, \mn@doi [arXiv e-prints] {10.48550/arXiv.2409.16053}, \href {https://ui.adsabs.harvard.edu/abs/2024arXiv240916053S} {p. arXiv:2409.16053}

\bibitem[\protect\citeauthoryear{{Thorne}}{{Thorne}}{1974}]{Thorne1974}
{Thorne} K.~S.,  1974, \mn@doi [\apj] {10.1086/152991}, \href {https://ui.adsabs.harvard.edu/abs/1974ApJ...191..507T} {191, 507}

\bibitem[\protect\citeauthoryear{{Weisskopf} et~al.,}{{Weisskopf} et~al.}{2022}]{Weisskopf22}
{Weisskopf} M.~C.,  et~al., 2022, \mn@doi [Journal of Astronomical Telescopes, Instruments, and Systems] {10.1117/1.JATIS.8.2.026002}, \href {https://ui.adsabs.harvard.edu/abs/2022JATIS...8b6002W} {8, 026002}

\bibitem[\protect\citeauthoryear{{Wilkins}, {Reynolds}  \& {Fabian}}{{Wilkins} et~al.}{2020}]{Wilkins20}
{Wilkins} D.~R.,  {Reynolds} C.~S.,   {Fabian} A.~C.,  2020, \mn@doi [\mnras] {10.1093/mnras/staa628}, \href {https://ui.adsabs.harvard.edu/abs/2020MNRAS.493.5532W} {493, 5532}

\bibitem[\protect\citeauthoryear{{Wilms}, {Allen}  \& {McCray}}{{Wilms} et~al.}{2000}]{Wilms00}
{Wilms} J.,  {Allen} A.,   {McCray} R.,  2000, \mn@doi [\apj] {10.1086/317016}, \href {https://ui.adsabs.harvard.edu/abs/2000ApJ...542..914W} {542, 914}

\bibitem[\protect\citeauthoryear{{Zhao} et~al.,}{{Zhao} et~al.}{2021}]{Zhao21}
{Zhao} X.,  et~al., 2021, \mn@doi [\apj] {10.3847/1538-4357/abbcd6}, \href {https://ui.adsabs.harvard.edu/abs/2021ApJ...908..117Z} {908, 117}

\bibitem[\protect\citeauthoryear{{Zhu}, {Davis}, {Narayan}, {Kulkarni}, {Penna}  \& {McClintock}}{{Zhu} et~al.}{2012}]{Zhu12}
{Zhu} Y.,  {Davis} S.~W.,  {Narayan} R.,  {Kulkarni} A.~K.,  {Penna} R.~F.,   {McClintock} J.~E.,  2012, \mn@doi [\mnras] {10.1111/j.1365-2966.2012.21181.x}, \href {https://ui.adsabs.harvard.edu/abs/2012MNRAS.424.2504Z} {424, 2504}

\makeatother
\end{thebibliography}

\appendix
\section{Comparison to simulations with different spin values}\label{app:sims}
\cite{Dhang25} ran a suite of simulations for different black hole spin parameters, namely $a = -0.9, 0, 0.3, 0.5, 0.9, 0.99$. This then represents an important data set with which to test the models of \cite{MummeryBalbus2023}.  

\begin{figure*}
    \centering
    \includegraphics[width=0.49\linewidth]{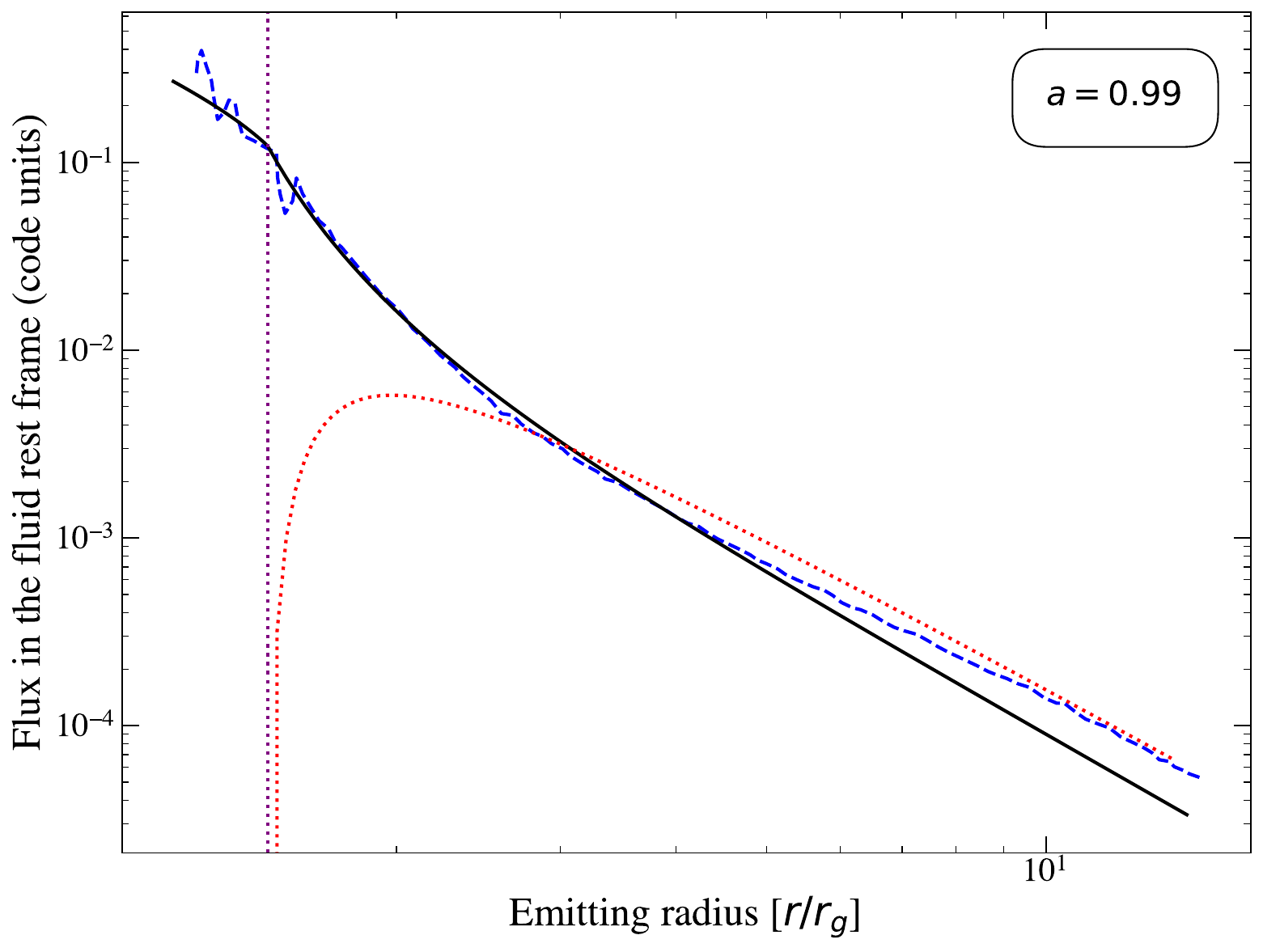}
    \includegraphics[width=0.49\linewidth]{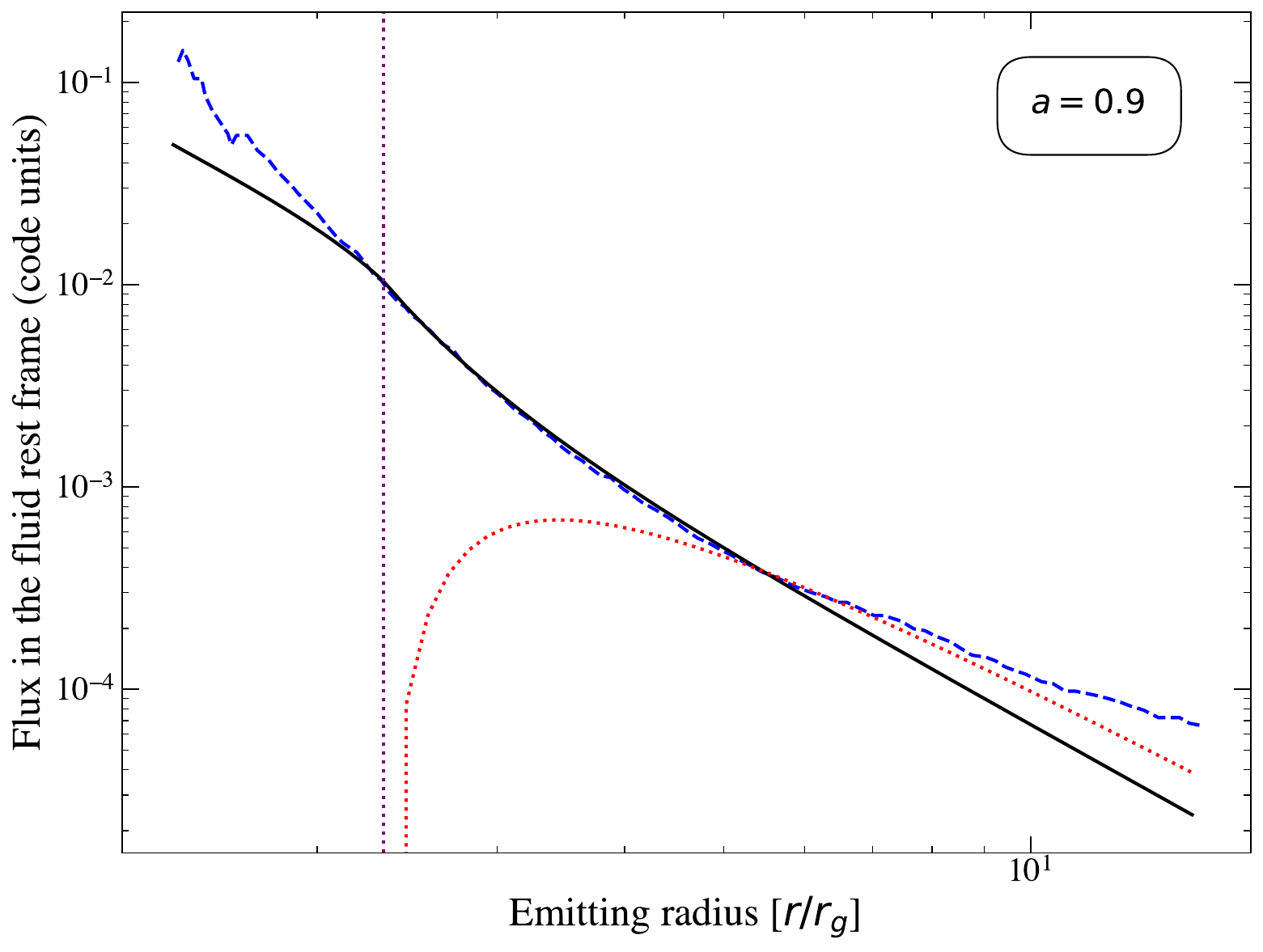}
    \includegraphics[width=0.49\linewidth]{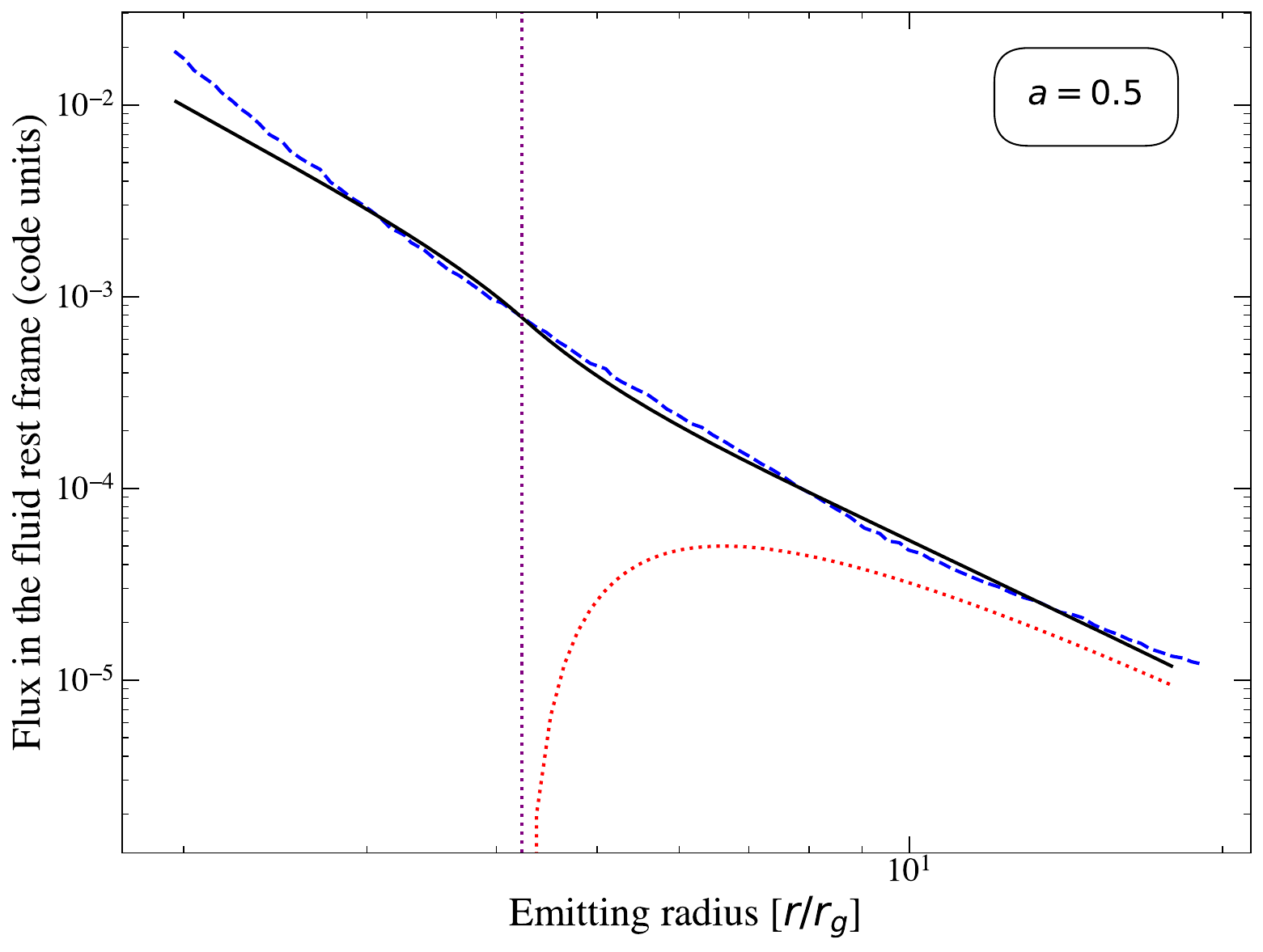}
    \includegraphics[width=0.49\linewidth]{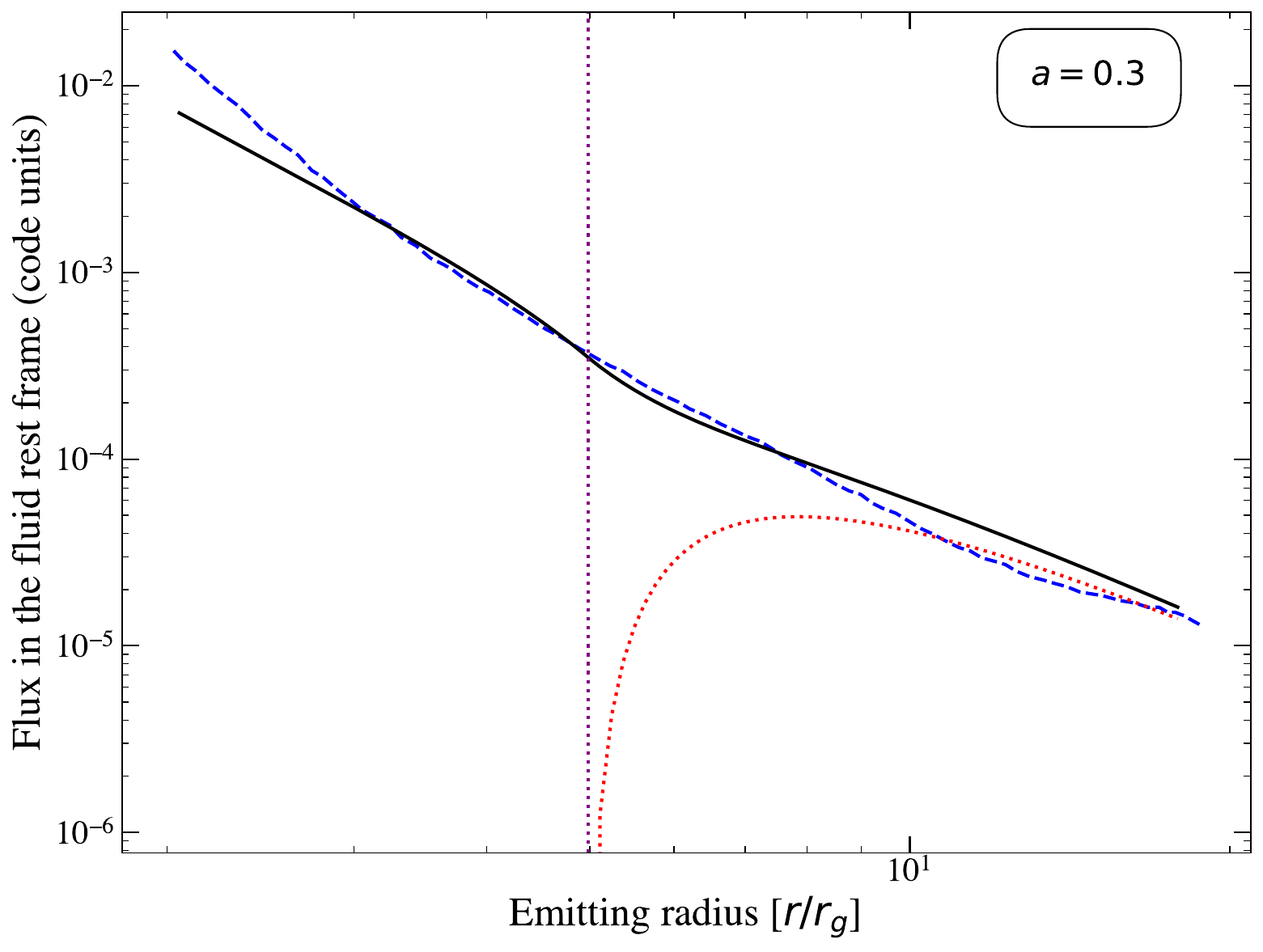}
    \includegraphics[width=0.49\linewidth]{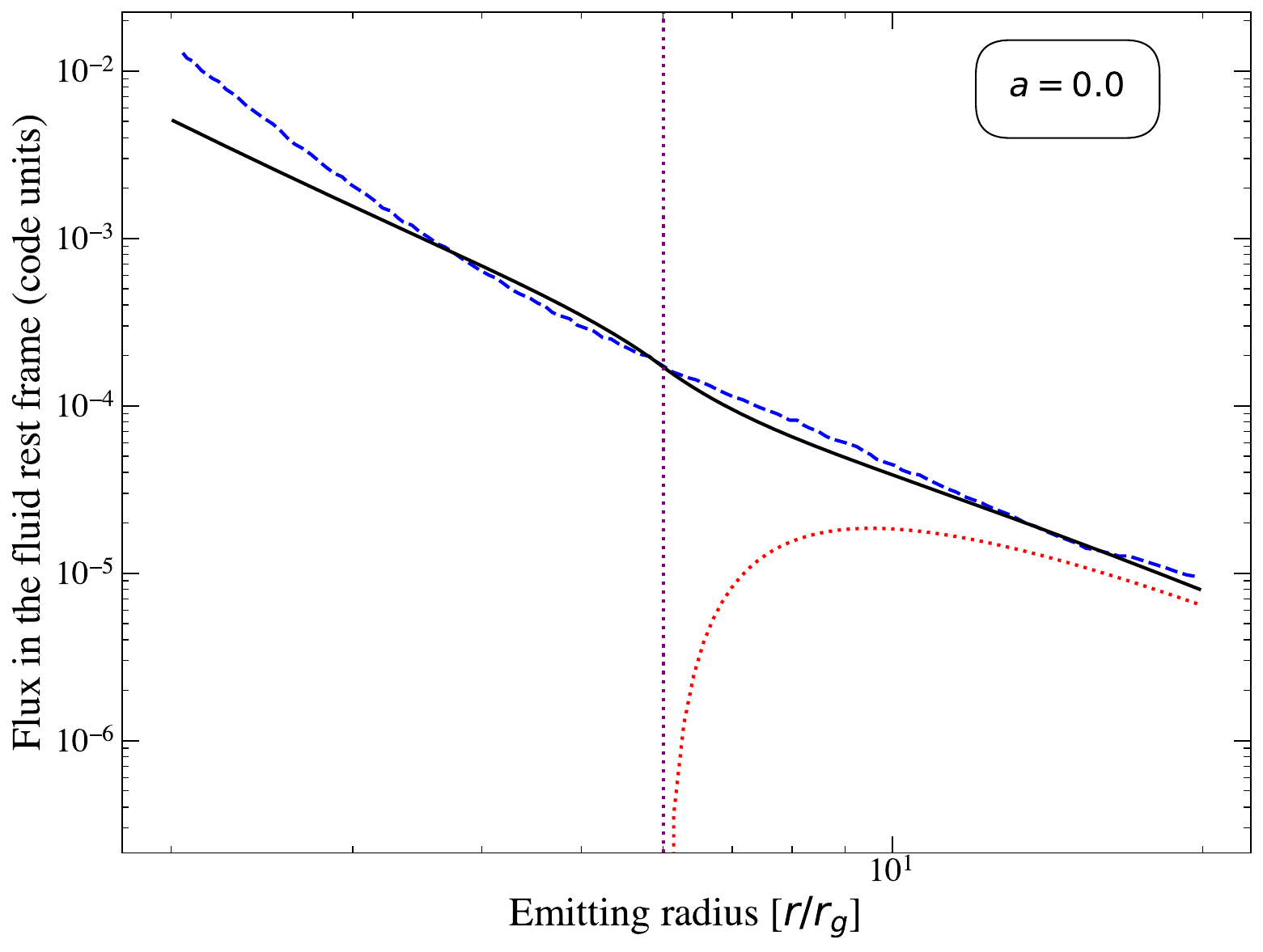}
    \includegraphics[width=0.49\linewidth]{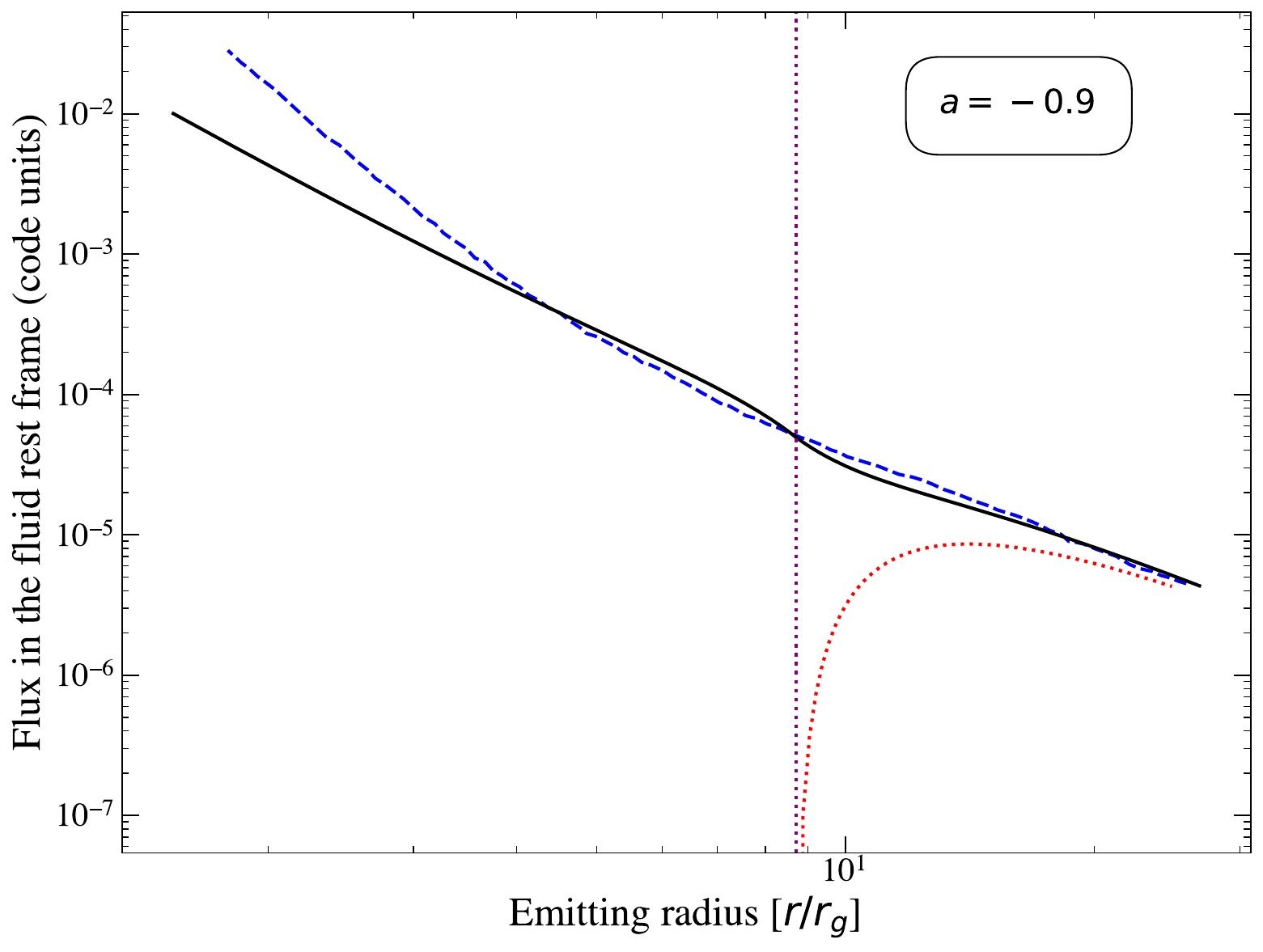}
    \caption{The results of the \citealt{Dhang25} simulations with different values of the black hole spin parameter (blue dashed curves), compared to classical models (red dotted curves) and the model of \citealt{MummeryBalbus2023} (black solid curves).  The flux shown here is that calculated in the fluid's rest frame (and so is an upper bound on the observable flux), and is plotted against radius. The ISCO radius (the beginning of the fluid's plunging region) is denoted on each plot by a vertical dotted purple curve.  Each simulation result is very poorly described by classical models, and is much better described by the new model of \citealt{MummeryBalbus2023}. See text for further details.  }
    \label{fig:all_spin_fluid}
\end{figure*}

In Figure \ref{fig:all_spin_fluid} we compare the  \cite{Dhang25} simulations (with different values of the black hole spin parameter; blue dashed curves), to classical \cite{NovikovThorne73} models (red dotted curves) and the model of \citealt{MummeryBalbus2023} (black solid curves).  The flux shown here is that calculated in the fluid's rest frame (and so is an upper bound on the observable flux), and is plotted against radius. The ISCO radius (the beginning of the fluid's plunging region) is denoted on each plot by a vertical dotted purple curve.  Each simulation result is very poorly described by classical models, and is much better described by the new model of \cite{MummeryBalbus2023}. Again, in reproducing this data we explored regions of parameter space described by $\delta_{\cal J} \sim 0.05-0.1$ and $u_I \sim 0.1c-0.2c$, and a range of parameter values in this broad area were adequate. 

The flux in the simulations of \cite{Dhang25} is even higher than predicted in the model of \cite{MummeryBalbus2023} on near-horizon $(r-r_+ \sim 1r_g)$ scales. Inspecting the simulation data we believe this is because the near horizon magnetic field strength is strong enough to push the fluid off the near-geodesic plunge assumed in the model of \cite{MummeryBalbus2023} on horizon scales. This slowing down of the flow reduces the radial expansion of the fluid, and reduces adiabatic losses, resulting in a hotter (and therefore larger fluid-frame flux) flow on small scales. 

\begin{figure*}
    \centering
    \includegraphics[width=0.49\linewidth]{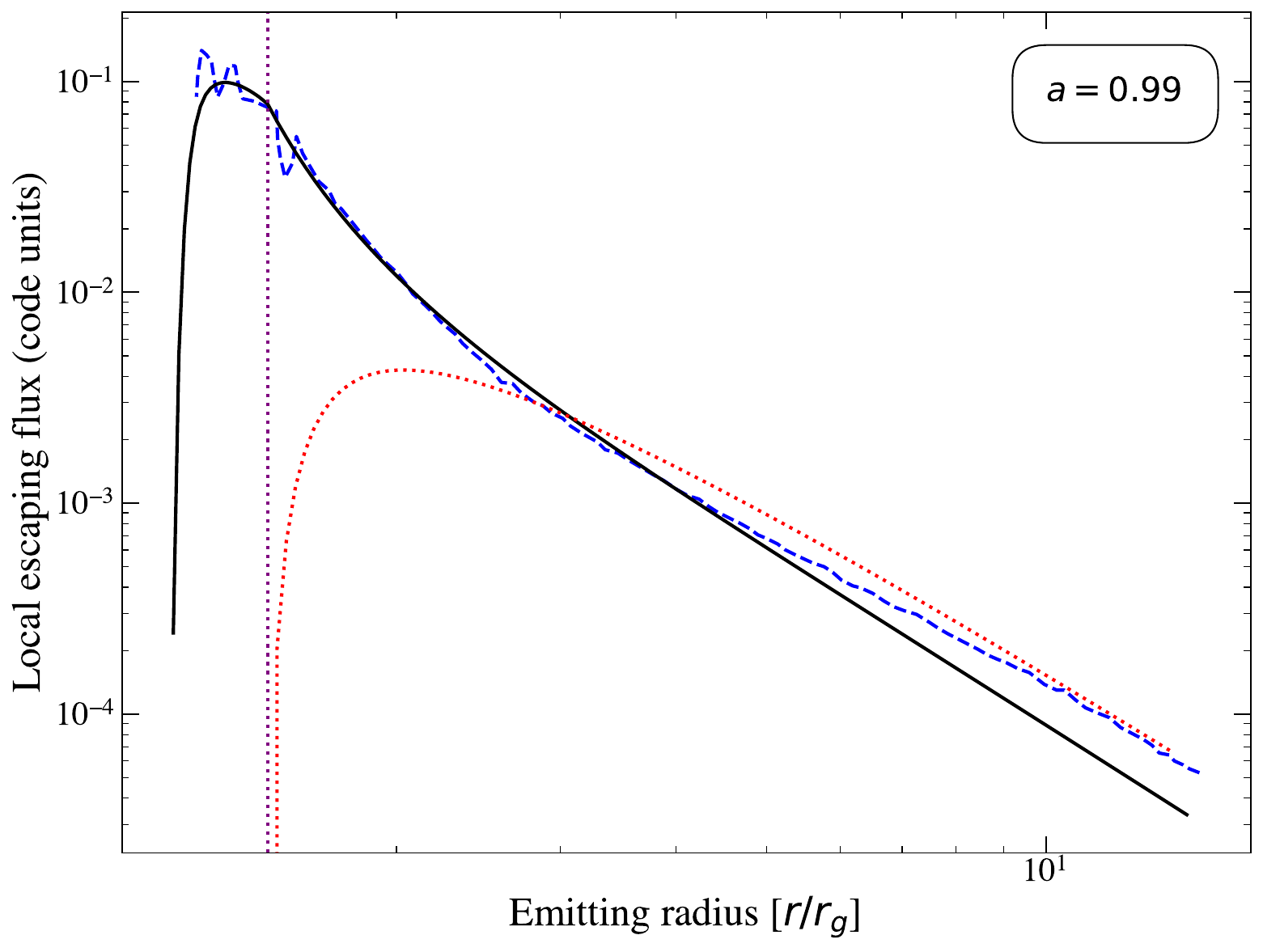}
    \includegraphics[width=0.49\linewidth]{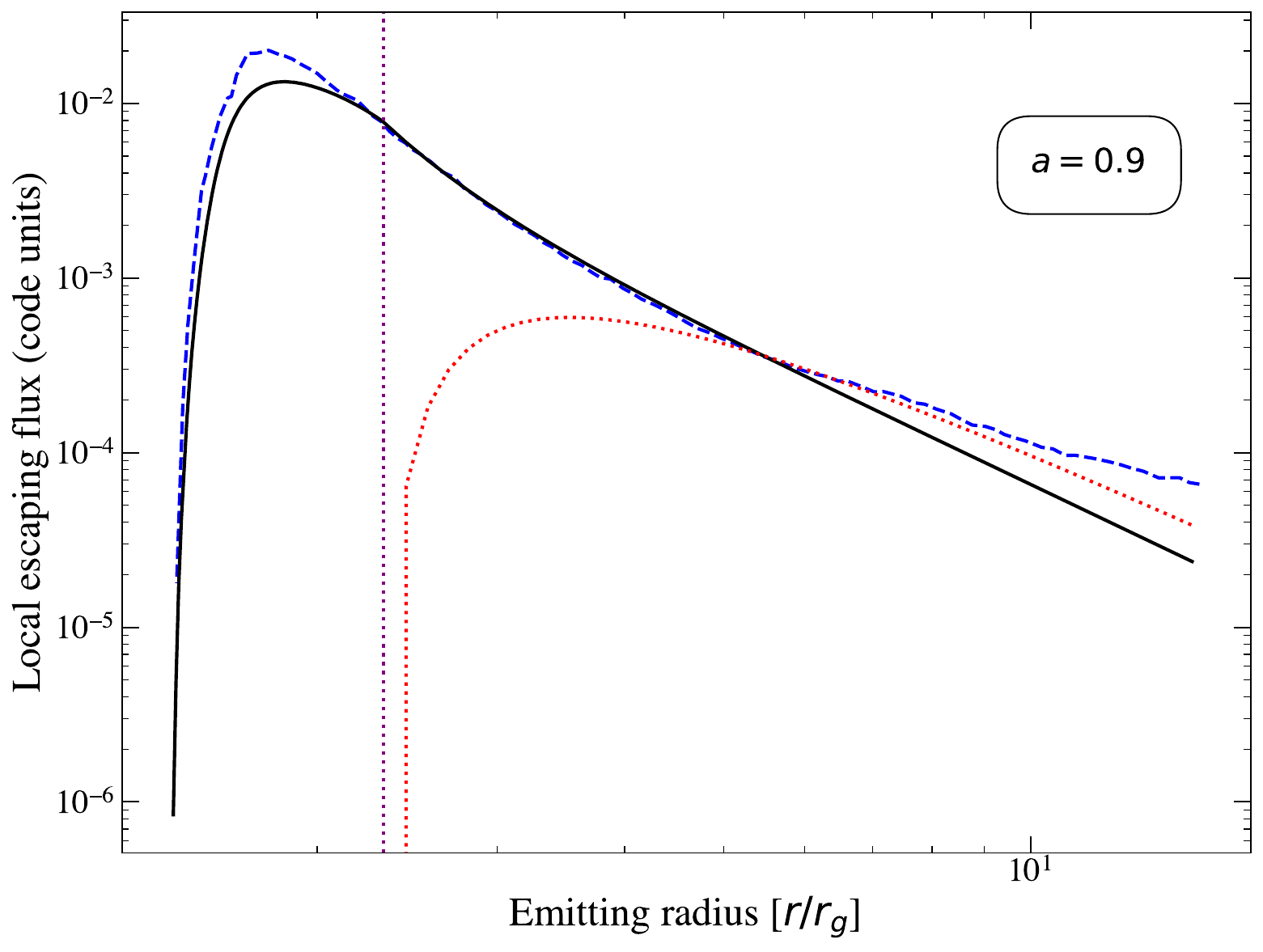}
    \includegraphics[width=0.49\linewidth]{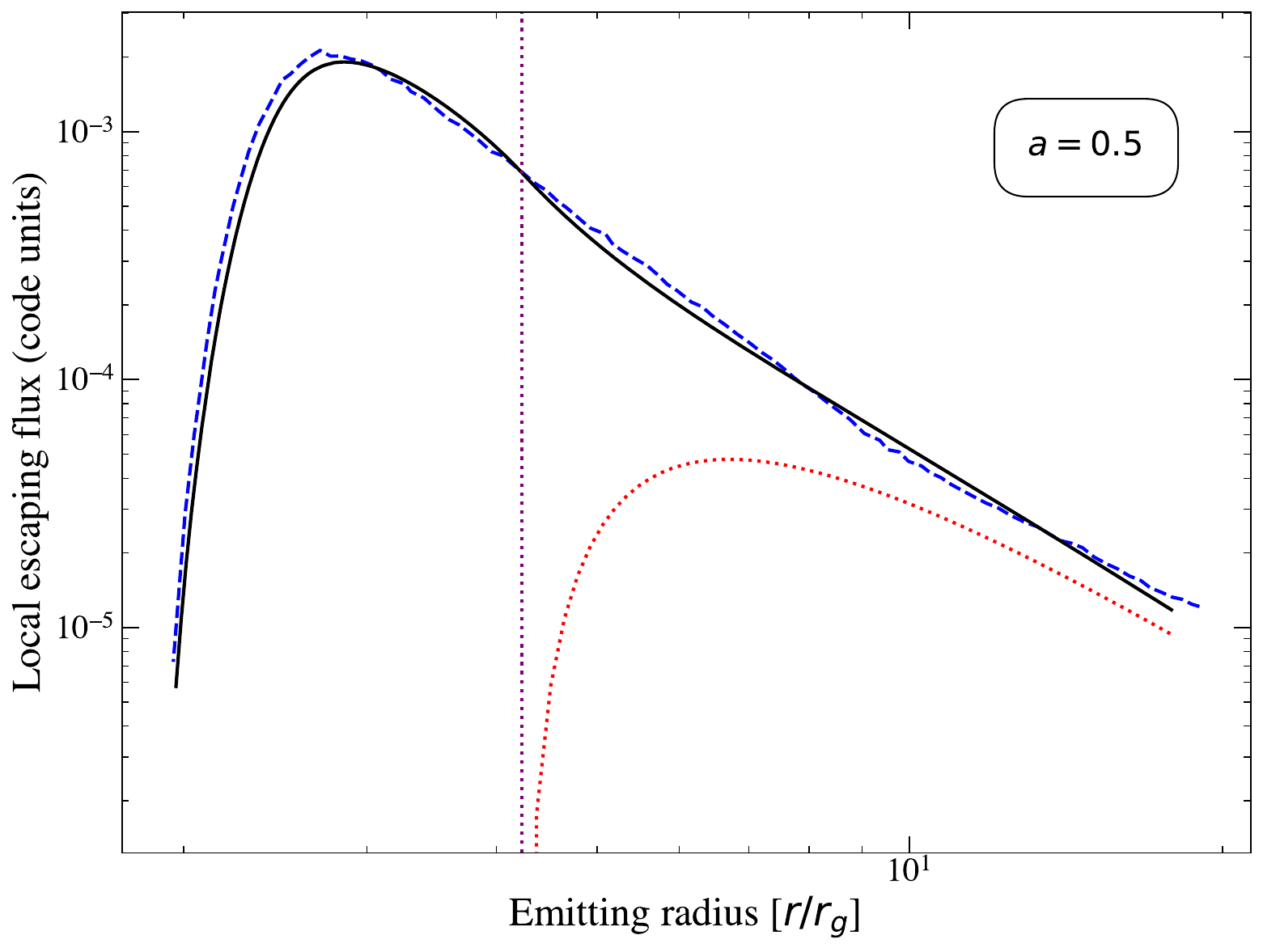}
    \includegraphics[width=0.49\linewidth]{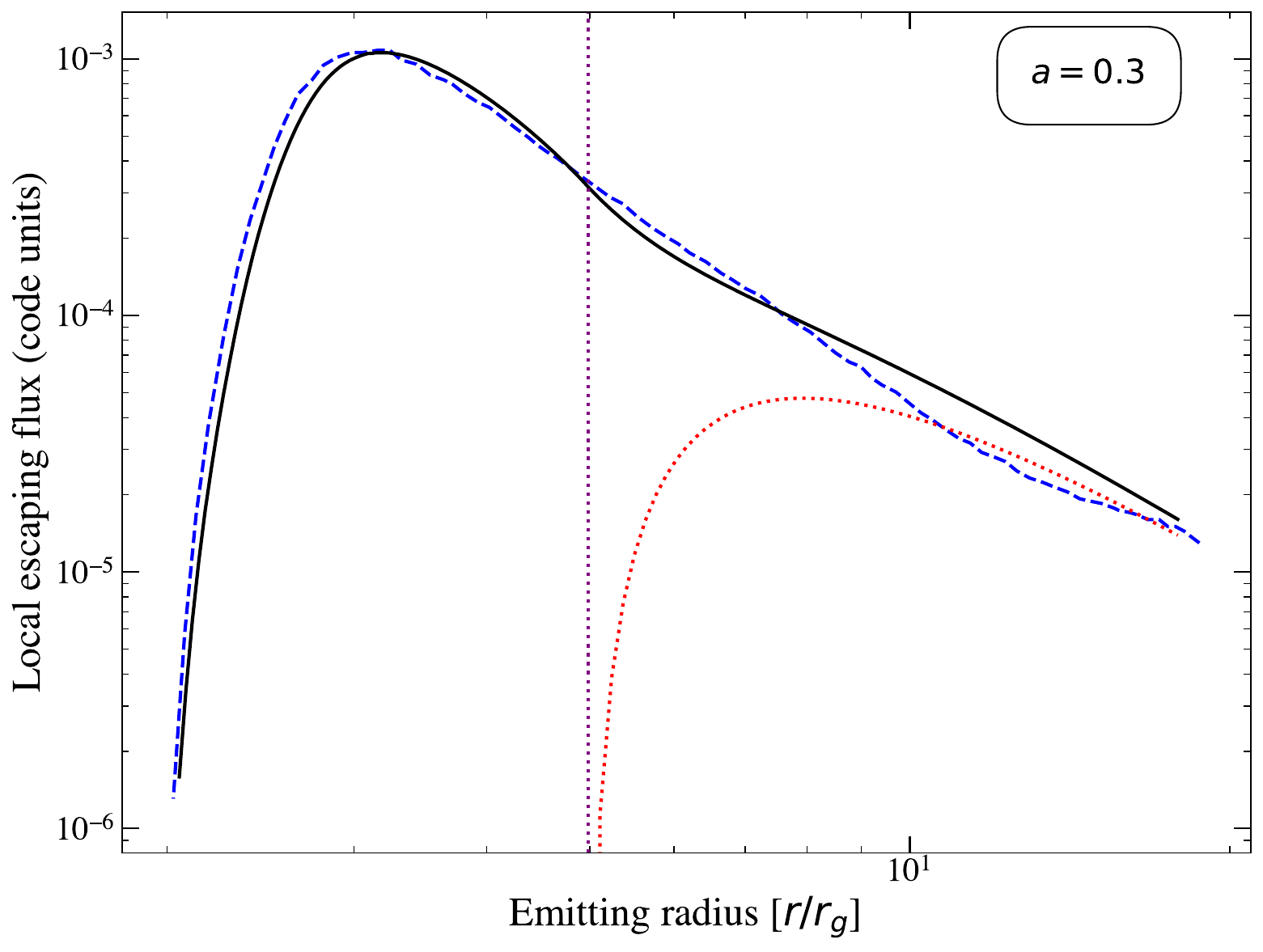}
    \includegraphics[width=0.49\linewidth]{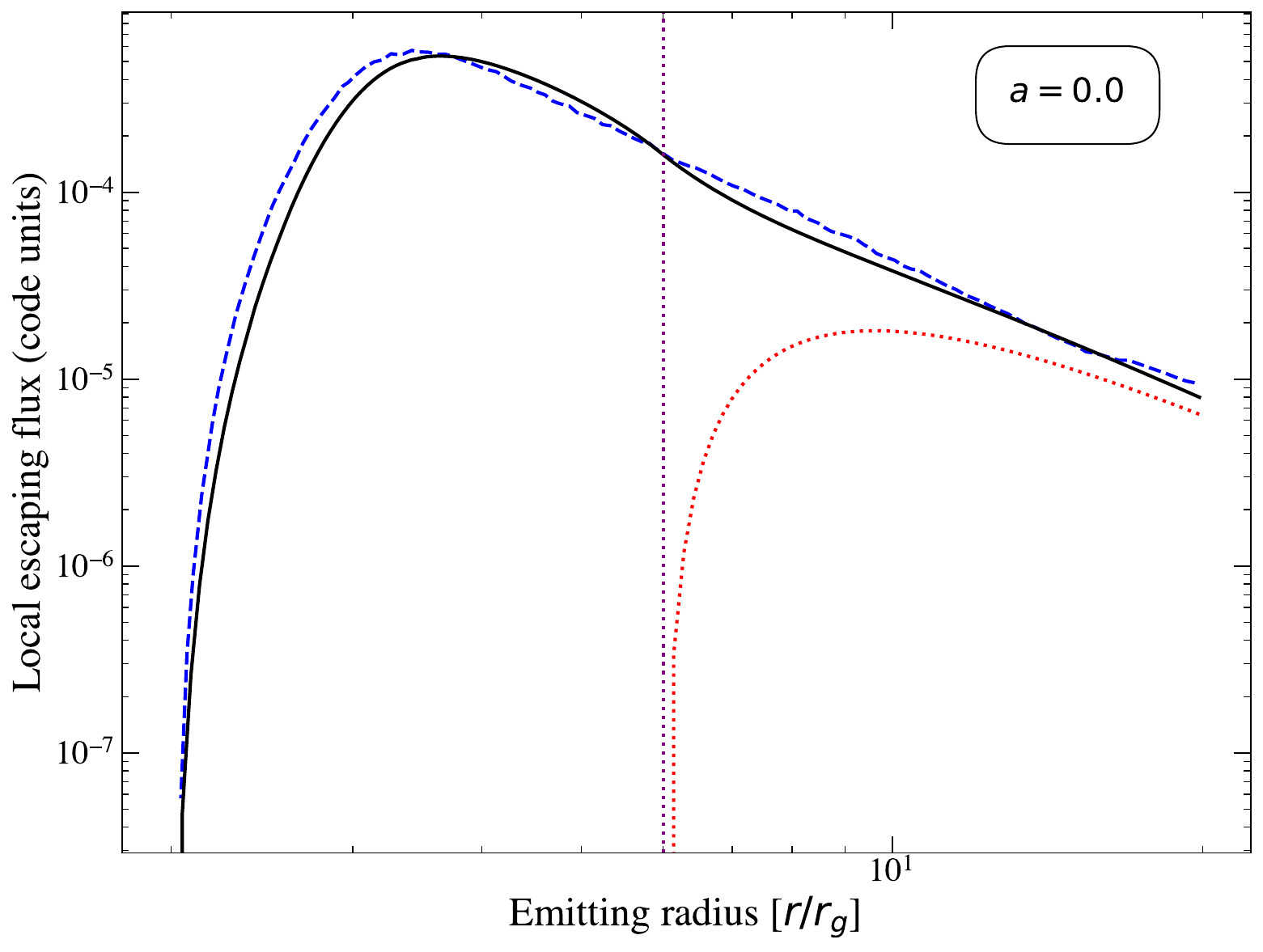}
    \includegraphics[width=0.49\linewidth]{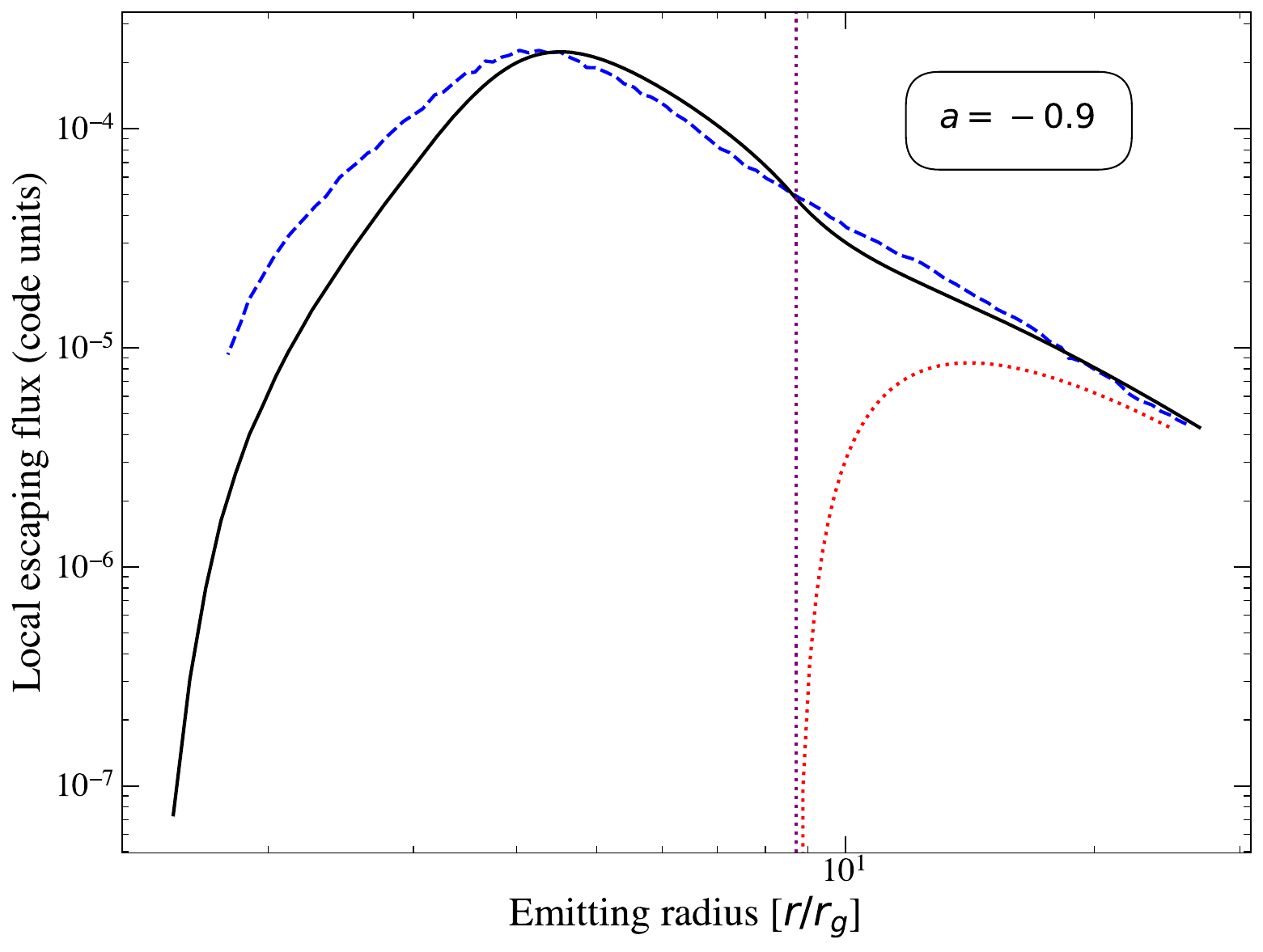}
    \caption{The flux profiles of Figure \ref{fig:all_spin_fluid} but now taking into account the capture of photons by the black hole. Despite the fluid's rest frame flux growing down to the horizon, the increasing fraction of photons captured by the black hole as the fluid accelerates towards the horizon results in a turn-over in the escaping flux at small $r\lesssim 3r_g$ radii.  }
    \label{fig:all_spin_escape}
\end{figure*}

Very little of this final flux escapes the black hole's spacetime however, and is predominantly captured by the black hole. In Figure \ref{fig:all_spin_escape} we display the radial profiles of the escaping flux for the six different spin parameters. Despite the fluid's rest frame flux growing down to the horizon, the increasing fraction of photons captured by the black hole as the fluid accelerates towards the horizon results in a turn-over in the escaping flux at small $r\lesssim 3r_g$ radii. 

\end{document}